\documentclass[prb,twocolumn,superscriptaddress,eqsecnum,a4,footinbib]{revtex4-1}

\usepackage{amsmath}
\usepackage{amssymb}
\usepackage[pdftex]{graphicx}
\usepackage{booktabs}
\usepackage{multirow}
\usepackage{eurosym}
\usepackage{hyperref}
\usepackage{color}
\usepackage{ifthen}
\usepackage[normalem]{ulem}
\usepackage{cancel,soul}

\newcommand{\vv}[1]{\boldsymbol{#1}}

\def\:={\,\raisebox{0.85pt}{.}\hspace{-2.78pt}\raisebox{2.85pt}{.}\!\!=\,}
\def\=:{\,=\!\!\raisebox{0.85pt}{.}\hspace{-2.78pt}\raisebox{2.85pt}{.}\,}

\begin{document}

\title{
Spiral order from orientationally correlated random bonds in classical
$XY$ models
      }

\author{Andrea Scaramucci}
\affiliation{Laboratory for Scientific Development and Novel Materials, Paul Scherrer Institut, 5235, Villigen PSI, Switzerland}
\author{Hiroshi Shinaoka}
\affiliation{Institute for Theoretical Physics, ETH Z\"{u}rich, CH-8093 Z\"{u}rich, Switzerland}
\affiliation{Department of Physics, University of Fribourg, 1700 Fribourg, Switzerland}
\affiliation{Department of Physics, Saitama University, 338-8570 Saitama, Japan}
\author{Maxim V. Mostovoy}
\affiliation{Zernike Institute for Advanced Materials,
University of Groningen, Nijenborgh 4,
9747 AG, Groningen, The Netherlands}
\author{Rui Lin}
\affiliation{Institute for Theoretical Physics, ETH Z\"{u}rich, CH-8093 Z\"{u}rich, Switzerland}
\author{Christopher Mudry}
\affiliation{Condensed  Matter  Theory  Group,  Paul  Scherrer  Institute,  CH-5232  Villigen  PSI,  Switzerland}
\affiliation{Institute of Physics,
\'Ecole Polytechnique F\'ed\'erale de Lausanne (EPFL), CH-1015
Lausanne, Switzerland}
\author{Markus M\"{u}ller}
\affiliation{Condensed  Matter  Theory  Group,  Paul  Scherrer  Institute,  CH-5232  Villigen  PSI,  Switzerland}
\affiliation{The Abdus Salam International Centre for Theoretical Physics, 34151, Trieste, Italy}

\begin{abstract} 
We discuss the stability of ferromagnetic long-range order in
three-dimensional classical $XY$ 
ferromagnets upon substitution of
a small subset of equally oriented bonds 
by impurity bonds, on which the ferromagnetic exchange $J^{\,}_{\perp}>0$  
is replaced by a strong antiferromagnetic coupling  $J^{\,}_{\mathrm{imp}}<0$.
In the impurity-free limit, the effective low-energy Hamiltonian is that
of spin waves. In the presence of a single, sufficiently strongly frustrating impurity bond, 
the ground state is two-fold degenerate, 
corresponding to either clockwise or anticlockwise
canting of the spins in the vicinity of the impurity bond.
For a small, but  finite concentration of impurity bonds, 
the effective low-energy Hamiltonian is that of Ising variables encoding 
the sense of rotation of the local canting around the impurities.
Those degrees of freedom interact pairwise through 
a dipolar interaction mediated by spin waves.
A  spatially random distribution of impurities leads to a ferromagnetic Ising ground state, which
indicates the instability of the $XY$ ferromagnet
towards a spiral state, with wave vector and transition temperature both proportional 
to the concentration of impurity bonds.
This mechanism of ``spiral order by disorder''
is relevant for magnetic materials such as YBaCuFeO$_5$, for which our theory
predicts a ratio between the spiral ordering temperature
and the modulus of the spiral wavevector close to the
measured ones.
\end{abstract}

\maketitle


\section{Introduction}
\label{sec: Introduction}

Insulating magnets supporting  {long-range magnetic} spiral order 
are of technological interest as they can display ``magnetically''
induced ferroelectricity
\cite{Katsura:2005cy,cheong_multiferroics:_2007,khomskii_trend:_2009,TokuraAdvMat2010}.
In prototypical spin-spiral multiferroics, e.g., RMnO$_3$ (R=Tb$^{3+}$,
Dy$^{3+}$, etc.)
\cite{kenzelmann_magnetic_2005,goto_ferroelectricity_2004}, 
a magnetic spiral phase can be stabilized by the competition 
between nearest-neighbor and further-neighbor magnetic
exchange interactions with opposite signs
\cite{mochizuki_microscopic_2009,lyons_method_1960}.  
However, the resulting frustration only induces spiral states 
if further-neighbor couplings are sufficiently strong 
as compared to nearest-neighbor couplings. The {latter}
are typically much bigger in magnitude, 
except under special circumstances that lead
to their suppression. In such exceptional cases, the characteristic exchange 
scale is set by the further-neighbor interactions and is thus very weak, 
entailing a low spiral ordering temperature.

In order to engineer magnetic insulators with 
magnetic spiral order establishing at
high temperatures, it is of fundamental interest to investigate
analogous mechanisms.  
An interesting route was suggested by  the study of Ivanov et {al.}
\cite{ivanov_square-lattice_1996} 
who considered a Heisenberg antiferromagnet 
on a square lattice, in which every other horizontal nearest-neighbor bond 
in a staggered pattern was replaced by a ferromagnetic coupling.
Sufficiently strongly frustrating bonds were shown 
to induce a magnetic spiral order.
From the experimental side, there are interesting hints that a similar
mechanism might be tied to the presence of disorder. 
Indeed, certain insulating compounds containing some degree
of chemical disorder were reported to stabilize magnetic spiral order
\cite{cox_magnetic_1963,utsumi_superexchange_2007,%
morin_incommensurate_2015, MorinNatComm2016, kundys_multiferroicity_2009} 
at high temperatures. For example, {the} transition
temperatures to the magnetic spiral phase were found to range from
180 K to 310 K \cite{kundys_multiferroicity_2009,%
morin_incommensurate_2015,caignaert_crystal_1995,%
kawamura_high-temperature_2010,%
ruiz-aragon_low-temperature_1998}
{in YBaCuFeO$_5$},
whereby several further characteristics of the spiral depend
on the degree of disorder.
This empiric observation suggests the possibility that, for some
materials, a magnetic spiral order might be induced by some ``impurity bonds'' 
formed by nearest-neighbor magnetic ions whose exchange
coupling frustrates the order that would establish in their
absence. Recent Monte Carlo simulations have confirmed this conjecture
in a model describing YBaCuFeO$_5$ 
with disorder in the spatial location of the magnetic Cu and Fe ions
\cite{Scaramucci_2016}.
The latter was assumed to result in a small concentration of locally
frustrating bonds along the $c$-direction, which indeed was shown to
induce magnetic spiral order in an experimentally relevant window of
parameters.

\begin{figure}[!]
\includegraphics[width=0.45\textwidth]{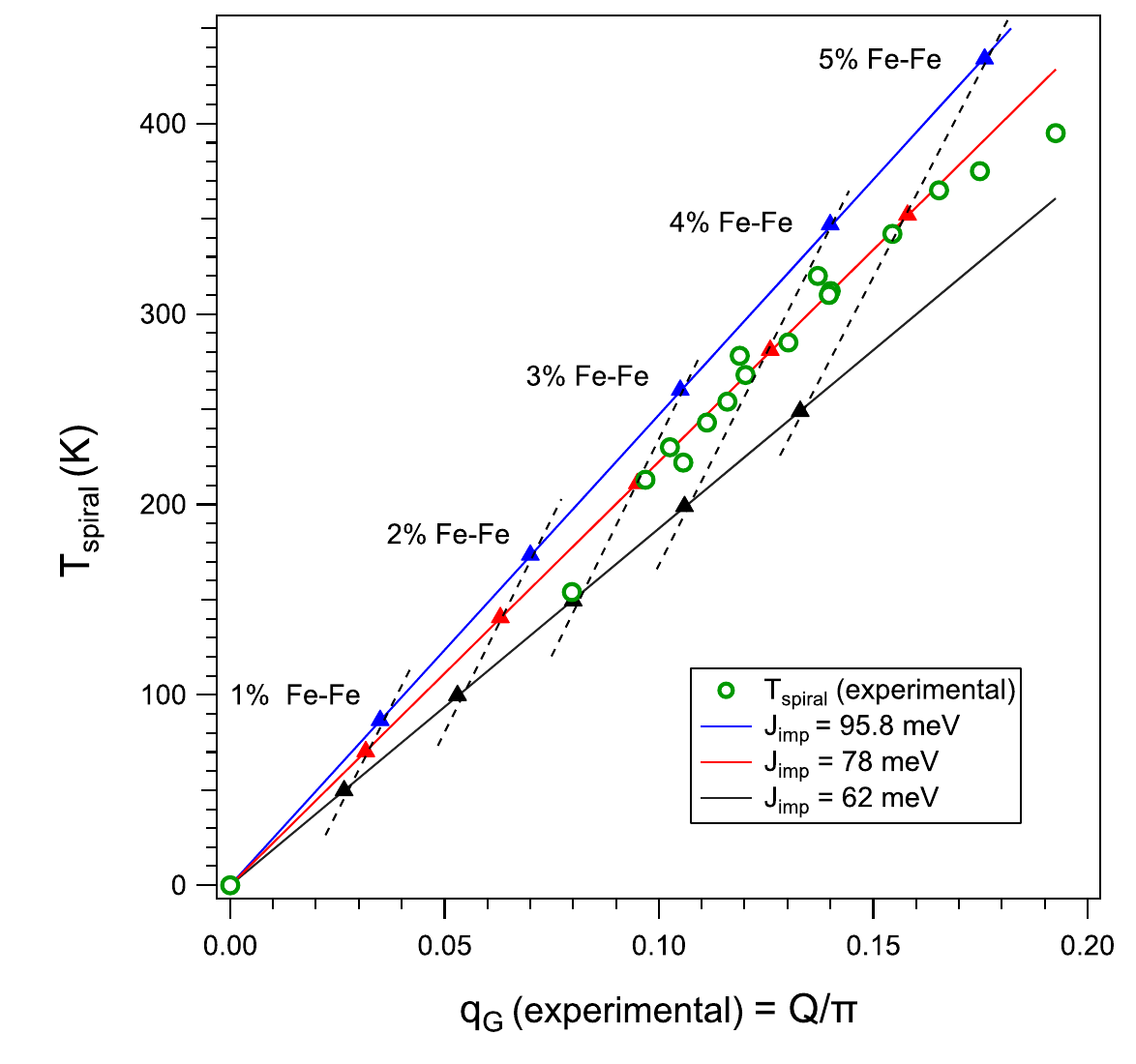}
\caption{(Color online)
Experimental data (circles) from Ref.\ \onlinecite{MorinNatComm2016}
for the compound YBaCuFeO$_5$, obtained by different annealing rates
that control the fraction of bipyramids hosting magnetically
frustrating Fe-Fe pairs instead of low energy Fe-Cu pairs. The
transition temperature to the spiral state is plotted versus the
spiral wavevector measured at low temperature. Our theory explains the
linear relationship between the two and predicts the slope as a
function of the three exchange couplings. We plot the predicted slope
for different values of the exchange on the driving frustrating bonds,
fixing the other two exchanges to the values estimated in
Ref.~\onlinecite{Scaramucci_2016}. The larger the frustrating
coupling, and the higher the fraction of frustrating Fe-Fe bonds in
the bipyramids of YBaCuFeO$_5$, the larger the transition temperature
and the spiral wavevector. Each triplet of triangular symbols joined by
a dashed line corresponds to the same concentration of
 frustrating bonds in our theory.
        }
\label{Fig:expth}
\end{figure}

In this paper, we describe and study the general mechanism that renders 
the ferromagnetic long-range order of classical $XY$ spins 
 unstable towards spiral order, 
{when} a finite fraction of the ferromagnetic
interactions is replaced by sufficiently strong 
antiferromagnetic exchange couplings.
In the end, we will confront the theory with experimental data,
as shown in Fig.~\ref{Fig:expth}, with good quantitative agreement.

The general physical mechanism at work is the following.  We consider
a geometrically unfrustrated lattice $\Lambda$ in $d>2$ dimensions,
hosting isotropic spins with a continuous symmetry. The symmetry is broken
spontaneously at low temperatures, which implies the existence of
Goldstone modes.  Dilute but strong impurity bonds embedded in this
lattice can induce local cantings which behave as ``dipole type''
defects with an Ising degree of freedom associated to them. The
Goldstone modes mediate an interaction between the defects, 
decaying as $r^{-d}$ for large separation.
Correlations in the distribution of such impurity bonds 
(e.g., a restriction to bonds that point in a single direction) 
may ensure a sufficiently non-frustrated pairwise interaction
between these defects so as to favor long-range ferromagnetic order in 
the orientation of the local cantings.
Such long-range Ising order entails a global twist 
of the ferromagnetic order parameter density, and thus a magnetic spiral,
as the local magnetization twists in the same sense across
every impurity bond. The wave vector of the resulting magnetic spiral
is proportional to the magnetization density of the Ising degrees of
freedom, and thus, to the density of impurity bonds.  We will show
that such a spiral state is the ground state of the
$XY$ system for a rather wide range of parameters  of the impurity
bond distribution. 

For simplicity, we consider a cubic host lattice $\Lambda$
embedded in three-dimensional
Euclidean space with the Cartesian coordinates $x$, $y$, and $z$.
We impose a tetragonal symmetry by choosing the 
ferromagnetic nearest-neighbor exchange to be
$J^{\,}_{\parallel}>0$ for couplings in the {$x$-$y$} plane and 
$J^{\,}_{\perp}>0$ for bonds oriented along the $z$-axis.
We further consider a set of impurity bonds, which form a dilute
subset of the nearest-neighbor bonds that are directed along the
{$c$}-direction of the cubic host lattice $\Lambda$.
For
each impurity bond, the ferromagnetic $J^{\,}_{\perp}>0$
is replaced by the antiferromagnetic
exchange coupling $J^{\,}_{\mathrm{imp}}<0$.

A single impurity bond does not destroy the ferromagnetic 
long-range order of the ground state. However, it does result
in a canting of the classical $XY$ spins in the vicinity of the
impurity bond, provided that the local frustration is sufficiently strong, 
i.e., $|J^{\,}_{\mathrm{imp}}|\geq J^{\,}_{\mathrm{c}}$ for some
threshold value $J^{\,}_{\mathrm{c}}>0$. 
Under these conditions (and with fixed boundary conditions at infinity)
the ground state is two-fold degenerate,
exhibiting either a clockwise or counter-clockwise sense of the local
canting.  At low concentration we can thus associate a corresponding
low-energy Ising degree of freedom to every impurity bond.  Apart from
these discrete soft degrees of freedom, the background ferromagnet hosts
low-energy spin wave excitations. They mediate an effective
interaction between the Ising degrees of freedom, which results in an
effective classical Ising model with effective two-body interactions
of dipolar type. Their algebraic decay at large distance is a direct
consequence of the gaplessness of the spin waves. A similar effective
interaction results in any system that spontaneously breaks a
continuous symmetry, and thus hosts gapless Goldstone modes mediating
algebraic interactions between impurity degrees of freedom.
The specific case where the impurity bonds form a
Bravais superlattice{,} with a unit cell that is large compared to that
of the cubic host lattice $\Lambda${,} is analytically tractable,
and for certain classes of superlattices we are
able to rigorously establish the presence of spiral order.

Although the analysis in this paper assumes ferromagnetic interactions
for the host lattice $\Lambda$,
we note that our results can be readily extended
to any unfrustrated $XY$ magnet.  For example, if the lattice is
bipartite and $J^{\,}_{\mathrm{imp}}$ has sign opposite to
$J^{\,}_{\perp}$, the system can be mapped to the above described
ferromagnet as follows. For every spin, a reference frame is chosen
such that the unfrustrated ground state of the impurity-free system
corresponds to a ferromagnetic configuration. By virtue of this mapping, 
the low energy effective theory presented below extends 
to this larger class of magnetic insulators.

We emphasize that for the establishment of ferromagnetic order it is
central that the impurity bonds be not randomly oriented. Otherwise,
the pair-wise interactions between the associated Ising degrees of
freedom would be strongly random in sign, which would most likely lead
to spin glass order, as observed in models of dilute, randomly
oriented Ising dipoles \cite{Alonso2010,Fernandez2009}.
Since an Ising glass state generally carries no net magnetization, it would not
induce a spiral state of the original $XY$ spins.  Also, in the limit of a
high density of randomly oriented impurity bonds, one  expects
long-range spin-glass order (observed directly at the level of the $XY$
spins), since the model becomes that of a random-bond $XY$ gauge
glass, as was studied by Villain
\cite{villain_theory_1975,villain_two-level_1977,villain_two-level_1978}.

The remainder of this paper is structured as follows.  In
Section~\ref{Sec:TheModel}, we define the spin lattice model.
Section~\ref{Sec:Mapping to an Ising system} begins with the case of
a single impurity bond.  We then consider a small concentration of
impurity bonds and derive a mapping to an effective Ising model for
low energies.  Section~\ref{Sec: Superlattices of impurity bonds}
describes how to find the ground state of the effective Ising model
when the impurity bonds realize a superlattice.  The effective Ising
model is solved by analytical and numerical means.  Its solution is
then compared to Monte Carlo simulations of a model with the same
network of exchange interactions, but in which the classical $XY$
spins are replaced by classical Heisenberg spins with an additional
easy-plane anisotropy. The latter allows for close contact with
experimentally realized magnets, such as YBaCuFeO$_5$, which are
believed to embody the physical ingredients and mechanisms discussed
above. In Section~\ref{sec: Random impurities: dilute limit},
we study the analytically tractable case of dilute,
randomly placed impurity bonds,
and conclude that random samples undergo a ferromagnetic ordering of cantings,
and thus form a spiral phase.
Section~\ref{sec:finiteT} addresses the onset temperature
of the spiral phase and predicts it to be proportional
to the spiral wavevector.
We compare our quantitative predictions with experimental observations.
Section~\ref{Sec:Conclusions} summarizes our findings
and discusses how the general mechanism  identified here applies
to other systems.

\section{Lattice Hamiltonian for classical $XY$ spins}
\label{Sec:TheModel}

\subsection{Definition of the $XY-$model}
\label{subsec: Definition of the XY-model}

We consider a magnet of classical spins, described by two-dimensional
unit vectors $\widehat{\vv{S}}^{\,}_{\vv{r}}$ with
$\widehat{\vv{S}}^{2}_{\vv{r}}=1$. They are located at the sites
$\vv{r}=x\,\vv{x}+y\,\vv{y}+z\,\vv{z}$ ($x,y,z\in\mathbb{Z}$) of a
cubic lattice $\Lambda$ made of $|\Lambda|$ sites
spanned by the orthonormal unit vectors $\vv{x}$,
$\vv{y}$, and $\vv{z}$ of $\mathbb{R}^{3}$.
In most cases we will impose periodic
boundary conditions on these classical spins. However, as usual,
the choice of boundary conditions does not affect the bulk properties. 

We consider a  classical Hamiltonian 
\begin{subequations}
\label{eq: def 3d XY model}
\begin{equation}
H^{\,}_{\mathcal{L}}\:=
H^{\,}_{0}
+
H^{\,}_{\mathrm{imp}},
\label{eq: def 3d XY model a}
\end{equation}
containing only nearest-neigbor interactions between spins.
Here, ${\mathcal{L}}$ denotes the set of impurity bonds,
as we will describe below.

{The exchange Hamiltonian in Eq.~(\ref{eq: def 3d XY model a}) 
\begin{equation}
H^{\,}_{0}\:= 
-
\frac{1}{2} 
\sum_{
\vv{r},\vv{r}^{\prime}\in\Lambda
     } 
J^{(0)}_{\vv{r},\vv{r}^{\prime}}\,
\widehat{\vv{S}}^{\,}_{\vv{r}} 
\cdot 
\widehat{\vv{S}}^{\,}_{\vv{r}^{\prime}}
\label{eq: def 3d XY model b}
\end{equation}
 possesses the translation symmetries of the cubic lattice, 
since its nearest-neighbor 
ferromagnetic Heisenberg exchange couplings depend only on the relative position of the spins, }
\begin{equation}
J^{(0)}_{\vv{r},\vv{r}^{\prime}}\:= 
J^{\,}_{\parallel} 
\sum_{\vv{\alpha}=\pm\vv{x},\pm\vv{y}}\, 
\delta^{\,}_{\vv{r},\vv{r}^{\prime}+\vv{\alpha}}  
+ 
J^{\,}_{\perp} 
\sum_{\vv{\alpha}=\pm\vv{z}}\, 
\delta^{\,}_{\vv{r},\vv{r}^{\prime}+\vv{\alpha}}=
J^{(0)}_{\vv{r}^{\prime},\vv{r}}.
\label{eq: def 3d XY modelc}
\end{equation}
The in-plane ($J^{\,}_{\parallel}$)
and out-of-plane ($J^{\,}_{\perp}$) couplings
are ferromagnetic but can be different,
$0<J^{\,}_{\parallel} \neq J^{\,}_{\perp}$,
in which case the cubic point-group symmetry is reduced
to the tetragonal one.

The contribution from the disorder in Eq.~(\ref{eq: def 3d XY model a})
\begin{equation}
H^{\,}_{\mathrm{imp}}\:=
(|J^{\,}_{\mathrm{imp}}|+J^{\,}_{\perp}) 
\sum_{
\tilde{\vv{r}}\in\mathcal{L}} 
\widehat{\vv{S}}^{\,}_{\tilde{\vv{r}}} 
\cdot 
\widehat{\vv{S}}^{\,}_{\tilde{\vv{r}}+\vv{z}}
\label{eq: def 3d XY model d}
\end{equation}
\end{subequations}
describes the presence of antiferromagnetic impurity bonds. We label
the bonds by the coordinate of the end point with the smaller $z$-coordinate.
These end points form a subset $\mathcal{L}$
of the points of the cubic host lattice  $\Lambda$.
This term breaks the lattice translation symmetry.
On all impurity bonds the ferromagnetic
$J^{\,}_{\perp}>0$ is replaced by the antiferromagnetic coupling
$J^{\,}_{\mathrm{imp}}<0$, inducing local frustration.

Hamiltonian (\ref{eq: def 3d XY model a}) is invariant under any
rotation of all spins by the same orthogonal $2\times2$ matrix,
i.e., $H^{\,}_{\mathcal{L}}$ has a global $O(2)$ symmetry.

\subsection{Impurity-free case}

Here we consider an impurity-free system, i.e., an empty set $\mathcal{L}$,
\begin{equation}
H^{\,}_{\mathcal{L}}=
H^{\,}_{0}.
\label{eq: def 3d XY model no disorder}
\end{equation}
The ground state is ferromagnetic with all spins parallel.
We choose the polar parametrization
\begin{equation}
\widehat{\vv{S}}^{\,}_{\vv{r}}\:=
\cos\theta^{\,}_{\vv{r}}\,
\widehat{\vv{x}} 
+
\sin\theta^{\,}_{\vv{r}}\,
\widehat{\vv{y}}
\label{eq: def polar representation O(2) spins}
\end{equation}
with the orthonormal basis $\widehat{\vv{x}}$ and $\widehat{\vv{y}}$
of $\mathbb{R}^{2}$. In this polar representation, 
\begin{equation}
H^{\,}_{0}=
-
\frac{1}{2} 
\sum_{\vv{r},\vv{r}^{\prime}\in\Lambda} 
J^{(0)}_{\vv{r},\vv{r}^{\prime}}\,
\cos
\left(
\theta^{\,}_{\vv{r}}
-
\theta^{\,}_{\vv{r}'}
\right)
\label{eq: def 3d XY model no disorder bis}
\end{equation}
has a ferromagnetic ground state defined by
\begin{equation}
\theta^{\mathrm{Ferro}}_{\vv{r}}\equiv {\mathrm{const.}}
\label{eq: def ferro state 3d XY model}
\end{equation}
for all lattice sites $\vv{r}$. 
The invariance of
$H^{\,}_{\mathcal{L}}$ under any global $O(2)$ symmetry
then becomes the invariance under the symmetry transformation
\begin{equation}
\theta^{\,}_{\vv{r}}\mapsto
\epsilon\,\theta^{\,}_{\vv{r}}
+
\Theta,
\label{eq: def global O(2) in polar rep}
\end{equation}
where $\epsilon=\pm1$ and $\Theta\in[0,2\pi[$ are arbitrary numbers
independent of $\vv{r}$.
The angle $\Theta\in[0,2\pi[$ parametrizes a proper rotation
in the connected Lie group $SO(2)$. The choice $\epsilon=-1$
corresponds to an improper rotation, i.e., an orthogonal matrix in $O(2)$
with negative determinant.

At low temperatures, $T\ll J^{\,}_{\perp},J^{\,}_{\parallel}$, we can
use the spin-wave approximation, which assumes that the deviations
from the ferromagnetic ground state 
(\ref{eq: def ferro state 3d XY model}) are small.  
In that case, the Hamiltonian (\ref{eq: def 3d XY model no disorder bis}) 
can be expanded to quadratic order in the angle differences,
\begin{subequations}
\label{eq: def SW 3d XY model no disorder}
\begin{align}
H^{\,}_{0}\approx&\,
E^{\,}_{\mathrm{FM}} 
+ 
\frac{1}{4} 
\sum_{\vv{r},\vv{r}^{\prime}\in\Lambda} 
J^{(0)}_{\vv{r},\vv{r}^{\prime}}\, 
\left(\theta^{\,}_{\vv{r}}-\theta^{\,}_{\vv{r}^{\prime}}\right)^{2} 
\nonumber\\
=&\,
E^{\,}_{\mathrm{FM}} 
+
\frac{1}{2} 
\sum_{\vv{r},\vv{r}^{\prime}\in\Lambda} 
\theta^{\,}_{\vv{r}}\,
D^{(0)}_{\vv{r},\vv{r}^{\prime}}\,
\theta^{\,}_{\vv{r}^{\prime}},
\label{eq: def SW 3d XY model no disorder a}
\end{align}
where
\begin{align}
E^{\,}_{\mathrm{FM}}\equiv
-
\frac{1}{2} 
\sum_{\vv{r},\vv{r}^{\prime}\in\Lambda} 
J^{(0)}_{\vv{r},\vv{r}^{\prime}}\,
\label{eq: def SW 3d XY model no disorder bb}
\end{align}
is the energy of the ferromagnetic ground state, and
\begin{align}
D^{(0)}_{\vv{r},\vv{r}^{\prime}}\:=&\,
\left(
\sum_{\vv{r}^{\prime\prime}\in\Lambda}
J^{(0)}_{\vv{r},\vv{r}^{\prime\prime}}
\right)
\delta^{\,}_{\vv{r}, \vv{r}^{\prime}} 
-
J^{(0)}_{\vv{r},\vv{r}^{\prime}}
\nonumber\\
=&\,
\left(
4 J^{\,}_{\parallel} 
+ 
2 J^{\,}_{\perp}
\right)\,
\delta^{\,}_{\vv{r}, \vv{r}^{\prime}} 
- 
J^{(0)}_{\vv{r},\vv{r}^{\prime}}
\nonumber\\
=&\,
D^{(0)}_{\vv{r}^{\prime},\vv{r}} 
\equiv D^{(0)}_{\vv{r}^{\prime}-\vv{r}}
\equiv D^{(0)}_{\vv{r}-\vv{r}^{\prime}}
\label{eq: def SW 3d XY model no disorder c}
\end{align}
\end{subequations}
is the symmetric spin-wave kernel. It only depends on the difference
$\vv{r}^{\prime}-\vv{r}$, which we henceforth use as the only
subscript. 

We close Sec.~\ref{Sec:TheModel} by establishing a few important properties
obeyed by the spin-wave kernel
$D^{(0)}$.
We observe that $D^{(0)}_{\vv{r}}$  obeys
\begin{equation}
\sum_{\vv{r}^{\prime}\in\Lambda} 
D^{(0)}_{\vv{r}^{\prime}}=0.
\label{eq: zero mode}
\end{equation}
This is a consequence of spin rotational symmetry, which implies that
any global orthogonal transformation
(\ref{eq: def global O(2) in polar rep})
leaves the bilinear form 
(\ref{eq: def SW 3d XY model no disorder a})
invariant. Moreover, if we impose that the angles $\theta^{\,}_{\vv{r}}$
obey periodic boundary conditions, we then have the Fourier transform
\begin{align}
\label{eq: def D(0)k}
D^{(0)}_{\vv{k}}\:=&\,
\frac{1}{|\Lambda|}
\sum_{\vv{r}\in\Lambda}
e^{-\mathrm{i}\vv{k}\cdot\vv{r}}
D^{(0)}_{{\vv{r}}}
\\
=&\,
2
J^{\,}_{\parallel}
\left(
2
-
\cos k^{\,}_{x}
-
\cos k^{\,}_{y}
\right)
+
2
J^{\,}_{\perp}
\left(
1
-
\cos k^{\,}_{z}
\right)\nonumber
\end{align}
for any $\vv{k}$ belonging to the Brillouin zone of the
host cubic lattice $\Lambda$. We shall denote this Brillouin zone
by $\mathrm{BZ}(\Lambda)$. Finally, it is convenient
to introduce the inverse of the spin-wave kernel
$D^{(0)}$ as the Green's function $G^{(0)}$, which satisfies
\begin{subequations}
\label{eq: inverting saddle point with G for one impurity} 
\begin{equation}
\sum_{\vv{r}^{\prime}\in\Lambda} 
G^{(0)}_{\vv{r}-\vv{r}^{\prime}}\, 
D^{(0)}_{\vv{r}^{\prime}-\vv{r}^{\prime\prime}}= 
\delta^{\,}_{\vv{r},\vv{r}^{\prime\prime}}, \quad \forall \vv{r}, \vv{r}^{\prime\prime}.
\label{eq: inverting saddle point with G for one impurity a}  
\end{equation}
Due to the zero mode (\ref{eq: zero mode}),
$G^{(0)}_{\vv{r}-\vv{r}^{\prime}}$
is defined  up to a constant, which we fix by requiring that
\begin{equation}
\sum_{\vv{r}\in\Lambda}G^{(0)}_{\vv{r}}=0,
\end{equation}
such that both $G^{(0)}$ and $D^{(0)}$ annihilate constant functions.
As the inverse of a symmetric kernel, $G^{(0)}_{\vv{r}-\vv{r}^{\prime}}$ is symmetric,
\begin{equation}
G^{(0)}_{\vv{r}-\vv{r}^{\prime}}=G^{(0)}_{\vv{r}^{\prime}-\vv{r}}.
\label{eq: inverting saddle point with G for one impurity c}  
\end{equation}
Imposing periodic boundary conditions, we have
\begin{equation}
G^{(0)}_{\vv{r}}= 
\frac{1}{|\Lambda|} 
\sum_{\vv{k}\in\mathrm{BZ}(\Lambda)\setminus\{\vv{0}\}}
\frac{
e^{
\mathrm{i}\vv{k}\cdot\vv{r}
  }
     }
     {
     D^{(0)}_{\vv{k}}
     }.
\label{eq: momentum rep G0}
\end{equation}
\end{subequations}

The asymptotic large distance behavior of the Green's function is

\begin{align}
G^{(0)}_{\vv{r}} 
\underset{|\vv{r}|\to\infty}\approx&\,  
\int 
\frac{\mathrm{d}^{3}\vv{k}}{(2\pi)^{3}}\, 
\frac{
e^{
\mathrm{i}\vv{k}\cdot\vv{r}}
  }
    { 
J^{\,}_{\parallel}\,
\left(  
k^{2}_{x} 
+  
k^{2}_{y} 
\right) 
+ 
J^{\,}_{\perp}\, 
k^{2}_{z}
    }
\nonumber\\
=&\,
\frac{
1
     }
     {
4\pi\sqrt{J^{\,}_{\parallel}}
     }\, 
\frac{
1
     }
     {
\sqrt{
J^{\,}_{\perp}\,
(x^{2}+y^{2}) 
+ 
J^{\,}_{\parallel}\,
z^{2}
     }
     }.
\label{eq: G(0) long distance}
\end{align}	
On the right-hand side of Eq.~(\ref{eq: G(0) long distance}),
we recognize  the three-dimensional Coulomb potential
for the rescaled coordinates 
\begin{equation}
\bar{x}=\sqrt{J^{\,}_{\perp}}\, x,
\qquad
\bar{y}=\sqrt{J^{\,}_{\perp}}\, y,
\qquad
\bar{z}=\sqrt{J^{\,}_{\parallel}}\, z.
\label{xyz_with_bars}
\end{equation} 

We will see in the next section that impurities couple to each other
through the combination of Green's functions
\begin{subequations}
\label{eq: def Gammas and hat Gammas}
\begin{align}
\Gamma^{(0)}_{{\vv{r}}}\:=&\,
2G^{(0)}_{{\vv{r}}}
-
G^{(0)}_{{\vv{r}}+\vv{z}}
-
G^{(0)}_{{\vv{r}}-\vv{z}}
\nonumber\\
=&\,
\frac{1}{|\Lambda|} 
\sum_{\vv{k}\in\mathrm{BZ}(\Lambda)\setminus\{\vv{0}\}}
\hat{\Gamma}^{(0)}_{{\vv{k}}}\,
e^{
\mathrm{i}\vv{k}\cdot\vv{r}
  }
\label{eq: def Gammas and hat Gammas a}
\end{align}
with the Fourier transform of $\Gamma^{(0)}_{{\vv{r}}}$,
\begin{align}
&
\hat{\Gamma}^{(0)}_{{\vv{k}}{\neq\vv{0}}}\:=
\frac{
  (1-\cos k^{\,}_{z})
     }
     {
     J^{\,}_{\parallel}
\left(
2
-
\cos k^{\,}_{x}
-
\cos k^{\,}_{y}
\right)
+
J^{\,}_{\perp}
\left(
1
-
\cos k^{\,}_{z}
\right)
     }.
\nonumber\\
&
\label{eq: def Gammas and hat Gammas b}
\end{align}
Note that $\hat{\Gamma}^{(0)}_{{\vv{k}}=\vv{{0}}}$
does not enter the sum. It is therefore convenient to define 
\begin{eqnarray}
\hat{\Gamma}^{(0)}_{{\vv{k}}=\vv{0}}\:=0.
\label{eq: def Gammas and hat Gammas c}
\end{eqnarray}
\end{subequations}
Asymptotically, $\Gamma^{(0)}_{{\vv{r}}}$ decays algebraically, like a
dipolar interaction with opposite sign,
\begin{align}
\label{dipolarint}
\Gamma^{(0)}_{{\vv{r}}}
\underset{|\vv{r}|\to\infty}\approx
-\partial^{2}_{z} G^{(0)}_{\vv{r}}\approx
\frac{
\sqrt{J^{\,}_{\parallel}}
     }
     {
4\pi
     }\, 
 \frac{|\bar{\vv{r}}|^{2}-3\bar{z}^{2}}{|\bar{\vv{r}}|^{5}},
\end{align}  
where we use the notation of Eq. (\ref{xyz_with_bars}).

\begin{figure*}
\centerline{\includegraphics[width=\textwidth]{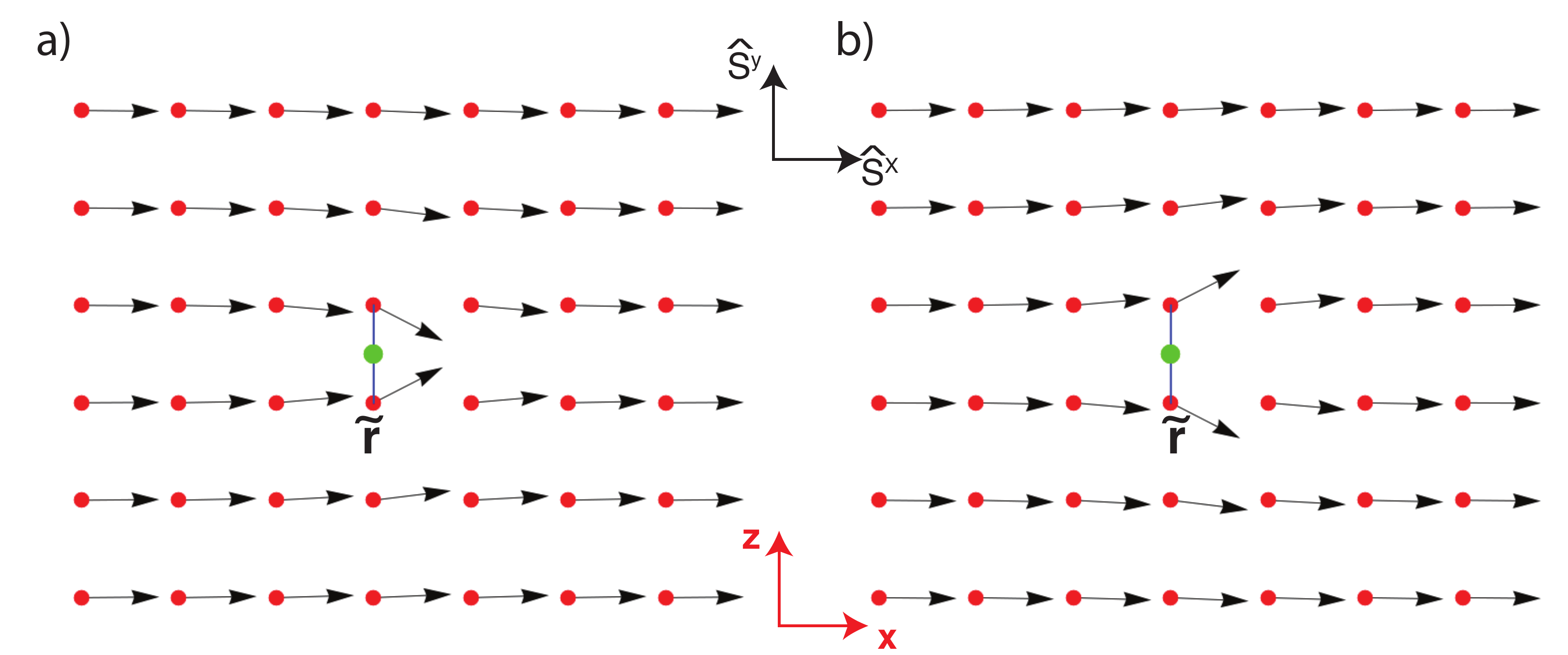}}
\caption{(Color online)
Cantings of the spins induced by a  single, strong impurity bond with
$|J^{\,}_{\mathrm{imp}}|>J^{\,}_{\mathrm{c}}$.
The two degenerate ground
states correspond to the two solutions
$\sigma^{\,}_{\tilde{\vv{r}}}=+1$ (a) and
$\sigma^{\,}_{\tilde{\vv{r}}}=-1$ (b) of
Eq.~(\ref{eq: saddle point}).  
Red dots indicate sites in the
$x$-$z$ plane (red reference frame).  The black arrows represent the
$\widehat{S}^x$ and $\widehat{S}^y$ components of the spins (black
reference frame). The blue line indicates a frustrating
antiferromagnetic bond embedded in the network of ferromagnetic
couplings.  The green dots indicate the inversion center with respect
to which the Hamiltonian is symmetric. The operation of inversion maps
the spin configurations (a) and (b) into each other.
        }
\label{Fig: single-impurity bond spin conf}
\end{figure*}

\section{Mapping to an effective Ising Hamiltonian}
\label{Sec:Mapping to an Ising system}

\subsection{Periodic boundary conditions}

We are ultimately interested in describing states with a spiralling
magnetic order, where the angles of the local magnetization grow
linearly with distance. However, before dealing with this possibility,
we first analyze a situation anticipating no spiralling. In this case
we can impose periodic boundary conditions on the spin angles.

We recall that the set of all directed impurity bonds
defines the set $\mathcal{L}\subset\Lambda$
consisting of all the sites $\tilde{\vv{r}}\in\Lambda$
such that $\langle\tilde{\vv{r}},\tilde{\vv{r}}+\vv{z}\rangle$
is a directed impurity bond. We anticipate that the twist angles
across bonds are relatively small, except on the impurity bonds.
We therefore make the approximation
\begin{subequations}
\label{eq: setup for mapping to an Ising system}
\begin{equation}
\begin{split}
H^{\,}_{\mathcal{L}}\approx&\,
E^{\,}_{\mathrm{FM}}
+
\frac{1}{2} 
\sum_{\vv{r},\vv{r}^{\prime}\in\Lambda} 
\theta^{\,}_{\vv{r}}\,
D^{(0)}_{\vv{r}-\vv{r}^{\prime}}\,
\theta^{\,}_{\vv{r}^{\prime}} 
\\
&\,
-
\sum_{
\tilde{\vv{r}}\in\mathcal{L}
     }
\left[
\frac{J^{\,}_{\perp}}{2}\,
\left(\Delta\theta^{\,}_{\tilde{\vv{r}}}\right)^{2}
-
|J^{\,}_{\mathrm{imp}}|\,
\cos\Delta\theta^{\,}_{\tilde{\vv{r}}}
\right],
\end{split}
\label{eq: setup for mapping to an Ising system b}
\end{equation}
where $\Delta\theta^{\,}_{\tilde{\vv{r}}}$ 
is the canting angle across the impurity bond,
\begin{equation}
\Delta\theta^{\,}_{\tilde{\vv{r}}}\equiv
\theta^{\,}_{\tilde{\vv{r}}}-\theta^{\,}_{\tilde{\vv{r}}+\vv{z}}.
\label{eq: setup for mapping to an Ising system c}
\end{equation}
\end{subequations}
This approximation is certainly valid on the majority of bonds in the limit 
of a dilute concentration of impurity bonds,
\begin{equation} 
n^{\,}_{\mathrm{imp}}\:=
\frac{|\mathcal{L}|}{|\Lambda|}\ll 1,
\label{eq: dilute limit}
\end{equation}
which we will assume from now on.

We are going to establish under what conditions there is a configuration
of angles other than the ferromagnetic one
that minimize the energy (\ref{eq: setup for mapping to an Ising system b})
in the presence of impurity bonds anchored on the set $\mathcal{L}$.

First, we fix all the angles
$\theta^{\,}_{\tilde{\vv{r}}}$
and
$\theta^{\,}_{\tilde{\vv{r}}+\vv{z}}$
with $\tilde{\vv{r}}\in\mathcal{L}$
and integrate out the angles on all other sites, i.e., on all sites from
$\Lambda\setminus(\mathcal{L}\cup\mathcal{L}+\vv{z})$. 
{Within our spin wave approximation, and treating the angles as non-compact variables, this can be done exactly (at any temperature where the spin wave approximation is justified)}
since those angular variables enter the Hamiltonian quadratically. 
Here, we carry out the calculation at $T=0$
by solving the saddle-point
equations for all angles on the sites
$\Lambda\setminus(\mathcal{L}\cup\mathcal{L}+\vv{z})$.
In this way, we shall find the angles $\theta^{\,}_{\vv{r}}$, as well as 
an effective Hamiltonian, expressed solely in terms of the angles
$
\{\theta^{\,}_{\tilde{\vv{r}}},
\tilde{\vv{r}}\in\mathcal{L} \cup\mathcal{L}+\vv{z}\}$.

The minimization over all
$\vv{r}\in\Lambda\setminus(\mathcal{L}\cup\mathcal{L}+\vv{z})$
requires
\begin{subequations}
\label{eq: saddle point for clean sites}
\begin{equation}
\begin{split}
\sum_{\vv{r}^{\prime}\in\Lambda}
D^{(0)}_{\vv{r}-\vv{r}^{\prime}}\,
\theta^{\,}_{\vv{r}^{\prime}}=&\,0.
\end{split}
\label{eq: saddle point for clean sites a}
\end{equation}
We supplement the set
(\ref{eq: saddle point for clean sites a})
of linear equations by
\begin{equation}
\begin{split}
\sum_{\vv{r}^{\prime}\in\Lambda}
D^{(0)}_{\tilde{\vv{r}}-\vv{r}^{\prime}}\,
\theta^{\,}_{\vv{r}^{\prime}}=&\,
\chi^{\,}_{\tilde{\vv{r}}},
\end{split}
\label{eq: saddle point for clean sites b}
\end{equation}
\end{subequations}
whereby
$\tilde{\vv{r}}\in\mathcal{L}\cup\mathcal{L}+\vv{z}$
runs over all sites in $\Lambda$ that are end points of an impurity
bond (two sites per impurity bond). Here,
$\chi^{\,}_{\tilde{\vv{r}}}$ has to be chosen in such a way that
the saddle-point equations
(\ref{eq: saddle point for clean sites})
are satisfied as
$\theta^{\,}_{\tilde{\vv{r}}}$
and
$\theta^{\,}_{\tilde{\vv{r}}+\vv{z}}$
take their prescribed values.

Inverting Eqs.\
(\ref{eq: saddle point for clean sites a})
and
(\ref{eq: saddle point for clean sites b})
for any $\vv{r}\in\Lambda$, we obtain
\begin{equation}
\theta^{\,}_{\vv{r}}=
\theta^{\,}_{0}
+
\sum_{\tilde{\vv{r}}^{\prime}\in\mathcal{L}\cup\mathcal{L}+\vv{z}}
G^{(0)}_{\vv{r},\tilde{\vv{r}}^{\prime}}\,
\chi^{\,}_{\tilde{\vv{r}}^{\prime}},
\label{eq: saddle point solution for clean angles}
\end{equation}
where $\theta^{\,}_{0}\in[0,2\pi[$
is the angle of the magnetization far away from impurities.
Restricting ourselves to the subset
$\tilde{\vv{r}}\in\mathcal{L}\cup\mathcal{L}+\vv{z}$,
Eq.~(\ref{eq: saddle point solution for clean angles})
can be inverted to yield
\begin{equation}
\chi^{\,}_{\tilde{\vv{r}}}=
\sum_{\tilde{\vv{r}}^{\prime}\in\mathcal{L}\cup\mathcal{L}+\vv{z}}
\widetilde{G}^{(0)-1}_{\tilde{\vv{r}},\tilde{\vv{r}}^{\prime}}\,
\left(\theta^{\,}_{\tilde{\vv{r}}^{\prime}}-\theta^{\,}_{0}\right),
\label{eq: saddle point solution for dirty angles}
\end{equation}
where $\widetilde{G}^{(0)}$ is the
$2|\mathcal{L}|\times2|\mathcal{L}|$
matrix obtained by restricting $G^{(0)}_{{\vv{r}},{\vv{r}}^{\prime}}$
to the sites belonging to the impurity bonds
[$\widetilde{G}^{(0)}_{\tilde{{\vv{r}}},\tilde{{\vv{r}}}^{\prime}}\equiv
G^{(0)}_{\tilde{{\vv{r}}},\tilde{{\vv{r}}}^{\prime}}$].
Combining Eqs.\
(\ref{eq: saddle point solution for clean angles})
and
(\ref{eq: saddle point solution for dirty angles}),
we obtain for any $\vv{r}\in\Lambda$
\begin{equation}
\theta^{\,}_{\vv{r}}=
\theta^{\,}_{0}
+
\sum_{\tilde{{\vv{r}}}^{\prime},\tilde{{\vv{r}}}^{\prime\prime}\in\mathcal{L}\cup\mathcal{L}+\vv{z}}
G^{(0)}_{\vv{r},\tilde{{\vv{r}}}^{\prime}}\,
\widetilde{G}^{(0)-1}_{\tilde{{\vv{r}}}^{\prime},\tilde{{\vv{r}}}^{\prime\prime}}\,
\left(\theta^{\,}_{\tilde{{\vv{r}}}^{\prime\prime}}-\theta^{\,}_{0}\right).
\label{eq: expressing all theta's in terms of dirty theta's only}
\end{equation}
The algebraic decay of $G^{(0)}_{\vv{r}}$ ensures that the distortions in
the angular pattern also have algebraic tails.

\begin{widetext}
Second, we evaluate the energy
(\ref{eq: setup for mapping to an Ising system b})
at the saddle point
(\ref{eq: expressing all theta's in terms of dirty theta's only}).
We thereby obtain the effective energy
\begin{equation}
H^{\,}_{\mathrm{eff}}\!
\left(
\left\{
\theta^{\,}_{\tilde{\vv{r}}},
\theta^{\,}_{\tilde{\vv{r}}+\vv{z}},
\tilde{{\vv{r}}}\in\mathcal{L}
\right\}
\right)\:=
\frac{1}{2}
\sum_{\tilde{{\vv{r}}},\tilde{{\vv{r}}}^{\prime}\in\mathcal{L}\cup\mathcal{L}+\vv{z}}
\theta^{\,}_{\tilde{\vv{r}}}\,
\widetilde{G}^{(0)-1}_{\tilde{\vv{r}},\tilde{\vv{r}}^{\prime}}\,
\theta^{\,}_{\tilde{\vv{r}}^{\prime}}
-
\sum_{\tilde{\vv{r}}\in\mathcal{L}}
\left[
\frac{J^{\,}_{\perp}}{2}\,
\left(\theta^{\,}_{\tilde{\vv{r}}+\vv{z}}-\theta^{\,}_{\tilde{\vv{r}}}\right)^{2}
-
|J^{\,}_{\mathrm{imp}}|\,
\cos\left(\theta^{\,}_{\tilde{\vv{r}}+\vv{z}}-\theta^{\,}_{\tilde{\vv{r}}}\right)
\right],
\label{eq: effective energy when all theta's are fct of dirty theta's}
\end{equation}
where we have assumed that the magnetization far away from the
impurities are oriented along $\theta^{\,}_{0}=0$,
making use of the fact that {the} constant $\theta^{\,}_{0}$ can be
chosen freely, since rotating all spins by the same angle does not
affect the energy. In what follows, we will drop such additive constants.
Note that the derivation of the effective action
(\ref{eq: effective energy when all theta's are fct of dirty theta's}),
starting from the Hamiltonian
(\ref{eq: setup for mapping to an Ising system b}),
is exact at any temperature, {if we use the Gaussian approximation and treat the domain of the angles $\theta^{\,}_{{\vv{r}}}$ as non-compact, ignoring that the energy is in fact $2\pi$-periodic in the angles.}
\end{widetext}

For the third and last step we assume $T=0$. We minimize the effective action
(\ref{eq: effective energy when all theta's are fct of dirty theta's})
with respect to the impurity bond angles
$\theta^{\,}_{\tilde{\vv{r}}}$
and
$\theta^{\,}_{\tilde{\vv{r}}+\vv{z}}$
with $\tilde{\vv{r}}\in\mathcal{L}$.
This can be done exactly for a single impurity bond, and approximately in the case of a dilute set of impurities.

\subsubsection{The case of a single impurity bond}
\label{subsec: The case of a single impurity bond}
 
In the case of a single impurity bond
$\langle\tilde{\vv{r}},\tilde{\vv{r}}+\vv{z}\rangle$
represented by the single site $\mathcal{L} = \{\tilde{\vv{r}}\}$, 
the inverse of the $2\times2$ symmetric matrix
with elements
$\widetilde{G}^{(0)}_{\tilde{{\vv{r}}},\tilde{{\vv{r}}}}$,
$\widetilde{G}^{(0)}_{\tilde{{\vv{r}}},\tilde{{\vv{r}}}+\vv{z}}$,
$\widetilde{G}^{(0)}_{\tilde{{\vv{r}}}+\vv{z},\tilde{{\vv{r}}}}$,
$\widetilde{G}^{(0)}_{\tilde{{\vv{r}}}+\vv{z},\tilde{{\vv{r}}}+\vv{z}}$,
is
\begin{equation}
\widetilde{G}^{(0)-1}_{{{\vv{r}}},{{\vv{r}}}^{\prime}}=
\frac{
G^{(0)}_{\vv{0}}\,\delta^{\,}_{{{\vv{r}}},{{\vv{r}}}^{\prime}}
-
G^{(0)}_{\vv{z}}\,
\left(
\delta^{\,}_{{{\vv{r}}},{{\vv{r}}}^{\prime}+\vv{z}}
+
\delta^{\,}_{{{\vv{r}}},{{\vv{r}}}^{\prime}-\vv{z}}
\right)
     }
     {
\left(G^{(0)}_{\vv{0}}\right)^{2}-\left(G^{(0)}_{\vv{z}}\right)^{2}
     },
\label{eq: g inverse on single impurity bond}
\end{equation}
where ${\vv{r}}$,  $\vv{r}^{\prime}\in\{\tilde{\vv{r}},\tilde{\vv{r}}+\vv{z}\}$.

\begin{widetext}
If we use Eq.~(\ref{eq: g inverse on single impurity bond})
in the Hamiltonian
(\ref{eq: effective energy when all theta's are fct of dirty theta's})
we find the effective energy 
\begin{equation}
H^{(1)}_{\mathrm{eff}}
\left(
\theta^{\,}_{\tilde{\vv{r}}},
\theta^{\,}_{\tilde{\vv{r}}+\vv{z}}
\right)=
\frac{1}{4}\,
\frac{
\left(
\theta^{\,}_{\tilde{\vv{r}}}
+
\theta^{\,}_{\tilde{\vv{r}}+\vv{z}}
\right)^{2}
     }
     {
G^{(0)}_{\vv{0}}
+
G^{(0)}_{\vv{z}}
     }
+
\frac{1}{4}\,
\frac{
\left(
\theta^{\,}_{\tilde{\vv{r}}}
-
\theta^{\,}_{\tilde{\vv{r}}+\vv{z}}
\right)^{2}
     }
     {
G^{(0)}_{\vv{0}}
-
G^{(0)}_{\vv{z}}
     } 
-
\frac{J^{\,}_{\perp}}{2}\,
\left(
\theta^{\,}_{\tilde{\vv{r}}}
-
\theta^{\,}_{\tilde{\vv{r}}+\vv{z}}
\right)^{2}
+  
|J^{\,}_{\mathrm{imp}}|\,
\cos\left(\theta^{\,}_{\tilde{\vv{r}}}-\theta^{\,}_{\tilde{\vv{r}}+\vv{z}}\right).
\label{eq: Heff_1imp}
\end{equation}
The center-of-mass angle
$\theta^{\,}_{\tilde{\vv{r}}}+\theta^{\,}_{\tilde{\vv{r}}+\vv{z}}$
and the relative angle
$\theta^{\,}_{\tilde{\vv{r}}}-\theta^{\,}_{\tilde{\vv{r}}+\vv{z}}$
are decoupled in  
$H^{(1)}_{\mathrm{eff}}(\theta^{\,}_{\tilde{\vv{r}}},\theta^{\,}_{\tilde{\vv{r}}+\vv{z}})$.
The effective Hamiltonian
(\ref{eq: Heff_1imp})
is invariant under the transformation
\begin{equation}
\theta^{\,}_{\tilde{\vv{r}}}+\theta^{\,}_{\tilde{\vv{r}}+\vv{z}}\mapsto
\theta^{\,}_{\tilde{\vv{r}}}+\theta^{\,}_{\tilde{\vv{r}}+\vv{z}},
\qquad
\theta^{\,}_{\tilde{\vv{r}}}-\theta^{\,}_{\tilde{\vv{r}}+\vv{z}}\mapsto
-
\left(\theta^{\,}_{\tilde{\vv{r}}}-\theta^{\,}_{\tilde{\vv{r}}+\vv{z}}\right).
\end{equation}
This symmetry is inherited from the fact that Hamiltonian
(\ref{eq: effective energy when all theta's are fct of dirty theta's})
is invariant under the inversion symmetry with respect to the bond
center $\vv{R}=\tilde{\vv{r}}+(\vv{z}/2)$.
\end{widetext}

{Minimization over
$\theta^{\,}_{\tilde{\vv{r}}}+\theta^{\,}_{\tilde{\vv{r}}+\vv{z}}$
imposes} the condition that the two angular distortions
away from the asymptotic $\theta^{\,}_{0}$ on either side of the
impurity bond are opposite,
\begin{equation}
\theta^{\,}_{\tilde{{\vv{r}}}}=
-\theta^{\,}_{\tilde{\vv{r}}+\vv{z}}.
\label{eq: mimimization center mass angle for 1 impurity bond}
\end{equation}

The remaining degree of freedom is the canting angle across the impurity bond,
\begin{equation}
\Delta\theta^{\,}_{\tilde{\vv{r}}}\equiv
\theta^{\,}_{\tilde{\vv{r}}}-\theta^{\,}_{\tilde{\vv{r}}+\vv{z}},
\end{equation}
with $-\pi < \Delta\theta^{\,}_{\tilde{\vv{r}}}\leq \pi$,
in terms of which the {effective} restricted {(eff/res)}
Hamiltonian becomes {
\begin{subequations}
\label{eq: Heff_1imp relative angle only}
\begin{equation}
H^{(1)}_{\mathrm{eff/res}}(\Delta\theta^{\,}_{\tilde{\vv{r}}})\:=
\frac{1}{2}\,
J^{\,}_{\mathrm{c}}\,
\left(\Delta\theta^{\,}_{\tilde{\vv{r}}}\right)^{2}
+
|J^{\,}_{\mathrm{imp}}|\,
\cos\left(\Delta\theta^{\,}_{\tilde{\vv{r}}}\right),
\label{eq: Heff_1imp relative angle only a}
\end{equation}
where we have introduced the short-hand notation
\begin{equation}
J^{\,}_{\mathrm{c}}\:=
\frac{1}{2\left(G^{(0)}_{\vv{0}}-G^{(0)}_{\vv{z}}\right)}
-
J^{\,}_{\perp}.
\label{eq: Heff_1imp relative angle only b}
\end{equation}
\end{subequations}
The coupling $J^{\,}_{\mathrm{c}}$ 
depends parametrically on $J^{\,}_{\parallel}$ and $J^{\,}_{\perp}$.
The rationale for the subscript in $J^{\,}_{\mathrm{c}}$ is
the following.} For small $|J^{\,}_{\mathrm{imp}}|${,
i.e., $|J^{\,}_{\mathrm{imp}}|<J^{\,}_{\mathrm{c}}$,
$H^{(1)}_{\mathrm{eff/res}}(\Delta\theta^{\,}_{\tilde{\vv{r}}})$}
has a single minimum at
\begin{equation}
\Delta\theta^{\,}_{\tilde{\vv{r}}}=0.
\label{eq: ferro saddle point}
\end{equation} 
However, for {$|J^{\,}_{\mathrm{imp}}|>J^{\,}_{\mathrm{c}}$,
$H^{(1)}_{\mathrm{eff/res}}(\Delta\theta^{\,}_{\tilde{\vv{r}}})$}
develops a double well with two degenerate minima.
The two degenerate minima occur at the relative canting angles
\begin{subequations}
\label{eq: saddle point}
\begin{equation}
\label{eq: saddle point a}
\Delta\theta^{\,}_{\tilde{\vv{r}}} =
\sigma^{\,}_{\tilde{\vv{r}}}\,
\Delta\theta,
\end{equation}
with
$\sigma^{\,}_{\tilde{\vv{r}}}=\pm1$ and $\Delta\theta$ being the
positive solution of
\begin{equation}
\label{eq: saddle point b}
\Delta\theta = 
\frac{|J^{\,}_{\mathrm{imp}}|}{J^{\,}_{\mathrm{c}}}
\sin\Delta\theta.
\end{equation}
\end{subequations} {\noindent
When $|J^{\,}_{\mathrm{imp}}|>J^{\,}_{\mathrm{c}}$
and the temperature $T$ is sufficiently small,
namely,
\begin{equation}
k^{\,}_{\mathrm{B}}\,T\ll
J^{\,}_{\mathrm{c}}
-
|J^{\,}_{\mathrm{imp}}|\,
\cos(\Delta\theta),
\label{eq: range T's for Ising approximation}
\end{equation}
thermal fluctuations around the minima of the double well are small.
We will thus call
$|\Delta\theta^{\,}_{\tilde{\vv{r}}}|$ a hard degree of freedom,
while we refer to the Ising variable 
$\mathrm{sgn}(\Delta\theta^{\,}_{\tilde{\vv{r}}})=\sigma^{\,}_{\tilde{\vv{r}}}$
as a soft degree of freedom.
{This terminology is motivated by expanding about the minimum of the double potential well, $\sigma^{\,}_{\tilde{\vv{r}}}\,\Delta\theta$, which is closest}
to $\Delta\theta^{\,}_{\tilde{\vv{r}}}$
\begin{eqnarray}
\label{eq: Taylor expansion H (1) eff/res}
&& H^{(1)}_{\mathrm{eff/res}}
(\Delta\theta^{\,}_{\tilde{\vv{r}}})-H^{(1)}_{\mathrm{eff/res}}(\Delta\theta)  \\
&& \qquad
=  
\frac{1}{2}
\left[
J^{\,}_{\mathrm{c}}
-
|J^{\,}_{\mathrm{imp}}|\,
\cos(\Delta\theta)
\right]
\left(
{|\Delta\theta^{\,}_{\tilde{\vv{r}}}|-
\Delta\theta}
\right)^{2}
+\cdots.
\nonumber
\end{eqnarray}
When the temperature is small compared
to the curvature at the two minima,
thermal fluctuations of the hard degree of freedom 
$|\Delta\theta^{\,}_{\tilde{\vv{r}}}|$
are much smaller than fluctuations due to the soft degree of freedom
$\sigma^{\,}_{\tilde{\vv{r}}}$.
We will thus ignore the latter in the range
(\ref{eq: range T's for Ising approximation})
of temperatures.

If we use the values of the relative angle
(\ref{eq: setup for mapping to an Ising system c})
at the pair of minima (\ref{eq: saddle point})
in combination with Eqs.\
(\ref{eq: mimimization center mass angle for 1 impurity bond})
and
(\ref{eq: expressing all theta's in terms of dirty theta's only})
with $\theta^{\,}_{0}=0$,
we obtain the two canting patterns shown in Fig.\
\ref{Fig: single-impurity bond spin conf}.

The combination
$2(G^{(0)}_{\vv{0}}-G^{(0)}_{\vv{z}})=\Gamma^{(0)}_{\vv{r}=\vv{0}}$
entering $J^{\,}_{\mathrm{c}}$ in
Eq.~(\ref{eq: Heff_1imp relative angle only b})
can be expressed with the help of
Eq.~(\ref{eq: def Gammas and hat Gammas a}) as
\medskip\medskip
\begin{widetext}
\begin{align}
\Gamma^{(0)}_{\vv{r}=\vv{0}}=
2\left(G^{(0)}_{\vv{0}}-G^{(0)}_{\vv{z}}\right)=
\frac{1}{J^{\,}_{\mathrm{c}}+J^{\,}_{\perp}}=
\frac{1}{|\Lambda|} 
\sum_{\vv{k}\in\mathrm{BZ}(\Lambda)\setminus\{\vv{0}\}}
\frac{
\left(1-\cos{k^{\,}_{z}}\right)
     }
     {
J^{\,}_{\parallel}
\left(
2
-
\cos k^{\,}_{x}
-
\cos k^{\,}_{y}
\right)
+
J^{\,}_{\perp}
\left(
1
-
\cos k^{\,}_{z}
\right)
}.
\label{eq: final saddle point for one impurity b}
\end{align}
\end{widetext}
\begin{figure}[!]
\centerline{\includegraphics[width=0.45\textwidth]{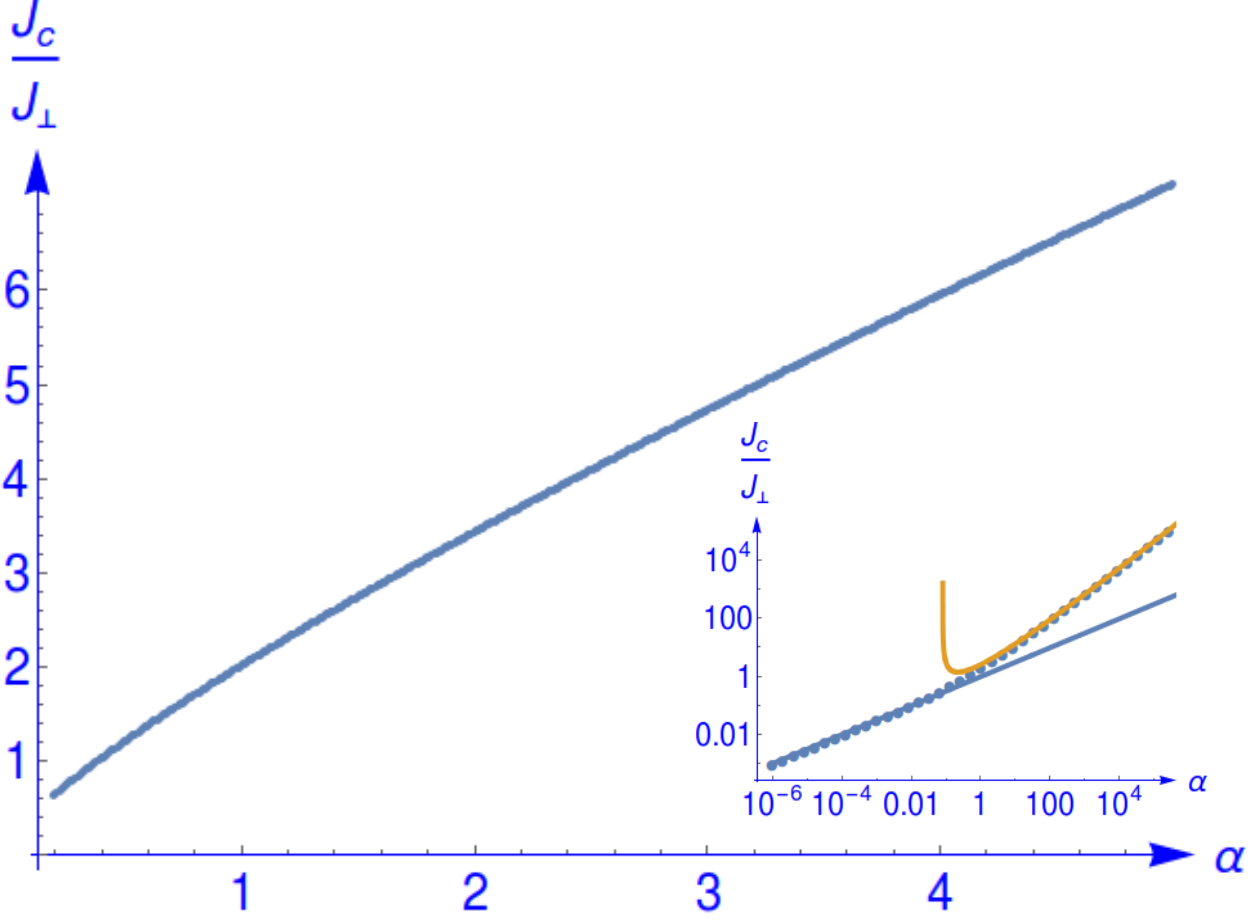}}
\caption{(Color online) Critical impurity bond strength $J^{\,}_{\mathrm{c}}$
(Eq.~\ref{eq: Heff_1imp relative angle only b}) (blue dots),
as a function of the ratio of couplings
$\alpha\equiv J^{\,}_{\parallel}/J^{\,}_{\perp}$.
The inset shows the same function on a logarithmic plot. For strong
anisotropies, one finds the asymptotics $J^{\,}_{\mathrm{c}}(\alpha\ll
1)/J^{\,}_{\perp}\approx C^{\,}_{1} \sqrt{\alpha}$ (blue line) and
$J^{\,}_{\mathrm{c}}(\alpha\gg 1)/J^{\,}_{\perp}\approx
2\pi \alpha/[\ln(\alpha) +C^{\,}_{2}]$ (yellow curve), where
$C^{\,}_{1}=\int^{\pi}_{0} \mathrm{d}x\mathrm{d}y
\sqrt{\sin^{2}x+\sin^{2}y}/\pi^{2}\approx0.958$ and
$C^{\,}_{2}=(5\ln 2-1)\approx 2.47$.}
\label{fig: critical Jc for single impurity}
\end{figure}

For isotropic couplings
$J^{\,}_{\parallel}= J^{\,}_{\perp}=J$,
this evaluates to
\begin{equation}
\Gamma^{(0)}_{\vv{0}}= \frac{1}{3J},
\label{eq: final saddle point for one impurity c}
\end{equation}
which yields the critical coupling 
\begin{equation}
J^{\,}_{\mathrm{c}}=2J.
\label{eq: Jc isotropic}
\end{equation}
For general couplings with ratio 
\begin{equation}
\alpha\equiv\frac{J^{\,}_{\parallel}}{J^{\,}_{\perp}},
\label{eq: def alpha}
\end{equation}
we plot the  threshold value of $J^{\,}_{\mathrm{c}}/J^{\,}_{\perp}$ in Fig.\
\ref{fig: critical Jc for single impurity} together with 
asymptotic expressions that become valid in the limit of strong anisotropy.

\subsection{The case of a dilute set of impurity bonds}
\label{subsec: Dilute density of impurity bonds}

When the density of impurity bonds is finite, we cannot rely on the explicit
representation
(\ref{eq: g inverse on single impurity bond}).
However,} at a small impurity concentration,
the interaction between the angles on
the same impurity bond is much stronger than the coupling between
angles on different impurity bonds. It thus makes sense to split
$\widetilde{G}^{(0)}$ into a bond-local term
[with inverse on every bond
given by Eq.~(\ref{eq: g inverse on single impurity bond})]
and a bond-off-diagonal term according to
\begin{equation}
\widetilde{G}^{(0)}\equiv
\widetilde{G}^{(0)}_{\mathrm{loc}}
+
\widetilde{G}^{(0)}_{\mathrm{nonloc}},
\end{equation}
and to approximate its inverse as
\begin{equation}
\widetilde{G}^{(0)-1}
\approx
\widetilde{G}^{(0)-1}_{\mathrm{loc}}
-
\widetilde{G}^{(0)-1}_{\mathrm{loc}}\,
\widetilde{G}^{(0)}_{\mathrm{nonloc}}\,
\widetilde{G}^{(0)-1}_{\mathrm{loc}}.
\end{equation}
\begin{widetext}
{This yields the Hamiltonian
\begin{subequations}
\begin{equation}
H^{\,}_{\mathrm{eff}}
\left(
\left\{
\theta^{\,}_{\tilde{\vv{r}}},
\theta^{\,}_{\tilde{\vv{r}}+\vv{z}},
\tilde{{\vv{r}}}\in\mathcal{L}
\right\}
\right)\approx
\sum_{\tilde{\vv{r}}\in\mathcal{L}}
H^{(1)}_{\mathrm{eff}}(\theta^{\,}_{\tilde{\vv{r}}},\theta^{\,}_{\tilde{\vv{r}}+\vv{z}})
-
\frac{1}{2}
\sum_{\tilde{{\vv{r}}},\tilde{{\vv{r}}}^{\prime}\in\mathcal{L}\cup\mathcal{L}+\vv{z}}
\theta^{\,}_{\tilde{\vv{r}}}\,
\left[
\widetilde{G}^{(0)-1}_{\mathrm{loc}}\,
\widetilde{G}^{(0)}_{\mathrm{nonloc}}\,
\widetilde{G}^{(0)-1}_{\mathrm{loc}}
\right]^{\,}_{\tilde{\vv{r}},\tilde{\vv{r}}^{\prime}}\,
\theta^{\,}_{\tilde{\vv{r}}^{\prime}},
\end{equation} 
\end{subequations}
{\noindent
where 
$H^{(1)}_{\mathrm{eff}}(\theta^{\,}_{\tilde{\vv{r}}},\theta^{\,}_{\tilde{\vv{r}}+\vv{z}})$
is given by Eq.~(\ref{eq: Heff_1imp}).}

Since 
this term}
is dominant at low impurity concentration, it is again reasonable to restrict
the angular configurations to the subspace given by
\begin{equation}
\theta^{\,}_{\tilde{{\vv{r}}}}=
-
\theta^{\,}_{\tilde{\vv{r}}+\vv{z}}=
\frac{\Delta\theta^{\,}_{\tilde{\vv{r}}}}{2}
\Longrightarrow
\theta^{\,}_{\tilde{{\vv{r}}}}
+
\theta^{\,}_{\tilde{\vv{r}}+\vv{z}}=0,
\qquad
\theta^{\,}_{\tilde{{\vv{r}}}}
-
\theta^{\,}_{\tilde{\vv{r}}+\vv{z}}=
\Delta\theta^{\,}_{\tilde{\vv{r}}},
\end{equation}
as in the single impurity-bond problem of Eq.~(\ref{eq: Heff_1imp}).
This leads to the effective {restricted} Hamiltonian
{[compare with Eq.~(\ref{eq: Heff_1imp relative angle only})]}
\begin{align}
H^{\,}_{\mathrm{eff/res}}\!
\left(\left\{\Delta\theta^{\,}_{\tilde{\vv{r}}}\right\}\right)\:=&\,
\sum_{\tilde{\vv{r}}\in\mathcal{L}}
H^{(1)}_{{\mathrm{eff/res}}}\left(\Delta\theta^{\,}_{\tilde{\vv{r}}}\right)
-
\frac{1}{2}
\sum_{\tilde{\vv{r}}\neq \tilde{\vv{r}}^{\prime}\in\mathcal{L}}
\frac{
\Delta\theta^{\,}_{\tilde{\vv{r}}}
     }
     {
2\left(G^{(0)}_{\vv{0}}-G^{(0)}_{\vv{z}}\right)
     }\,
\left(
2G^{\,}_{\tilde{\vv{r}},\tilde{\vv{r}}^{\prime}}
-
{G}^{\,}_{\tilde{\vv{r}}+\vv{z},\tilde{\vv{r}}^{\prime}}
-
{G}^{\,}_{\tilde{\vv{r}}, \tilde{\vv{r}}^{\prime}+\vv{z}}
\right)\,
\frac{
\Delta\theta^{\,}_{\tilde{\vv{r}}^{\prime}}
     }
     {
2\left(G^{(0)}_{\vv{0}}-G^{(0)}_{\vv{z}}\right)
     }
\nonumber\\
=&\,
\sum_{\tilde{\vv{r}}\in\mathcal{L}}
H^{(1)}_{{\mathrm{eff/res}}}(\Delta\theta^{\,}_{\tilde{\vv{r}}})
-
\frac{1}{2\left(\Gamma^{(0)}_{\vv{0}}\right)^{2}}
\sum_{\tilde{\vv{r}}\neq\tilde{\vv{r}}^{\prime}\in\mathcal{L}}
\Delta\theta^{\,}_{\tilde{\vv{r}}}\,
\Gamma^{(0)}_{\tilde{\vv{r}}-\tilde{\vv{r}}^{\prime}}\,
\Delta\theta^{\,}_{\tilde{\vv{r}}^{\prime}},
\label{eq: Heff2} 
\end{align}
\end{widetext}
where the interaction
$\Gamma^{(0)}_{\tilde{\vv{r}}-\tilde{\vv{r}}^{\prime}}$
is seen to be mediated by the combination of
Green's functions introduced in Eq.~(\ref{eq: def Gammas and hat Gammas a}) {
that scales like an anti-dipolar interaction at long distances}.

As it should be,
the effective Hamiltonian  is invariant under both the global rotation
$\theta^{\,}_{\tilde{\vv{r}}}\mapsto\theta^{\,}_{\tilde{\vv{r}}}+\Theta$,
$\Theta\in[0,2\pi[$ 
and the global Ising symmetry
$\theta^{\,}_{\tilde{\vv{r}}}\mapsto-\theta^{\,}_{\tilde{\vv{r}}}$ for all
$\tilde{\vv{r}}\in\mathcal{L}$.

It remains to minimize
$H^{\,}_{\mathrm{eff/res}}
{\left(\left\{\Delta\theta^{\,}_{\tilde{\vv{r}}}\right\}\right)}$
with respect to the canting angles
$\Delta\theta^{\,}_{\tilde{\vv{r}}}$ on the impurity bonds.  The
corresponding saddle point equation reads,
\begin{subequations}
\label{eq: saddle point sw with density impurities bis}
\begin{equation}
\Delta\theta^{\,}_{\tilde{\vv{r}}}=
\frac{
|J^{\,}_{\mathrm{imp}}|
     }
     {
J^{\,}_{\mathrm{c}}
     }
\sin\Delta\theta^{\,}_{\tilde{\vv{r}}}
+
\Xi^{\,}_{\tilde{\vv{r}}},
\label{eq: saddle point sw with density impurities bis a}
\end{equation}
where
\begin{equation}
\Xi^{\,}_{\tilde{\vv{r}}}\:=
\frac{1}{J^{\,}_{\mathrm{c}}}
\sum_{
\tilde{\vv{r}}^{\prime} \in\mathcal{L}\setminus\{\tilde{\vv{r}}\}
     } 
\frac{
\Gamma^{(0)}_{\tilde{\vv{r}}-\tilde{\vv{r}}^{\prime}}
     }
     {
\left(\Gamma^{(0)}_{\vv{0}}\right)^{2}
     }\,
\Delta\theta^{\,}_{\tilde{\vv{r}}^{\prime}}.
\label{eq: saddle point sw with density impurities bis b}
\end{equation}
\end{subequations}

As before, the ferromagnetic state (\ref{eq: ferro saddle point}) 
is a solution of Eqs.\
(\ref{eq: saddle point sw with density impurities bis a})
and
(\ref{eq: saddle point sw with density impurities bis b}),
but it becomes unstable for sufficiently
large $|J^{\,}_{\mathrm{imp}}| $.

The term $\Xi^{\,}_{\tilde{\vv{r}}}$ is expected to be dominated by
the closest neighbor bonds, since the sum is over a set of decreasing
terms, which, for a valid saddle point, contribute with alternating signs.
Using Eqs.~(\ref{dipolarint}) and
(\ref{eq: saddle point sw with density impurities bis b}),
one expects that $\Xi^{\,}_{\tilde{\vv{r}}}$
scales as $n^{\,}_{\mathrm{imp}}$.  In that case, the condition
\begin{equation}
\frac{
|J^{\,}_{\mathrm{imp}}| }
     {J^{\,}_{\mathrm{c}}     }  
\gg \Xi^{\,}_{\tilde{\vv{r}}}
\label{eq: condition on Xir}
\end{equation}
will be met for sufficiently large values of $J_{\mathrm{imp}}$ 
and sufficiently small values of $n^{\,}_{\mathrm{imp}}$.
Under the condition (\ref{eq: condition on Xir}),
local minima of the Hamiltonian
(\ref{eq: setup for mapping to an Ising system b}),
i.e., solutions to
Eq.~(\ref{eq: saddle point sw with density impurities bis}),
exist, which locally look like the solution for
a single antiferromagnetic impurity bond.

In the dilute impurity regime{,} a low energy state will have a canting
of angles across the impurities bonds close to the single impurity
case, i.e.,
\begin{subequations}
\begin{equation}
\Delta\theta^{\,}_{\tilde{\vv{r}}}\approx
\sigma^{\,}_{\tilde{\vv{r}}}\,\Delta \theta.
\end{equation}
Inserting this Ansatz into the effective Hamiltonian
(\ref{eq: Heff2}),
we obtain the effective Ising model
\begin{equation}
\begin{split}
\label{eq: many impurity saddle point estimate H cal L}
H^{\,}_{\mathcal{L}}[\sigma^{\,}_{\tilde{\vv{r}}}]\:=&\,
E^{\,}_{\mathrm{FM}}
+
E(\Delta\theta)\,
|\mathcal{L}|
\\
&\,
-
\frac{1}{2}\,
\left(\frac{\Delta\theta}{\Gamma_{\vv{0}}^{(0)}}\right)^{2}
\sum_{
\tilde{\vv{r}}\neq\tilde{\vv{r}}^{\prime}\in\mathcal{L}
     }
\sigma^{\,}_{\tilde{\vv{r}}}\,
\Gamma^{(0)}_{\tilde{\vv{r}}-\tilde{\vv{r}}^{\prime}}\,
\sigma^{\,}_{\tilde{\vv{r}}^{\prime}}.
\end{split}
\end{equation}
\end{subequations}
This effective Ising model is invariant under the global Ising symmetry
$\sigma^{\,}_{\tilde{\vv{r}}}\mapsto-\sigma^{\,}_{\tilde{\vv{r}}}$
for all $\tilde{\vv{r}}\in\mathcal{L}$.
Equation (\ref{eq: many impurity saddle point estimate H cal L})
is the main result of Sec.\ \ref{subsec: Dilute density of impurity bonds}.

\begin{figure}[t!]
\centerline{\includegraphics[width=0.25\textwidth]{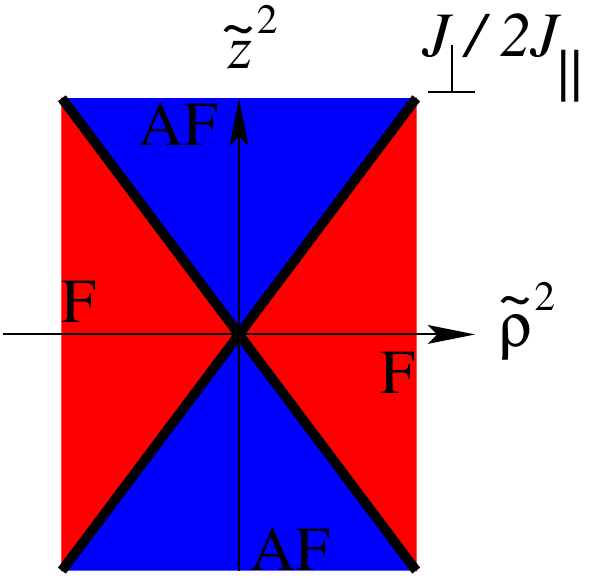}}
\caption{(Color online)
Sign of the {(anti-)}dipolar interaction {$\Gamma^{(0)}_{\tilde{\vv{r}}}$}
(\ref{dipolarint})
between cantings, in the 
$\tilde{\rho}^{2}-\tilde{z}^{2}$-plane
(with $\tilde{\rho}^{2}\equiv\tilde{x}^{2}+\tilde{y}^{2}$).
Along the $z$-axis the interactions are antiferromagnetic,
while for separation along the $xy-$plane
they are ferromagnetic.
The sign change occurs on the cone described by
$J^{\,}_{\perp}\tilde{\rho}^{2}=2J^{\,}_{\parallel}\tilde{z}^{2}$. 
In the quasi-one-dimensional limit $J^{\,}_{\parallel}/J^{\,}_{\perp}\to0$,
the  interaction  {$\Gamma^{(0)}_{\tilde{\vv{r}}}$} is ferromagnetic
for any $\tilde{\rho}^{2}>0$.
In the quasi-two-dimensional limit $J^{\,}_{\perp}/J^{\,}_{\parallel}\to0$,
 {$\Gamma^{(0)}_{\tilde{\vv{r}}}$} is antiferromagnetic
for any $\tilde{z}^{2}>0$. 
        }
\label{Fig: cone}
\end{figure}

The interaction
$\Gamma^{(0)}_{\tilde{\vv{r}}}$
between a pair of Ising variables 
a distance $\tilde{\vv{r}}$ apart
that enters on the right-hand side of
Eq.~(\ref{eq: many impurity saddle point estimate H cal L})
was derived in  Eq.~(\ref{dipolarint}). 
Its asymptotic behavior is that of Ising dipoles ($\sigma^{\,}_{\tilde{\vv{r}}}$
oriented along the direction $\vv{z}$),
albeit with the opposite sign as compared to the usual dipolar interaction.
In the dilute impurity limit, the sign of the two-body interaction
$\Gamma^{(0)}_{\vv{r}}$ depends on the relative position
$\tilde{\vv{r}}$ (see Fig.~\ref{Fig: cone}). It is
ferromagnetic when
\begin{subequations}
\label{eq: cone defined by Gamma}
\begin{equation}
\tilde{x}^{2}+\tilde{y}^{2}>
\frac{2J^{\,}_{\parallel}}{J^{\,}_{\perp}}\,
\tilde{z}^{2},
\end{equation}
vanishing on the conical surface 
\begin{equation}
\tilde{x}^{2}+\tilde{y}^{2}=
\frac{2J^{\,}_{\parallel}}{J^{\,}_{\perp}}\,
\tilde{z}^{2},
\end{equation}
and antiferromagnetic when
\begin{equation}
\tilde{x}^{2}+\tilde{y}^{2}<
\frac{2J^{\,}_{\parallel}}{J^{\,}_{\perp}}\,
\tilde{z}^{2}.
\end{equation}
\end{subequations}
\medskip

\subsection{{Boundary} conditions along the $\vv{z}$ axis}
\label{Sec:ScrewBC}

So far, we have considered the  ferromagnetic state,
$\theta^{\,}_{\vv{r}}=0$ for all $\vv{r}\in\Lambda$, of $H^{\,}_{0}$
defined in Eq.~(\ref{eq: def 3d XY model no disorder bis})
and performed a spin-wave expansion about it,
imposing periodic boundary conditions.
However, this precludes the possibility that the reference state
for the spin wave expansion is non-collinear,
and moreover, the periodic boundary conditions on the angles
$\theta^{\,}_{\vv{r}}$ preclude a spiral state. To overcome these limitations,
we allow for a linear growth of 
\begin{equation}
\theta^{\,}_{\vv{r}}=
\phi^{\,}_{\vv{r}}
+
Q\, 
(\vv{r}\cdot\vv{z}),
\label{eq: def TBC}
\end{equation}
along the $\vv{z}$ direction.
Here, the global degree of freedom $Q\in[-\pi,\pi[$
describes a constant twist rate,
while the local degrees of freedom $\phi^{\,}_{\vv{r}}$
obey periodic boundary conditions.
In finite systems, the twist rate $Q$ along the $\vv{z}$ direction
should be an integer multiple of $2\pi/L^{\,}_{z}$
if we impose periodic boundary conditions on the original spins
$\widehat{\vv{S}}^{\,}_{\vv{r}}=
\cos\theta^{\,}_{\vv{r}}\,\widehat{\vv{x}}
+\sin\theta^{\,}_{\vv{r}}\,\widehat{\vv{y}}$,
but this discrete constraint is irrelevant in the thermodynamic limit.
With the change of variables (\ref{eq: def TBC}), the spin-wave
approximation (\ref{eq: setup for mapping to an Ising system b})
becomes
\begin{subequations}
\label{eq: sw approx with nonvanishing density impurities twist Q}
\begin{equation}
\begin{split}
H^{\,}_{\mathcal{L}}\approx&\,
E^{\,}_{\mathrm{FM}}
+
\frac{1}{2} 
\sum_{
\vv{r},\vv{r}'\in\Lambda
     } 
\phi^{\,}_{\vv{r}}\,  
D^{(0)}_{\vv{r}-\vv{r}^{\prime}}\,
\phi^{\,}_{\vv{r}^{\prime}} 
+  
\frac{J^{\,}_{\perp}\,Q^{2}}{2}\,|\Lambda|
\\
&\,
-
\sum_{
\tilde{\vv{r}}\in\mathcal{L}
     }
\left(
\frac{J^{\,}_{\perp}}{2}(\Delta\phi^{\,}_{\tilde{\vv{r}}}-Q)^{2}
-
|J^{\,}_{\mathrm{imp}}| \cos(\Delta\phi^{\,}_{\tilde{\vv{r}}}-Q)
\right),
\end{split}
\label{eq: def H cal L with Q}
\end{equation}
where we recall that $|\Lambda|$ is the number of sites in the host cubic
lattice $\Lambda$, and we again denote by
\begin{equation}
\Delta\phi^{\,}_{\tilde{\vv{r}}}\:=\phi^{\,}_{\tilde{\vv{r}}}-\phi^{\,}_{\tilde{\vv{r}}+z}
\end{equation}
\end{subequations}
the  twist of $\phi$ across the impurity bond labelled by $\tilde{\vv{r}}$.

For a  low impurity
concentration $0<n^{\,}_{\mathrm{imp}}\ll1$, we assume and will
verify \textit{a posteriori}, that
{
\begin{equation}
|Q|\sim n^{\,}_{\mathrm{imp}}\ll\Delta\theta,
\label{eq: assumption on |Q|}
\end{equation} }\noindent
where
$\Delta\theta{\geq0}$ is the modulus of
the canting angle across an isolated impurity
bond. The leading effect of the emerging spiral order will appear at order
$\mathcal{O}(n^{2}_{\mathrm{imp}})$ in the energy per spin.
We therefore expand the impurity bond terms
in Eq.~(\ref{eq: def H cal L with Q}),
i.e., the second line on the right-hand side of
Eq.~(\ref{eq: def H cal L with Q}),
up to linear order in $Q$.
In this approximation,
the saddle-point values for $\Delta\phi^{\,}_{\tilde{\vv{r}}}$
are therefore again given by {
\begin{equation}
\label{eq:delta_phi_screw}
\Delta\phi^{\,}_{\tilde{\vv{r}}}\approx
\sigma^{\,}_{\tilde{\vv{r}}}\,\Delta\theta
\end{equation} 
up to corrections of order $\mathcal{O}(Q)$}.
We can neglect those since they
lead to corrections to the energy per spin
of order $\mathcal{O}(n^{3}_{\mathrm{imp}})$.
The angles of spins that do not belong to impurity bonds
are again given by
Eq.~(\ref{eq: saddle point sw with density impurities bis}),
with $\Delta\phi^{\,}_{\tilde{\vv{r}}}$ replacing
$\Delta\theta^{\,}_{\tilde{\vv{r}}}$.  However, minimizing over the
twist rate $Q$ {after linearization with respect to $Q$ of 
the second line on the right-hand side of
Eq.~(\ref{eq: def H cal L with Q})},
yields the non-trivial
saddle point value
\begin{align}
Q=&\,
-
\frac{1}{|\Lambda|}
\sum_{
\tilde{\vv{r}}\in\mathcal{L}
     } 
\left(
\frac{|J^{\,}_{\mathrm{imp}}|}{J^{\,}_{\perp}}\,
\sin\left(\Delta\phi^{\,}_{\tilde{\vv{r}}}\right)
+
\Delta\phi^{\,}_{\tilde{\vv{r}}}
\right)
\nonumber\\
=&\,-
\Delta\theta\,
\frac{J^{\,}_{\mathrm{c}}+J^{\,}_{\perp}}{J^{\,}_{\perp}}\,
\frac{1}{|\Lambda|}
\sum_{
\tilde{\vv{r}}\in\mathcal{L}
     }
\sigma^{\,}_{\tilde{\vv{r}}},
\label{Qsaddle}
\end{align}
where we have used the saddle point equation (\ref{eq: saddle point})
{to reach the second equality}.
A spontaneous net winding ($Q\neq0$) of the spins
along the $\vv{z}$ direction thus occurs for canting configurations $
\{ \sigma^{\,}_{\tilde{\vv{r}}} \}$ with a net bias. An Ising
configuration $\{\sigma^{\,}_{\tilde{\vv{r}}}\}$ with a net uniform
magnetization thus corresponds to a spiral state for the $XY$ spins.
Note that the wave vector $Q$ of the spiral is proportional to the
magnetization density of the Ising variables.

We can now verify \textit{a posteriori} the validity of the assumption
 (\ref{eq: assumption on |Q|}).
The maximal value of $|Q|$ is given by
\begin{align}
|Q|^{\,}_{\mathrm{max}}=&\,
n^{\,}_{\mathrm{imp}}\,
\frac{
J^{\,}_{\mathrm{c}}+J^{\,}_{\perp}
 }
 {
J^{\,}_{\perp}
 }
\,
\Delta\theta.
\end{align} 
Thus, for 
\begin{equation}
n^{\,}_{\mathrm{imp}}\,
\ll
\frac{
J^{\,}_{\perp}
 }
 {
J^{\,}_{\mathrm{c}}+J^{\,}_{\perp}
 },
\end{equation}
our assumption is certainly self-consistent.
In the opposite regime, as $n^{\,}_{\mathrm{imp}}$ {approaches 1 from below,}
the interaction between the planes starts to be dominated
by the impurity bonds, which may induce an entirely different ground state
with no spiral order.

Injecting the saddle point value of $Q$, Eq.~(\ref{Qsaddle}), into
Eq.~(\ref{eq: sw approx with nonvanishing density impurities twist Q})
and expressing the energy as a function of the Ising variables
$\sigma^{\,}_{\tilde{\vv{r}}}$ leads to the same effective Hamiltonian
as in Eq.~(\ref{eq: many impurity saddle point estimate H cal L}),
except for an additional term
$-\frac{J^{\,}_{\perp}\,Q^{2}}{2}\,|\Lambda|$, which expresses the
lowering of the total energy due to the coupling of the canting
pattern to the spiral order,
\begin{subequations}
\label{eq: final effective Ising Hamiltonian}
\begin{align}
H^{\,}_{\mathcal{L}}[\sigma^{\,}_{\tilde{\vv{r}}}]\:=&\,
E^{\,}_{\mathrm{FM}}
+
E(\Delta\theta)\,
|\mathcal{L}|
\nonumber\\
&\,
-
\frac{\gamma}{2}
\sum_{
\tilde{\vv{r}}\neq\tilde{\vv{r}}^{\prime}\in\mathcal{L}
     }
\sigma^{\,}_{\tilde{\vv{r}}}\,
\Gamma^{(0)}_{\tilde{\vv{r}}-\tilde{\vv{r}}^{\prime}}\,
\sigma^{\,}_{\tilde{\vv{r}}^{\prime}}
-
\frac{J^{\,}_{\perp}\,Q^{2}}{2}\,
|\Lambda|
\nonumber\\
=&\,
E^{\,}_{\mathrm{FM}}
+
E(\Delta\theta)\,
|\mathcal{L}|
\nonumber\\
&\,-
\frac{1}{2}
\sum_{
\tilde{\vv{r}}\neq\tilde{\vv{r}}'\in\mathcal{L}
     }
\sigma^{\,}_{\tilde{\vv{r}}}\,
J^{(\mathrm{I})}_{\tilde{\vv{r}}-\tilde{\vv{r}}'}\,
\sigma^{\,}_{\tilde{\vv{r}}'} 
-
\frac{
\gamma\,n^{\,}_{\mathrm{imp}}
     }
     {
2J^{\,}_{\perp}
     },
\label{eq: final effective Ising Hamiltonian a}
\end{align}
where we have introduced the constant
\begin{equation}
\gamma\:=
\left(
\Delta\theta
\right)^{2}\,
\left(J^{\,}_{\mathrm{c}}+J^{\,}_{\perp}\right)^{2}= 
\left(\frac{\Delta\theta}{\Gamma_{\vv{0}}^{(0)}}\right)^{2},
\label{eq: final effective Ising Hamiltonian b}
\end{equation}
and the effective Ising interaction
\begin{equation}
\begin{split}
J^{(\mathrm{I})}_{\tilde{\vv{r}}-\tilde{\vv{r}}'}\:=&\,
\gamma\,
\left(
\Gamma^{(0)}_{\tilde{\vv{r}}-\tilde{\vv{r}}^{\prime}}\,
+
\frac{1}{J^{\,}_{\perp}\,|\Lambda|}
\right).
\end{split}
\label{eq: final effective Ising Hamiltonian c}
\end{equation}  
\end{subequations}
Note that the last term in (\ref{eq: final effective Ising Hamiltonian a})
is non-extensive and thus irrelevant {in the thermodynamic limit}.

There are two additive contributions
to the Ising exchange coupling $J^{(\mathrm{I})}_{\tilde{\vv{r}}-\tilde{\vv{r}}'}$.
The contribution $\Gamma^{(0)}_{\tilde{\vv{r}}-\tilde{\vv{r}}^{\prime}}$
represents an (anti)-dipolar two-body interaction between the effective
Ising degrees of freedom $\sigma^{\,}_{\tilde{\vv{r}}}$
associated with the dilute antiferromagnetic bonds.
This interaction is long ranged 
and frustrated, owing to the indefinite sign of the kernel
$\Gamma^{(0)}_{\tilde{\vv{r}}-\tilde{\vv{r}}^{\prime}}$.
The coupling of the canting pattern to the spiral order instead favors a net
(ferromagnetic) bias of the cantings
$\sigma^{\,}_{\tilde{\vv{r}}}$ and contributes an all-to-all interaction of
strength $\frac{1}{J^{\,}_{\perp}\,|\Lambda|}$,
proportional to the inverse volume.

According to the saddle-point equation
(\ref{eq: saddle point sw with density impurities bis a}),
a uniform magnetization of the Ising spins
$\sigma^{\,}_{\tilde{\vv{r}}}$ favors a spiral state with a nonvanishing $Q$
for the original $O(2)$ spin degrees of freedom. Hence, if
the ground state of Eq.~(\ref{eq: final effective Ising Hamiltonian})
supports a non-vanishing magnetization $\sigma^{\,}_{\tilde{\vv{r}}}$,
 the frustration
induced by the dilute impurity bonds
turns the pristine ferromagnetic order of the
impurity-free ground state into spiral order. 
If instead the ground state supports no net bias
of the Ising spins
$\sigma^{\,}_{\tilde{\vv{r}}}$, 
a state with $Q=0$ is favored, and thus no net winding of the $O(2)$ spins
is induced. An example of a possible resulting ground state is a fan-like
magnetic state where the Ising spins order in a layered antiferromagnetic
pattern. This translates into a pattern of the original $O(2)$ spins
ordering essentially ferromagnetically, but with orientations
that alternate slightly between successive layers.

The Ising Hamiltonian with anti-dipolar coupling
(\ref{eq: final effective Ising Hamiltonian})
is structurally very similar to Ising systems
with standard dipolar couplings, as are
realized, e.g., in rare earth compounds with strongly localized
magnetic moments, such as LiHo$_x$Y$_{1-x}$F$_4$ at moderate to low
dilution $x$ ~\cite{BabkevichRonnow2016}. It has been theoretically
and experimentally well established that such random dipolar Ising
systems exhibit a glass transition towards an amorphous magnetic order
at low temperature~\cite{Alonso2010,Alonso2015}. It may thus come as
a surprise that in our case we will find that a change of sign of the
dipolar term, in conjunction with the additional term arising from the
coupling to the spiral, suffices to induce ferromagnetic Ising order,
{in spite} of the positional randomness of the Ising spins. This
difference is presumably largely due to the additional mean-field like
interaction mediated by the formation of the spiral, which stabilizes
the ferromagnetic phase. That type of interaction is absent in systems
of elementary magnetic dipoles, which therefore fall much more easily
into a glassy phase.

\section{Superlattices of impurity bonds}
\label{Sec: Superlattices of impurity bonds}

As Eq.~(\ref{eq: final effective Ising Hamiltonian}) involves long-range
two-body interactions whose sign depends 
on the relative positions of the impurity bonds,
 the ground state cannot be found explicitly
for an arbitrary choice of $\mathcal{L}$, so that in general one has to  
resort to numerical methods, or to approximate treatments.

However, if the set of impurity bonds $\mathcal{L}$
realizes certain Bravais superlattices,
it is possible to establish a sufficient condition
for the ground state of the effective Ising Hamiltonian 
(\ref{eq: final effective Ising Hamiltonian}) 
to be ferromagnetic and, thus, for the ground-state spin configuration of the 
Hamiltonian (\ref{eq: setup for mapping to an Ising system b}) 
to sustain spiral order.

\subsection{Analytical considerations}
\label{subsec: Analytical considerations}

We consider the case when
the subset $\mathcal{L}$ of the cubic host lattice
$\Lambda$
forms a Bravais lattice with the basis vectors
$\vv{A}$, $\vv{B}$, and $\vv{C}$
given by three independent linear combinations 
with integer-valued coefficients
of $\vv{a}\equiv(1,0,0)^{\mathsf{T}}$, 
$\vv{b}\equiv(0,1,0)^{\mathsf{T}}$, 
and 
$\vv{c}\equiv(0,0,1)^{\mathsf{T}}$.
The concentration of the impurity bonds is
\begin{equation}
n^{\,}_{\mathrm{imp}}\equiv
\frac{1}{|\vv{A}\cdot(\vv{B}\wedge\vv{C})|}.  
\end{equation}
In reciprocal space, the superlattice $\mathcal{L}$
defines a small first Brillouin zone $\mathrm{BZ}(\mathcal{L})$,
which is $1/n^{\,}_{\mathrm{imp}}$ times smaller than 
the first Brillouin zone $\mathrm{BZ}(\Lambda)$
of the cubic host lattice $\Lambda$. 

For the $|\mathcal{L}|$ Ising degrees of freedom 
$\sigma^{\,}_{\tilde{\vv{r}}}$ associated with  impurity bonds anchored at 
$\tilde{\vv{r}}\in\mathcal{L}$, 
{we use the Fourier representation
\begin{equation}
\sigma^{\,}_{\tilde{\vv{r}}}= 
\frac{1}{|\mathcal{L}|}
\sum_{
\vv{q}\in\mathrm{BZ}(\mathcal{L})
     }
e^{
\mathrm{i}\vv{q}\cdot\tilde{\vv{r}}
  }\,
\sigma^{\,}_{\vv{q}},
\end{equation}
where  
$\mathrm{BZ}(\mathcal{L})$ is the first  Brillouin zone associated with the lattice $\mathcal{L}$.}
For any $\vv{p}\in\mathbb{R}^{3}$,
we shall make use of the identity 
\begin{equation}
\frac{1}{|\mathcal{L}|}
\sum_{\tilde{\vv{r}}\in\mathcal{L}}
e^{\mathrm{i}\vv{p}\cdot\tilde{\vv{r}}}= 
\sum_{
\vv{\mathsf{G}}\in\mathcal{L}^{\star}
     } 
\delta^{\,}_{\vv{p},\vv{\mathsf{G}}},
\label{eq: Fourier identity}
\end{equation}
where $\mathcal{L}^{\star}$ denotes the reciprocal lattice of 
$\mathcal{L}$.
{The effective Ising Hamiltonian
(\ref{eq: final effective Ising Hamiltonian}) 
 can now be written in  reciprocal space as} 
\begin{subequations}    
\label{eq: effective theory if cal L is Bravais lattice}
\begin{align}
H^{\,}_{\mathcal{L}}=&\,
E^{\,}_{\mathrm{FM}}
\!+\!
E({\Delta\theta})\,
|\mathcal{L}|
\nonumber\\
&\,
+
\frac{\gamma}{2|\mathcal{L}|} 
\sum_{
\vv{q}\in\mathrm{BZ}(\mathcal{L})
     } 
\Upsilon^{\,}_{\vv{q}}\,
\sigma^{\,}_{+\vv{q}}\, 
\sigma^{\,}_{-\vv{q}},
\label{eq: effective theory if cal L is Bravais lattice a}
\end{align}
where $\gamma$ is defined by
Eq.~(\ref{eq: final effective Ising Hamiltonian b})
and
\begin{equation}
\Upsilon^{\,}_{\vv{q}}\:=
-
\sum_{\tilde{\vv{r}}\in\mathcal{L}\setminus\{\vv{0}\}}  
e^{-\mathrm{i}\vv{q}\cdot\tilde{\vv{r}}}\,
\Gamma^{(0)}_{\tilde{\vv{r}}} 
-
\frac{n^{\,}_{\mathrm{imp}}}{J^{\,}_{\perp}}\,
\delta^{\,}_{\vv{q},\vv{0}}.
\end{equation}
Using Eq.~(\ref{eq: Fourier identity}), we obtain
\begin{equation}
\Upsilon^{\,}_{\vv{q}} 
=
-n^{\,}_{\mathrm{imp}}\!
\sum_{
\substack{
\vv{\mathsf{G}}\in\mathcal{L}^{\star}
\\
\vv{q}+\vv{\mathsf{G}}\in\mathrm{BZ}(\Lambda)
         }
     }\!
\Gamma^{(0)}_{
\vv{q} +\vv{\mathsf{G}}
            }
+
\Gamma^{(0)}_{
\tilde{\vv{r}}=\vv{0}}      
- 
\frac{n^{\,}_{\mathrm{imp}}}{J^{\,}_{\perp}}
\delta^{\,}_{\vv{q},\vv{0}}. 
\label{eq: effective theory if cal L is Bravais lattice c}
\end{equation}
\end{subequations}
For a generic choice of the Bravais lattice $\mathcal{L}$ 
and of the couplings
$J^{\,}_{\perp}$, $J^{\,}_{\parallel}$, and $J^{\,}_{\mathrm{imp}}$,
the ground state of the {Ising} Hamiltonian 
(\ref{eq: effective theory if cal L is Bravais lattice})
cannot be found in closed form. However, an analytical solution is available
in certain cases. For instance, if over the reduced Brillouin zone 
$\mathrm{BZ}(\mathcal{L})$, the kernel
$\Upsilon^{\,}_{\vv{q}}$ assumes its global minimum at
a unique momentum $\vv{q}^{\,}_{\mathrm{min}}$,
such that for all $\tilde{\vv{r}}\in\mathcal{L}$
\begin{equation}
\sigma^{\mathrm{min}}_{\tilde{\vv{r}}}=
e^{\mathrm{i}\vv{q}^{\,}_{\mathrm{min}}\cdot\tilde{\vv{r}}}=\pm1,
\label{eq: condition for global minimum to be OK}
\end{equation}
the ground state of the Ising Hamiltonian is then given by the configuration
described by 
Eq.~(\ref{eq: condition for global minimum to be OK}). 
Inserting Eq.~(\ref{eq: condition for global minimum to be OK})
into Eqs.~(\ref{eq: def TBC}),
(\ref{eq:delta_phi_screw}),
and (\ref{Qsaddle}), and using the Fourier representation
(\ref{eq: def Gammas and hat Gammas a},\ref{eq: def Gammas and hat Gammas b}),
one finds
\begin{widetext}
\begin{subequations}
\label{eq: solution if sigma min}
\begin{equation}
\theta^{\mathrm{min}}_{\vv{r}}=
\phi^{\mathrm{min}}_{\vv{r}}
+
Q^{\mathrm{min}}\, 
(\vv{r}\cdot\vv{z}),
\label{eq: solution if sigma min a}
\end{equation}
where
\begin{align}
\phi^{\mathrm{min}}_{\vv{r}}\:=
   {
n^{\,}_{\mathrm{imp}}
\sqrt{\gamma}
\sum_{
\left\{
\vv{k}\in\mathrm{BZ}(\Lambda)|\vv{k}-\vv{q}^{\,}_{\mathrm{min}}\in\mathcal{L}^{\star}\setminus\{ \vv{0}\}
\right\}
     }
\frac{
\left(
1-e^{-\mathrm{i}k^{\,}_{z}}
\right)\,
e^{
\mathrm{i}
\vv{k}\cdot\vv{r}
  }
     }
     { 
2 
J^{\,}_{\parallel}
\left(
2-\cos k^{\,}_{x}-\cos k^{\,}_{y}
\right) 
+ 
2
J^{\,}_{\perp} 
\left(
1-\cos k^{\,}_{z}
\right)
     },
     }
\label{eq: solution if sigma min b}
\end{align}
with $\gamma$ defined by
Eq.~(\ref{eq: final effective Ising Hamiltonian b})
and
\begin{align}
Q^{\mathrm{min}}\:=
-
\frac{\sqrt{\gamma}}{J^{\,}_{\perp }}\,
n^{\,}_{\mathrm{imp}}\,
\delta^{\,}_{\vv{q}^{\,}_{\mathrm{min}},\vv{0}}.
\label{eq: solution if sigma min c}
\end{align}
\end{subequations}
\end{widetext}
Examples of 
$\vv{q}^{\,}_{\mathrm{min}}$
for which Eq.~(\ref{eq: condition for global minimum to be OK}) 
holds are
\begin{subequations}
\begin{equation}
\vv{q}^{\,}_{\mathrm{min}}=\vv{0},
\label{Eq:BZPoints1}
\end{equation}
and
\begin{equation}
\vv{q}^{\,}_{\mathrm{min}}\in 
\left\{
\pm
\frac{1}{2}\,\vv{A}^{\star},
\pm
\frac{1}{2}\,\vv{B}^{\star},
\pm
\frac{1}{2}\,\vv{C}^{\star}
\right\},
\label{Eq:BZPoints2}
\end{equation}
\end{subequations}
where $\vv{A}^{\star}$, $\vv{B}^{\star}$, and $\vv{C}^{\star}$ are the
basis vectors of the reciprocal lattice $\mathcal{L}^{\star}$.  In the
former case {(\ref{Eq:BZPoints1})}, $Q^{\mathrm{min}}\neq0$ and the
ground state is an $O(2)$ magnetic spiral, while in the latter case
(\ref{Eq:BZPoints2}), $Q^{\mathrm{min}}=0$ and thus there is no
spiral.

\begin{figure*}
\centerline{\includegraphics[width=1.1\textwidth]{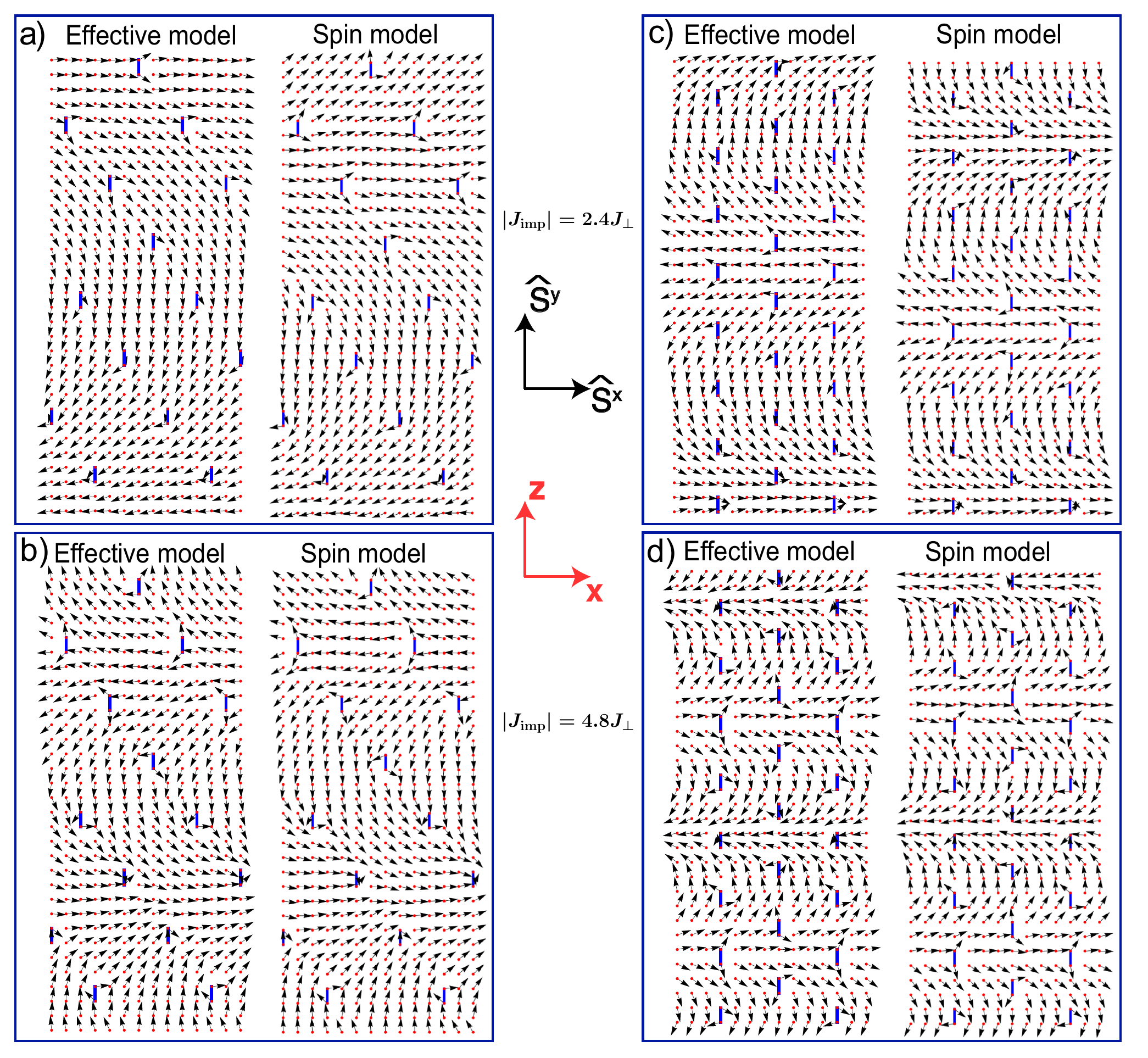}}
\caption{(Color online)
Spin configurations of the ground state in an $x$-$z$ plane of
the cubic host lattice $\Lambda$. Black arrows represent the $O(2)$ ($XY$)
spins defined in Eq.~(\ref{eq: def polar representation O(2) spins})
(black reference frame). Red dots represent the sites 
in a cross section of the cubic host lattice $\Lambda$
(red reference frame). 
Blue lines represent the impurity bonds. 
Results are obtained for 
$J^{\,}_{\parallel}/J^{\,}_{\perp}=1$.  
Panels (a) and (b) show the comparison between 
approximate analytical results
(as described in 
Sec.\ \ref{subsec: Analytical considerations})
and MC simulations of Eq.~(\ref{eq: def O(3) lattice model}),
respectively,
for a superlattice with the basis
$\vv{A}=(5,3,2)^{\mathsf{T}}$, 
$\vv{B}=(3,4,4)^{\mathsf{T}}$,
and 
$\vv{C}=(4,5,2)^{\mathsf{T}}$.    
The superlattice in panels (c) and (d) has the basis
$\vv{A}=(3,3,2)^{\mathsf{T}}$, 
$\vv{B}=(0,4,2)^{\mathsf{T}}$,
and 
$\vv{C}=(4,0,2)^{\mathsf{T}}$.
The impurity strengths are 
$|J^{\,}_{\mathrm{imp}}/J^{\,}_{\perp}|=2.4$
for panels (a,c) and 
$|J^{\,}_{\mathrm{imp}}/J^{\,}_{\perp}|=4.8$ 
for panels (b,d).
In all cases the effective Ising Hamiltonian
(\ref{eq: effective theory if cal L is Bravais lattice})
correctly predicts a spiral state.
However, the accuracy of the predicted value of $Q$ significantly improves
with increasing $|J^{\,}_{\mathrm{imp}}/J^{\,}_{\perp}|$ away from the critical value
of $2$, where canting sets in. This can be seen by comparing the relative
orientations of the  spins in the lattice corners. 
   }
\label{Fig:SpinFields}
\end{figure*}

\subsection{Comparison between analytical and numerical results
for superlattices}
\label{Sec:CompNum}

To illustrate that the effective Ising Hamiltonian
(\ref{eq: final effective Ising Hamiltonian})
captures the low-energy physics of the microscopic Hamiltonian
(\ref{eq: def 3d XY model}), 
we consider several superlattices $\mathcal{L}$ 
of impurity bonds and compare their microscopic ground state 
to the ground state of the effective Ising Hamiltonian
(\ref{eq: effective theory if cal L is Bravais lattice}).

Instead of directly studying the ground state of the microscopic Hamiltonian
(\ref{eq: def 3d XY model}), we actually study the microscopic Hamiltonian
\begin{equation}
\begin{split}
H^{\,}_{\mathrm{Heis}}\:=&\,
-
\frac{1}{2} 
\sum_{
\vv{r},\vv{r}^{\prime}\in\Lambda
     } 
J^{(0)}_{\vv{r},\vv{r}^{\prime}}\, 
\vv{S}^{\,}_{\vv{r}} 
\cdot 
\vv{S}^{\,}_{\vv{r}^{\prime}} 
\\
&\,
+ 
\left(|J^{\,}_{\mathrm{imp}}|+J^{\,}_{\mathrm{\perp}}\right) 
\sum_{
\tilde{\vv{r}}\in\mathcal{L}
     }   
\vv{S}^{\,}_{\tilde{\vv{r}}} 
\cdot 
\vv{S}^{\,}_{\tilde{\vv{r}}+\vv{z}} 
\\
&\,
+ 
\Delta 
\sum_{\vv{r}} 
(S^{z}_{\vv{r}})^{2},
\end{split}
\label{eq: def O(3) lattice model}
\end{equation}
which is closer to experimental realizations.
Here, we have replaced the classical $XY$ spins
from Eq.~(\ref{eq: def 3d XY model})
with classical Heisenberg spins $\vv{S}^{\,}_{\vv{r}}$
(being unit vectors in $\mathbb{R}^{3}$).
A single-ion anisotropy $\Delta>0$ 
penalizes a spin orientation along the $\vv{z}$-axis. 
We apply open boundary conditions along
all principal directions of the cubic lattice.
To approximately find the ground state, 
we perform parallel tempering Monte Carlo (MC)
simulations. We use $140$ temperatures
$T^{\,}_{i}$ with a constant ratio
$T^{\,}_{i+1}/T^{\,}_{i}$,
covering a range from $T\le4\cdot10^{-3}$ $J^{\,}_{\parallel}$
up to temperatures well in the paramagnetic phase.  
The ground state is
obtained by keeping track of the minimal energy state visited during
the Monte Carlo evolution for  
the lowest temperature. The single-ion anisotropy 
was restricted to  $0<\Delta\le$ 0.02 $|J^{\,}_{\parallel}|$. 
For $T\ll\Delta$, the low-energy states are essentially coplanar
with spins lying in the $xy-$plane. At $T=0$ the
ground state of the $XY$ Hamiltonian (\ref{eq: def 3d XY model})
is identical to that of the anisotropic Heisenberg Hamiltonian 
(\ref{eq: def O(3) lattice model}).
The typical size of the cubic host lattice $\Lambda$ used in the 
MC simulations was $14\times14\times32$ lattice spacings.

To obtain the corresponding effective model in terms of cantings we
take the following steps.
First, we verify that the exchange couplings
allow for a non-trivial solution ${\Delta\theta}\neq 0$ of
Eq.~(\ref{eq: saddle point}).
Second, we solve for the ground state of the effective Ising Hamiltonian
(\ref{eq: effective theory if cal L is Bravais lattice}).
To this end we minimize the function
$\Upsilon^{\,}_{\vv{q}}$ in
Eq.~(\ref{eq: effective theory if cal L is Bravais lattice c})
with respect to $\vv{q}$.  If the absolute
minimum in the Brillouin zone $\mathrm{BZ}(\mathcal{L})$ occurs at
$\vv{q}=0$, we predict a spiral ground state ($Q\neq 0$).  If the
absolute minimum occurs at
$\vv{q}^{\,}_{\text{min}}=\vv{C}^{\star}/2$, a ground state
with $Q=0$ is predicted. If instead $\vv{q}^{\,}_{\text{min}}$ does not
satisfy Eq.~(\ref{eq: condition for global minimum to be OK}), the
ground state has more than one Fourier component and we would need to
solve the effective Ising model numerically.
Third, from the Ising ground state
of Hamiltonian (\ref{eq: effective theory if cal L is Bravais lattice}),
the microscopic pattern (\ref{eq: def polar representation O(2)
  spins}) of the $O(2)$ spins is finally obtained from Eq.~(\ref{eq:
  solution if sigma min}), using the value of
${\Delta\theta}$ obtained from solving Eq.~(\ref{eq: saddle point}).

\begin{figure*}[ht!]
\centerline{\includegraphics[width=\textwidth]{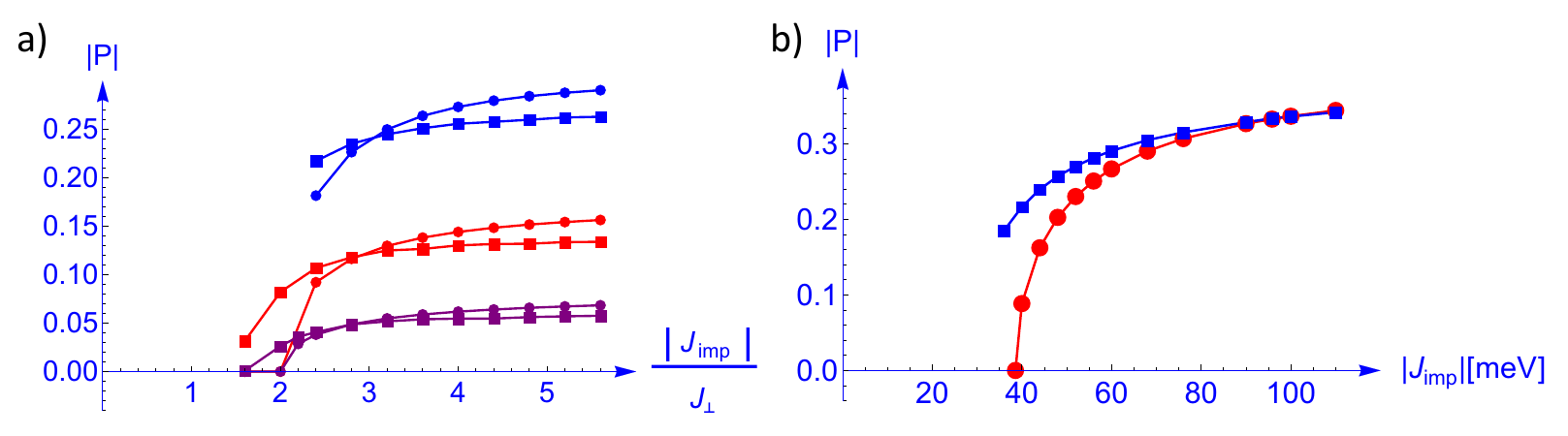}}
\caption{(Color online)
Comparison between the approximate analytical (dots) and
numerical (squares) values of the spiral order parameter $|P|$
[cf. Eq.~(\ref{eq: definition P})] for $XY-$spins on a cubic lattice with
$L^{\,}_{x}=14$, $L^{\,}_{y}=14$, and $L^{\,}_{z}=32$.  The impurity
bonds form a superlattice.  
Panel (a) shows the results for the
isotropic case $J^{\,}_{\parallel}/J^{\,}_{\perp}=1$ for three
superlattices with the following basis vectors:
(blue): $\vv{A}=(3,3,2)^{\mathsf{T}}$, 
$\vv{B}=(0,4,2)^{\mathsf{T}}$, 
and 
$\vv{C}=(4,0,2)^{\mathsf{T}}$, $n^{\,}_{\mathrm{imp}}=1/16$; (red): 
$\vv{A}=(5,3,2)^{\mathsf{T}}$, 
$\vv{B}=(3,4,4)^{\mathsf{T}}$, 
and
$\vv{C}=(4,5,2)^{\mathsf{T}}$,  $n^{\,}_{\mathrm{imp}}=1/32$;
(purple): 
$A=(4,3,0)^{\mathsf{T}}$,
$B=(0,4,3)^{\mathsf{T}}$, 
and 
$C=(5,0,2)^{\mathsf{T}}$, $n^{\,}_{\mathrm{imp}}=1/77$.
Note that the magnitude $|P|$ of
the spiral order parameter $P$ 
increases approximately linearly with $n^{\,}_{\mathrm{imp}}$.
Panel (b) shows the dependence of $|P|$ on  $|J^{\,}_{\mathrm{imp}}|$,
{for the parameters of YBaCuFeO$_5$ (cf.Eq.~{(\ref{values of J in YBaCuFeO5})}),}
and  a superlattice of density $n^{\,}_{\mathrm{imp}}=1/56$, with the basis
$\vv{A}=(4,3,0)^{\mathsf{T}}$, 
$\vv{B}=(0,4,2)^{\mathsf{T}}$, 
and 
$\vv{C}=(4,0,2)^{\mathsf{T}}$.
        }        
\label{Fig:Pol}
\end{figure*}

The effective 
Ising Hamiltonian (\ref{eq: effective theory if cal L is Bravais lattice})
is found to be very accurate once the impurity coupling
$|J^{\,}_{\mathrm{imp}}|$ sufficiently exceeds the critical value
$J^{\,}_{\mathrm{c}}$.
This is illustrated by Fig.\ \ref{Fig:SpinFields}. Its four panels compare
the approximate ground state obtained via the effective Ising Hamiltonian
(\ref{eq: effective theory if cal L is Bravais lattice a})
(shown on the left) with the 
ground state of the Hamiltonian (\ref{eq: def O(3) lattice model})
obtained via MC simulation (shown on the right).
This is done for two strengths of impurity couplings
and two different superlattices. We choose parameters such that both methods
yield a spiral state, with $\vv{q}^{\,}_{\text{min}}=0$ minimizing the
kernel $\Upsilon^{\,}_{\vv{q}}$. No coupling anisotropy
($J^{\,}_{\parallel}/J^{\,}_{\perp}=1$) was assumed in all these
cases. Figures~\ref{Fig:SpinFields}(a) and \ref{Fig:SpinFields}(b)
correspond to the same superlattice, but different impurity strengths,
$|J^{\,}_{\mathrm{imp}}|/J^{\,}_{\perp}= 2.4$ and $4.8$, respectively.
A spiral ground state is correctly predicted in both cases.
However, while the canting angle ${\Delta\theta}$ at the impurity bonds  
and especially the spiral wave vector $Q$ are rather accurately predicted
for strong impurity couplings $|J^{\,}_{\mathrm{imp}}|/J^{\,}_{\perp}=4.8$,
they are underestimated by the effective
theory when the impurity coupling
$|J^{\,}_{\mathrm{imp}}|/J^{\,}_{\perp}=2.4$ comes relatively close
to the threshold strength $J^{\,}_{\mathrm{c}}/J^{\,}_{\perp}=2$,
cf.\ Eq.~(\ref{eq: Heff_1imp relative angle only b}).
The agreement between the analytical approximation and the MC
simulations improves with increasing
$|J^{\,}_{\mathrm{imp}}|/J^{\,}_{\perp}$. This agrees with what one expects
from the considerations of Sec.\ 
\ref{subsec: Dilute density of impurity bonds}. 
Indeed, assuming that all canting angles take the same value
$\Delta\theta^{\,}_{\tilde{\vv{r}}} = \Delta\theta$, and assuming a
superlattice of impurities,
Eq.~(\ref{eq: saddle point sw with density impurities bis})
takes the form
\begin{equation}
\Delta\theta=
\frac{
|J^{\,}_{\mathrm{imp}}|
     }
     {
J^{\,}_{\mathrm{c}}
-
\left(\Gamma^{(0)}_{\vv{0}}\right)^{-2}
\sum\limits_{
\tilde{\vv{r}}
\in\mathcal{L}\setminus\{\vv{0}\}
 	    } 
\Gamma^{(0)}_{\tilde{\vv{r}}}
     }
\sin\Delta\theta.
\label{eq: saddle point for small j imp}
\end{equation}
As compared to the saddle point equation for a single impurity,
Eq.~(\ref{eq: saddle point}),
the denominator $J^{\,}_{\mathrm{c}}$
is shifted by the
small correction
\begin{equation}
\left(\Gamma^{(0)}_{\vv{0}}\right)^{-2}
\sum\nolimits_{\tilde{\vv{r}}\in\mathcal{L}\setminus\{\vv{0}\}}
\Gamma^{(0)}_{\tilde{\vv{r}}}\sim
n^{\,}_{\mathrm{imp}}\,
\times\,
\mathcal{O}(J^{\,}_{\perp},J^{\,}_{\parallel})
\ll
J^{\,}_{\mathrm{c}}.
\end{equation}
When $|J^{\,}_{\mathrm{imp}}|/J^{\,}_{\mathrm{c}}$ is large, this renormalization
has little effect on the solution of the saddle point equation,
$\Delta\theta$, which will be close to $\pi$ in any case. However,
when $|J^{\,}_{\mathrm{imp}}|/J^{\,}_{\mathrm{c}}$
is close to the threshold of 1, an effective
reduction of $J^{\,}_{\mathrm{c}}$
(which is expected for superlattice that favor
ferromagnetic Ising order) leads to an increase of $\Delta\theta$.
For example, in Fig.~\ref{Fig:SpinFields}(a), the value of the canting
angle would $\Delta\theta = 1.02$ according to Eq.~(\ref{eq: saddle point}),
but increaes to $\Delta\theta = 1.15$ after correcting it
by Eq.~(\ref{eq: saddle point for small j imp}).

Similar results are found for other superlattices.  For instance, panels (c)
and (d) show results for a denser superlattice, but with the same
exchange couplings as in panels (a) and (b), respectively.  In all
panels of Fig.\ \ref{Fig:SpinFields}, the deviations from the
local ferromagnetic order at non-impurity bonds are small,
justifying \textit{a posteriori} the {spin-wave}
approximation used to
derive the effective Ising Hamiltonian 
(\ref{eq: effective theory if cal L is Bravais lattice}). 

To quantify the quality of the approximations incurred 
when trading the microscopic Hamiltonian 
(\ref{eq: def O(3) lattice model}) 
for the effective Ising Hamiltonian
(\ref{eq: effective theory if cal L is Bravais lattice}),
{we compare the quantity 
\begin{equation}
P\:= 
\frac{1}{L^{\,}_{x}\,L^{\,}_{y}\,\left(L^{\,}_{z}-1\right)} 
\sum_{\substack{\vv{r}\in \Lambda\\1\le r^{\,}_{z} \le L^{\,}_{z}-1}}  
\sin
\left(
\phi^{\,}_{\vv{r}+\vv{z}}
-
\phi^{\,}_{\vv{r}}  
\right),
\label{eq: definition P}
\end{equation}
obtained from both Hamiltonians for several superlattices and various ratios 
$|J^{\,}_{\mathrm{imp}}|/J^{\,}_{\perp}$ in Fig. \ref{Fig:Pol}(a). {Here,}
$L^{\,}_{x}$, $L^{\,}_{y}$, and $L^{\,}_{z}$ in Eq.~(\ref{eq: definition P})
are the linear dimensions of the lattice{, while}
$P$ is an order parameter for the magnetic spiral phase.
On the right-hand side, the sine of the relative angle between
$\widehat{\vv{S}}^{\,}_{\vv{r}}$
and
$\widehat{\vv{S}}^{\,}_{\vv{r}+\vv{z}}$
is summed over all sites of the cubic host lattice $\Lambda$.


Figure\ref{Fig:Pol}(a) shows how the value of  $|P|$, evaluated on the minimal energy configuration,
increases with increasing $|J^{\,}_{\mathrm{imp}}|/J^{\,}_{\perp}\geq
J^{\,}_{\mathrm{c}}/J^{\,}_{\perp}$ for three superlattices of
impurity bonds in an isotropic cubic lattice 
($J^{\,}_{\parallel}/J^{\,}_{\perp}=1$){, whereby
all superlattices were chosen so that they induce a spiral state.}
At relatively
large $|J^{\,}_{\mathrm{imp}}|/J^{\,}_{\perp}$, the results for $|P|$
from the effective Ising Hamiltonian
(\ref{eq: effective theory if cal L is Bravais lattice})
(dots) are close to those obtained from the
microscopic simulation of Hamiltonian
(\ref{eq: def O(3) lattice model}) (squares),
up to corrections of order $n^{\,}_{\mathrm{imp}}$,
as anticipated in the discussion around
Eq.~(\ref{eq: condition on Xir}).
However, as $|J^{\,}_{\mathrm{imp}}|$ approaches $J^{\,}_{\mathrm{c}}$
from above, deviations become stronger, as we discussed after
Eq.~(\ref{eq: saddle point for small j imp}). In this regime the
double-well potential defining the Ising degrees of freedom associated
with the canting pattern becomes very shallow. Thus, even relatively
weak contributions from neighboring impurities,
$\Xi^{\,}_{\tilde{\vv{r}}}$, can stabilize and enhance the local
canting ${\Delta\theta}$ and strengthen the spiral wave vector beyond
the approximations we used to derive the effective model.  For the
same reason of mutual stabilization, we still find a finite spiral
order, $P\neq0$, even when
$J^{\,}_{\mathrm{imp}}/J^{\,}_{\perp}\lesssim
J^{\,}_{\mathrm{c}}/J^{\,}_{\perp}=2$.}

\subsection{Spiral phase in a realistic model for YBaCuFeO$_5$}

It was argued in Ref.\ \onlinecite{Scaramucci_2016}
that the magnetic degrees of freedom in the insulator YBaCuFeO$_5$
realize a close cousin of
Hamiltonian (\ref{eq: def O(3) lattice model}),
in that $J^{\,}_{\perp}>0$ is to be replaced by two distinct values 
$J^{\prime}_{\perp}>0$
and
$J^{\prime\prime}_{\perp}>0$
depending on the parity of the $z$ component of the coordinate $\vv{r}$
of the cubic lattice. From the estimates for
$J^{\,}_{\parallel}>0$,
$J^{\prime}_{\perp}>0$,
$J^{\prime\prime}_{\perp}>0$,
and
$J^{\,}_{\mathrm{imp}}<0$
made in Ref.~\onlinecite{Scaramucci_2016} we have borrowed
the values
\begin{eqnarray}
\label{values of J in YBaCuFeO5}
\begin{split}
&J^{\,}_{\parallel}=28.9\text{meV}, \\
&J^{\,}_{\mathrm{imp}}=-95.8\text{meV}, \\
&J^{\,}_{\perp}\equiv(J^{\prime}_{\perp}+J^{\prime\prime}_{\perp})/2=4.1\text{meV}.
\end{split}
\end{eqnarray}

Figure \ref{Fig:Pol}(b) compares the
dependence of the magnitude $|P|$ of the
spiral order parameter $P$ defined in Eq.~(\ref{eq: definition P})
on $J^{\,}_{\mathrm{imp}}$
for the microscopic Hamiltonian (\ref{eq: def O(3) lattice model})
(squares) with that for the effective Ising Hamiltonian
(\ref{eq: effective theory if cal L is Bravais lattice}) (dots)
for the case when the impurity bonds form a superlattice that stabilizes
a long-range spiral order.  Again, good agreement is found once the
the impurity bond strength is sufficiently stronger than the threshold,
in which case the two possible canting patterns form robust local minima
of the Hamiltonian. This is indeed the case in YBaCuFeO$_5$, where
$|J^{\,}_{\mathrm{imp}}|/J^{\,}_{\perp}\approx23>
J^{\,}_{\mathrm{c}}/J^{\,}_{\perp}\approx 9.4$.

\subsection{Dependence of the ground state on the superlattice
of impurity bonds}

We now illustrate how
the choice made for the superlattice of impurity bonds
affects the ground state.
We use again the values (\ref{values of J in YBaCuFeO5})
corresponding to ideal YBaCuFeO$_5$ when defining
the effective Ising Hamiltonian
(\ref{eq: effective theory if cal L is Bravais lattice})
and the microscopic Hamiltonian
(\ref{eq: def O(3) lattice model}).

\begin{figure*}
\centerline{\includegraphics[width=0.9\textwidth]{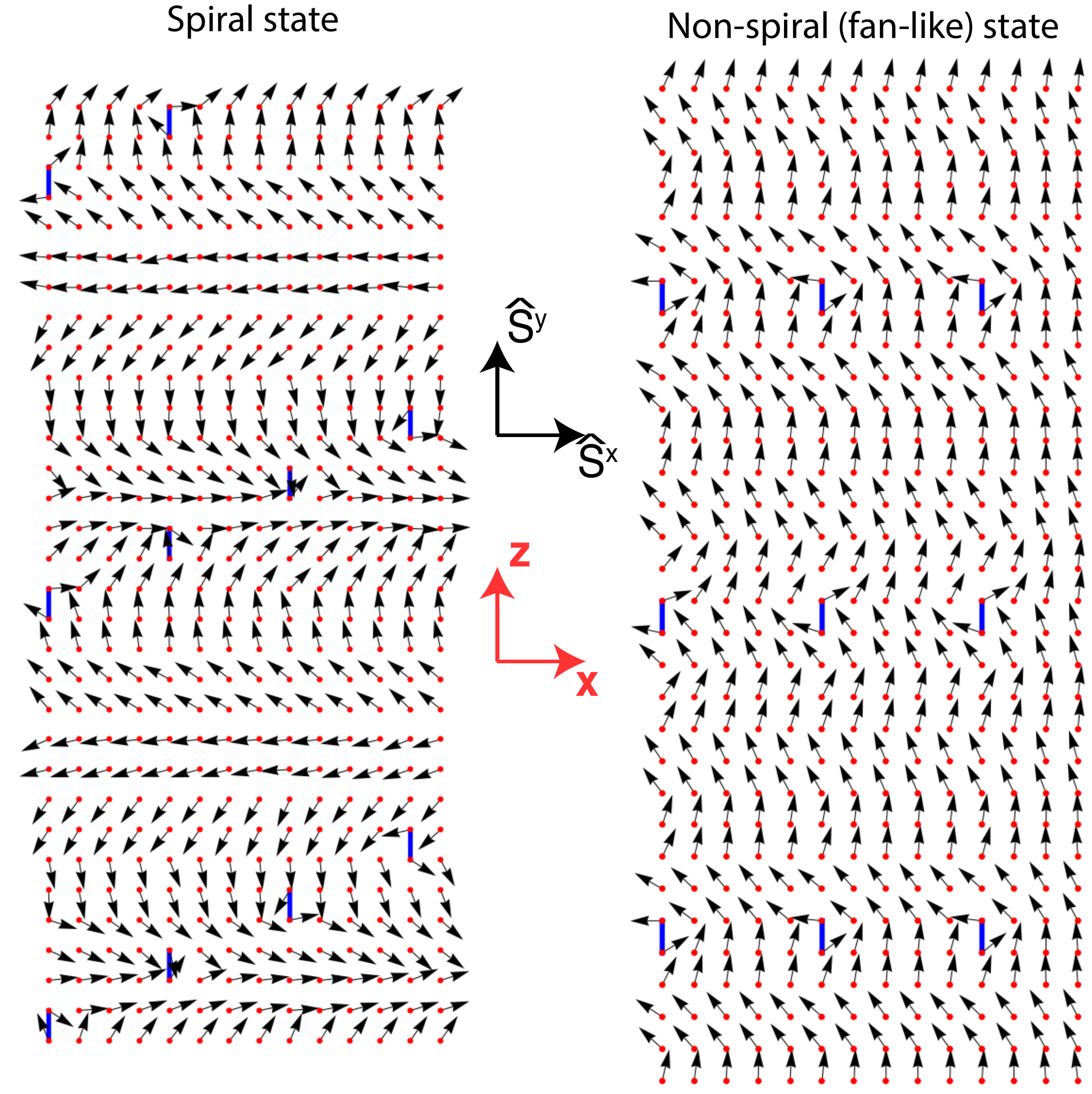}}
\caption{(Color online)
Ground states obtained 
from the microscopic Hamiltonian
(\ref{eq: def O(3) lattice model}).
We choose the values of the exchange couplings
motivated by those calculated for
YBaCuFeO$_5$,
cf.\ Eq.~(\ref{values of J in YBaCuFeO5}),
and with impurity bonds forming a regular superlattice.
The ground state depends qualitatively on the superlattice:
(a) A superlattice with basis vectors 
$\vv{A}=(4,3,0)^{\mathsf{T}}$, 
$\vv{B}=(0,4,2)^{\mathsf{T}}$, 
and
$\vv{C}=(4,0,2)^{\mathsf{T}}$ ($n^{\,}_{\mathrm{imp}}=1/56$)
results in a magnetic spiral.
(b) A superlattice with basis vectors  $\vv{A}=(5,0,0)^{\mathsf{T}}$,
$\vv{B}=(0,5,0)^{\mathsf{T}}$, 
and 
$\vv{C}=(0,1,2)^{\mathsf{T}}$ 
($n^{\,}_{\mathrm{imp}}=1/50$) results in a non-spiral (fan-like) ground state.
In this case,
$\Upsilon^{\,}_{\vv{q}}$
(Eq.~\ref{eq: effective theory if cal L is Bravais lattice c})
takes its minimum at
$\vv{q}^{\,}_{\text{min}}=\vv{C}^{\star}/2$,
which yields a non-spiral ground state (with $Q=0$). 
        }
\label{Fig:SpinFieldTexVal}
\end{figure*}

We consider two superlattices
of impurity bonds. They are chosen such that
$\Upsilon^{\,}_{\vv{q}}$ has a global minimum at 
$\vv{q}^{\,}_{\text{min}}=0$
for one, and at $\vv{q}^{\,}_{\text{min}}=\vv{C}^{\star}/2$ for the other,
cf.~Fig.~\ref{Fig:SpinFieldTexVal}).
Furthermore, both superlattices are chosen {such} that they share with
YBaCuFeO$_5$ the additional property that impurity bonds only occur
between every other plane.

The essential difference between the two superlattices lies in the
relative position of nearest-neighbor impurity bonds.  In the
first lattice, the majority of nearest-neighbor impurity bonds is
ferromagnetically coupled.  This favors a ferromagnetic Ising phase in
the effective Ising Hamiltonian
(\ref{eq: effective theory if cal L is Bravais lattice}),
i.e., a spiral magnetic phase.  In the second
lattice, the majority of nearest-neighbor impurity bonds is
antiferromagnetically coupled.  This favors a layered
antiferromagnetic Ising ground state of the effective Ising
Hamiltonian (\ref{eq: effective theory if cal L is Bravais lattice}),
and thus, a fan-like magnetic order with no net winding of the spins,
whereby the orientation of the magnetization of the layers alternate
between even and odd pairs of planes.

We have verified using MC simulations of the microscopic Hamiltonian
(\ref{eq: def O(3) lattice model}) that the long-range spiral order, which is
present when the impurity bonds are arranged in certain Bravais superlattices,
is robust to weak distortions of that superlattice, as expected on
theoretical grounds.

\subsection{ Limit of dilute impurities}
\label{Sec:RandImp}

Next we analyze the limits of very dilute
tetragonal, face-centered-cubic, and body-centered-cubic
superlattices of impurity bonds. These are tractable analytically.
We shall show that
ferromagnetic order prevails at the
level of the Ising degrees of freedom associated with local cantings
for dilute face-centered-cubic and body-centered-cubic
superlattices of impurity bonds.
This is to say that spiral order for the underlying $XY$
spin degrees of freedom prevails for these diluted superlattices of
impurity bonds.
The results obtained here
will also be useful for the study of random impurity bonds in
Sec.\ \ref{sec: Random impurities: dilute limit}.

\subsubsection{Cubic superlattices}

We assume that the impurity bonds occupy a cubic sublattice
$\mathcal{L}$ of the cubic host lattice $\Lambda$.  As it turns out,
this case supports antiferromagnetic order {in the Ising model}.
{Some results of this calculation will later help us to establish that
  the disordered case, in contrast, orders ferromagnetically.}

If the cubic host lattice $\Lambda$ and the superlattice $\mathcal{L}$
are finite and not too large,
it is possible to calculate the energy
(\ref{eq: final effective Ising Hamiltonian a})
for all Ising spin configurations  by
exact evaluation of the Ising kernel
$J^{(\mathrm{I})}_{\tilde{\vv{r}}-\tilde{\vv{r}}'}$
defined in Eq.~(\ref{eq: final effective Ising Hamiltonian c}).
In the thermodynamic limit, $|\Lambda|\to\infty$,
with $n^{\,}_{\mathrm{imp}}$ held fixed, this approach is not possible anymore. 
Instead, we shall restrict ourselves to a few
long range-ordered Ising configurations that are likely candidates
for the ground state, and compare their energies.

The ferromagnetic Ising configuration is described by
\begin{subequations}
\label{eq: def competing sigma's}
\begin{equation}
\sigma^{\mathrm{F}}_{\tilde{\vv{r}}}\:=1,
\qquad
\frac{1}{|\Lambda|}
\sum_{\tilde{\vv{r}}\in\mathcal{L}}\sigma^{\mathrm{F}}_{\tilde{\vv{r}}}=
n^{\,}_{\mathrm{imp}}.
\label{eq: def sigma F}
\end{equation}
The most relevant competing states have ferromagnetic order in plane
(as favored by the ferromagnetic interactions in the $xy-$plane), but
antiferromagnetic order along the $z$-axis. We consider the family of
states defined by [$\tilde{\vv{r}}=(\tilde{x},\tilde{y},\tilde{z})$]
\begin{equation}
\begin{split}
\sigma^{\mathrm{AF}(m)}_{\tilde{\vv{r}}}\:=
(-1)^{\lfloor\tilde{z}/m\,\ell\rfloor},
\qquad
\frac{1}{|\Lambda|}
\sum_{\tilde{\vv{r}}\in\mathcal{L}}
\sigma^{\mathrm{AF}(m)}_{\tilde{\vv{r}}}=
0,
\end{split}
\label{eq: def sigma AF}
\end{equation}
which describes a sequence of stacks of $m\geq 1$ layers, whose
magnetization alternates. Here,
\begin{equation}
\ell\equiv n^{-1/3}_{\mathrm{imp}}
\end{equation}
\end{subequations}
denotes the lattice spacing of the cubic superlattice, and
$\lfloor\tilde{z}/m\,\ell\rfloor$ returns the integer part of
the fraction $\tilde{z}/m\,\ell$.

After subtraction of the three constants
$E^{\,}_{\mathrm{FM}}$, $E({\Delta\theta})\,|\mathcal{L}|$, and
$-\gamma \, n^{\,}_{\mathrm{imp}}/ 2J^{\,}_{\perp}$
on the right-hand side of
Eq.~(\ref{eq: final effective Ising Hamiltonian a}),
the energy per impurity bond of the configurations
$\mathrm{C}\in\left\{\mathrm{F,AF}(m)\right\}$ is given by
\begin{subequations}
\label{eq: shifted energy per impurity bond}
\begin{equation}
\varepsilon^{\mathrm{C}}_{\mathcal{L}}=
-
\frac{\gamma}{2}
\left(
\sum_{\tilde{\vv{r}}\in \mathcal{L}\setminus\{\vv{0}\}}
\Gamma^{(0)}_{\tilde{\vv{r}}}\,
f^{\mathrm{C}}(\tilde{z})
+
\frac{n^{\,}_{\mathrm{imp}}}{J^{\,}_{\perp}}\,
\delta^{\,}_{\mathrm{C,F}}
\right),
\label{eq: shifted energy per impurity bond a}
\end{equation}
where the spin autocorrelation function
\begin{equation}
f^{\mathrm{C}}(\tilde{\vv{r}})\equiv
\langle
\sigma^{\,}_{\tilde{\vv{r}}'} \,
\sigma^{\,}_{\tilde{\vv{r}}'+\tilde{\vv{r}}}
\rangle^{\,}_{\tilde{\vv{r}}'}=
f^{\mathrm{C}}(\tilde{z})
\label{eq: shifted energy per impurity bond b}
\end{equation}
only depends on the difference in the $\tilde{z}$ coordinate, owing to
Eq.~(\ref{eq: def competing sigma's}). Here,
$\langle\dots\rangle_{\tilde{\vv{r}}'}$
denotes the average over the sites ${\tilde{\vv{r}}'}$
of the superlattice $\mathcal{L}$.
For configurations $\mathrm{F}$ and $\mathrm{AF(1)}$, it is given by
\begin{equation}
f^{\mathrm{F}}(\tilde{z})=1,
\qquad
f^{\mathrm{AF(1)}}(\tilde{z})=(-1)^{\tilde{z}/\ell}.
\end{equation}
\end{subequations}

In the dilute limit $n^{\,}_{\mathrm{imp}}\to 0$,
the typical distance between a pair of nearest-neighbor impurities is large.
Hence, the typical pair-wise interaction
$\Gamma^{(0)}_{\tilde{\vv{r}}}$ tends to the dipolar form
(\ref{dipolarint}) and can be safely used to evaluate
$\varepsilon^{\mathrm{AF}(m)}_{\mathcal{L}}$ up to  corrections which are subleading
in the limit $n^{\,}_{\mathrm{imp}}\to 0$. The case of the ferromagnetic
configuration is more subtle, however. Indeed{,} a naive use of
Eq.~(\ref{dipolarint}) would suggest that the first term in
the right-hand side of
Eq.~(\ref{eq: shifted energy per impurity bond a})
vanishes due to the sum over symmetry related directions,
while in fact it does not. This is due to
corrections to the dipolar interaction (\ref{dipolarint}) that
scale as the inverse of the volume, but  add up to a finite
contribution when summed with equal signs over the whole
superlattice. In the case of an isotropically shaped, cubic sample
with $L^{\,}_{x}=L^{\,}_{y}=L^{\,}_{z}$ and isotropic interactions
$J^{\,}_{\parallel}=J^{\,}_{\perp}\equiv J$, the computation can be done
exactly, using the fact that upon averaging over all the permutations
$k^{\,}_{x}\to k^{\,}_{y}\to k^{\,}_{z}\to k^{\,}_{x}$
the kernel $\hat\Gamma_{\vv{k}}^{(0)}$ (\ref{eq: def Gammas and hat Gammas b})
reduces to $1/3J$.
Using $f^{\mathrm{F}}(z)=1$ this allows us to evaluate the lattice sum
exactly for any impurity density as
\begin{align}
\sum_{\tilde{\vv{r}}\in\mathcal{L}\setminus\{\vv{0}\}}
\Gamma^{(0)}_{\tilde{\vv{r}}} f^{\mathrm{F}}(\tilde{z}) =&\,
\frac{1}{3J\,|\Lambda|}
\sum_{\tilde{\vv{r}}\in\mathcal{L}\setminus\{\vv{0}\}}
\sum_{\vv{k}\in\mathrm{BZ}(\Lambda)\setminus\{\vv{0}\}}
e^{\mathrm{i}\vv{k}\cdot \tilde{\vv{r}}}
\nonumber\\
=&\,
\frac{1}{3J\,|\Lambda|}
\sum_{\tilde{\vv{r}}\in\mathcal{L}\setminus\{\vv{0}\}}
\left(
|\Lambda|\,\delta^{\,}_{\tilde{\vv{r}},\vv{0}}-1
\right)
\nonumber\\
=&\,
-\frac{1}{3J}
\frac{|\mathcal{L}|-1}{|\Lambda|}
\nonumber\\
=&\,
-
\frac{n^{\,}_{\mathrm{imp}}}{3J}+\mathcal{O}\left(\frac{1}{|\Lambda|}\right).
\label{eq: demagnetization effect}
\end{align}
This finite, negative contribution disfavors the ferromagnet, in
analogy to demagnetizing factors known from standard magnetic dipolar
systems. Restricting ourselves to the isotropic case and inserting
Eq.~(\ref{eq: demagnetization effect}) into
Eq.~(\ref{eq: shifted energy per impurity bond a}),
we obtain the energy per impurity
\begin{equation}
\varepsilon^{\mathrm{F}}_{\mathcal{L}}=
-
\frac{1}{3}\,
\frac{\gamma\,n^{\,}_{\mathrm{imp}}}{J^{\,}_{\perp}}.
\label{eq: evaluation EFerro if isotropic dipolar}
\end{equation}

More generally, it is useful to cast the energy
(\ref{eq: shifted energy per impurity bond a})
for the ferromagnetic configuration (\ref{eq: def sigma F})
of the Ising variables
in a different form, namely,
\begin{align}
\frac{\varepsilon^{\mathrm{F}}_{\mathcal{L}}}{(\gamma/2)}=&\,
-
\sum_{\tilde{\vv{r}}\in\mathcal{L}}
\Gamma^{(0)}_{\tilde{\vv{r}}}
+
\Gamma^{(0)}_{\tilde{\vv{r}}=\vv{0}}
-
\frac{n^{\,}_{\mathrm{imp}}}{J^{\,}_{\perp}}
\nonumber\\
=&\,
-
\frac{1}{|\Lambda|}
\sum_{\tilde{\vv{r}}\in\mathcal{L}}
\sum_{\vv{k}\in\mathrm{BZ}(\Lambda)}
e^{\mathrm{i}\vv{k}\cdot\tilde{\vv{r}}}\,
\hat{\Gamma}^{(0)}_{\vv{k}}
+
\Gamma^{(0)}_{\tilde{\vv{r}}=\vv{0}}
-
\frac{n^{\,}_{\mathrm{imp}}}{J^{\,}_{\perp}}
\nonumber\\
=&\,
-
\frac{|\mathcal{L}|}{|\Lambda|}
\sum_{
\substack{
\vv{k}\in\mathrm{BZ}(\Lambda)
\\
\vv{k}\in\mathcal{L}^{\star}
         }
     }
\hat{\Gamma}^{(0)}_{\vv{k}}
+
\Gamma^{(0)}_{\tilde{\vv{r}}=\vv{0}}
-
\frac{n^{\,}_{\mathrm{imp}}}{J^{\,}_{\perp}},
\label{EFerro}
\end{align}
where
the reciprocal lattice $\mathcal{L}^{\star}$ of
$\mathcal{L}$ {enters through the identity
(\ref{eq: Fourier identity})},
$\hat{\Gamma}^{(0)}_{{\vv{k}}{\neq\vv{0}}}$ has been
defined in Eq.~(\ref{eq: def Gammas and hat Gammas b}) and we recall that
$\hat\Gamma^{(0)}_{\vv{k}=\vv{0}}=0$. The sum over $\vv{k}$ thus
contains $|\Lambda|/|{\cal L}|- 1=1/n^{\,}_{\mathrm{imp}} -1$ terms.
Note that the self-interaction $\Gamma^{(0)}_{\tilde{\vv{r}}=\vv{0}}$
[cf.\, Eq.~\ref{eq: final saddle point for one impurity b})],
which appears also in the single impurity energy, is subtracted on the
right-hand side of Eq.~(\ref{EFerro}).

We point out an important difference between the present effective
dipolar problem and genuine magnetic dipoles. Genuine dipolar
interactions are mediated by magnetic fields which extend everywhere
in space, beyond the boundaries of the sample. Therefore they only
depend on the relative position of two spins, irrespective of where
the spins are deep in the bulk, or close to a surface of a finite
sample. However, this is not so in our case where the dipolar
interactions arise through the mediation of spin waves, which are
confined to the sample. Accordingly, the interactions involving Ising
spins at the periphery of the sample are not exactly the same as those
for bulk Ising spins with the same relative position. More importantly
there are no magnetic stray fields beyond the sample.  In real dipolar
magnets those store a lot of magnetic energy, which is avoided in the
ground state by domain formation. The unavoidable presence of domains
complicates the computation of the energy density. In particular, the
evaluation for a homogeneously magnetized sample yields a shape
dependent result, a fact that is reflected in the ambiguity of the
value of the Fourier transform of the dipolar interactions
Eq.~(\ref{eq: def Gammas and hat Gammas b}) in the limit $\vv{k}\to
0$. In the present case, however, such problems do not arise, since
the spin-wave mediated interaction is such that
$\hat\Gamma^{(0)}_{\vv{k}=\vv{0}}=0$. This eliminates the potential
ambiguity and therefore eliminates the shape dependence. We also do
not expect the effective dipolar interactions to induce domains, in
contrast to genuine ferromagnets.

Performing an analogous calculation to the one above yields for the
antiferromagnet $\mathrm{AF}(1)$
\begin{align}
\label{AF1}
\frac{\varepsilon^{\mathrm{AF}(1)}_{\mathcal{L}}}{(\gamma/2)}=&\,
-\sum_{\tilde{\vv{r}}\in\mathcal{L}}
\Gamma^{(0)}_{\tilde{\vv{r}}}\,
(-1)^{\lfloor\tilde{z}/\ell\rfloor}
+
\Gamma^{(0)}_{\tilde{\vv{r}}=\vv{0}}
\nonumber\\
=&\,
-\frac{|\mathcal{L}|}{|\Lambda|}
\sum_{
\substack{
\vv{k}\in\mathrm{BZ}(\Lambda)
\\
\vv{k}+(0,0,\pi/\ell)^{\mathsf{T}}\in\mathcal{L}^\star
         }
     }
\hat\Gamma^{(0)}_{\vv{k}}
+
\Gamma^{(0)}_{\tilde{\vv{r}}=\vv{0}}.
\end{align}

Even though the quantitative mapping from the $XY$ Hamiltonian
(\ref{eq: def 3d XY model}) to the effective Ising Hamiltonian
(\ref{eq: final effective Ising Hamiltonian}) only holds for low
densities of impurity bonds, it is useful to study the effective Ising
Hamiltonian (\ref{eq: final effective Ising Hamiltonian}) in its own
right, i.e., without requiring the impurity bonds to be dilute.

A maximally dense superlattice is defined by
\begin{equation}
\ell=1,\qquad
\mathcal{L}=\Lambda,\qquad
n^{\,}_{\mathrm{imp}}=\frac{|\mathcal{L}|}{|\Lambda|}=1.
\end{equation}
For such a lattice,
one finds the ferromagnetic (F) and antiferromagnetic
[AF(1)] states to be degenerate,
\begin{equation}
\frac{
\varepsilon^{\mathrm{F}}_{\mathcal{L}}
-
\varepsilon^{\mathrm{AF}(1)}_{\mathcal{L}}
     }
     {
(\gamma/2)\, n^{\,}_{\mathrm{imp}}
     } 
=\,
\hat\Gamma^{(0)}_{\vv{k}=(0,0,{\pi})^{\mathsf{T}}}
-
\hat\Gamma^{(0)}_{\vv{k}=\vv{0}}
-
\frac{1}{J^{\,}_{\perp}}=0,
\end{equation}
since, cf.\ Eqs.~(\ref{eq: def Gammas and hat Gammas b})
{and (\ref{eq: def Gammas and hat Gammas c})},
\begin{equation}
\hat\Gamma^{(0)}_{(0,0,k^{\,}_{z})^{\mathsf{T}}}=
\frac{1-\delta^{\,}_{k^{\,}_{z},0}}{J^{\,}_{\perp}},
\label{eq: identity Gamma k along z axis}
\end{equation}
The identity (\ref{eq: identity Gamma k along z axis})
obeyed by the kernel (\ref{eq: def Gammas and hat Gammas b})
can be used together with the expression
(\ref{eq: effective theory if cal L is Bravais lattice a})
and the fact that only $\vv{q}$ of the form
$(0,0,k^{\,}_{z})^{\mathsf{T}}$
enter it, to show that for a maximally dense superlattice
\textit{all} antiferromagnetic
states $\mathrm{AF}(m)$ are degenerate with the ferromagnet.  More
generally, it is shown in appendix \ref{appsec: Lattice sum} that the
ferromagnet is degenerate with \textit{any} Ising configuration
in which the spins in every given plane at fixed
$z$ coordinate are ferromagnetically aligned,
irrespective of the relative orientation
of the magnetization of different planes. 

This degeneracy is, however, lifted at finite dilution, whereby the
way in which the dilution is realized is crucial.
\begin{figure}
\centerline{\includegraphics[width=0.5\textwidth]{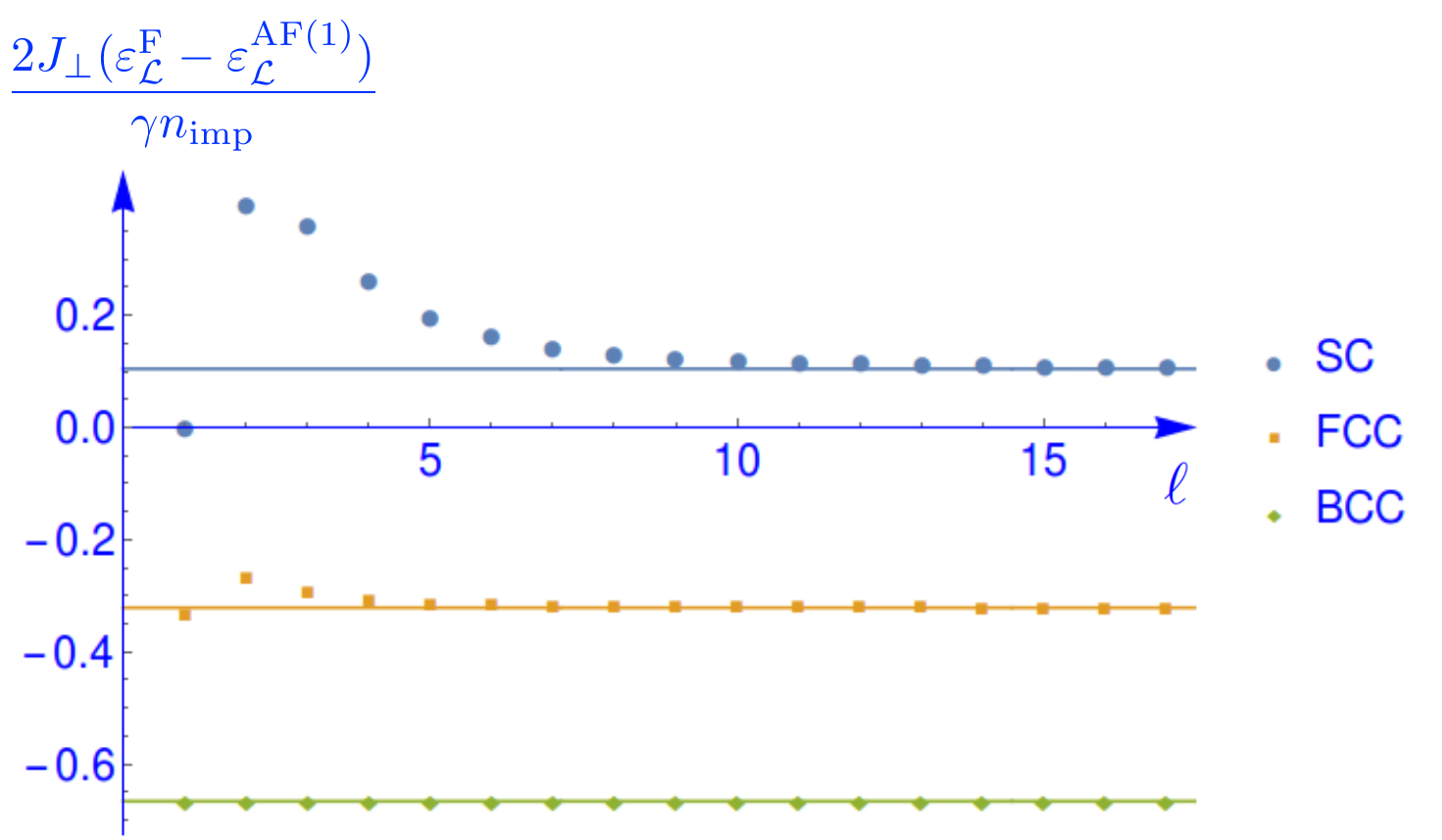}}
\caption{(Color online)
Dependence of the energy difference between the Ising ferromagnetic
(F) and antiferromagnetic [AF($1$)] states on the linear size $\ell$
of the unit cell of the superlattice for different classes of
superlattices. Blue dots represent a simple cubic (SC) superlattice
with the basis vectors $(\ell,0,0)$, $(0,\ell,0)$, and
$(0,0,\ell)$. Yellow squares represent a face centered cubic (FCC)
superlattice with lattice vectors $(\ell,\ell,0)$, $(\ell,0,\ell)$,
and $(0,\ell,\ell)$. Green diamonds represent body centered cubic
(BCC) superlattices with the lattice vectors $(\ell,\ell,\ell)$,
$(\ell,-\ell,\ell)$, and $(\ell,\ell,-\ell)$. The correspondingly
colored horizontal solid lines represent the dilute limit
$\ell\to\infty$ of these energy differences for each superlattice.
        }
\label{superlatticeFvsAF_nimp}
\end{figure}
For example, diluting the impurity density $n^{\,}_{\mathrm{imp}}$
by maintaining a cubic superlattice,
but increasing its integer lattice spacing $\ell$ disfavors the
ferromagnetic state. This is illustrated in
Fig.\ \ref{superlatticeFvsAF_nimp},
where we plot the energies per impurity
as a function of superlattice spacing $\ell$. For small $\ell$,
the energy difference is obtained from the representations
(\ref{EFerro}) and (\ref{AF1}). In the dilute limit,
$n^{\,}_{\mathrm{imp}}=\ell^{-3}\to 0$, the reciprocal lattice
$\mathcal{L}^\star$ only contains small wavevectors,
and we may replace $1-\cos k^{\,}_{i}$
(where $k^{\,}_{i}=2\pi\,n^{\,}_{i}/\ell$)
in the kernel (\ref{eq: def Gammas and hat Gammas b})
by $(2\pi\,n^{\,}_{i})^{2}/2\ell^{2}$
with $n^{\,}_{i}\in\mathbb{Z}$ for $i=x,y,z$, i.e.,
\begin{subequations}
\label{eq: diff energy between F AF(1) for cubic lattice}
\begin{equation}
\frac{
\varepsilon^{\mathrm{F}}_{\mathcal{L}}
-
\varepsilon^{\mathrm{AF}(1)}_{\mathcal{L}}
     }
     {
(\gamma/2)\,
n^{\,}_{\mathrm{imp}}
     } 
\to
\frac{\delta(\alpha)}{J^{\,}_{\perp}}
\label{eq: diff energy between F AF(1) for cubic lattice a}
\end{equation}
where the exchange anisotropy parameter $\alpha$ was defined in
Eq.~(\ref{eq: def alpha}),
and
{\begin{equation}
\begin{split}
\delta(\alpha)\:=&\,
-
1
-
\sum_{\vv{n}\in\mathbb{Z}^{3}\setminus{\{\vv{0}\}}}
\frac{
n^{2}_{z}
     }
     {
\alpha
\left(n^{2}_{x}+n^{2}_{y}\right)
+
n^{2}_{z}
     }
\\
&\,
+ \sum_{\vv{n}\in\mathbb{Z}^{3}}
\frac{
\left(n^{\,}_{z}-1/2\right)^{2}
     }
     {
\alpha
\left(n^{2}_{x}+n^{2}_{y}\right)
+
(n^{\,}_{z}-1/2)^{2}
     }.
\label{eq: diff energy between F AF(1) for cubic lattice c}
\end{split}
\end{equation}}
\end{subequations}

The sum over $n^{\,}_{z}$ can be carried out explicitly,
\begin{equation}
\delta(\alpha)=
\sum_{\vv{n}\in\mathbb{Z}^{2}\setminus\{\vv{0}\}}
\frac{
2\pi\,
\sqrt{\alpha\,(n^{2}_{x}+n^{2}_{y})}
     }
     {
\sinh\left(2\pi\sqrt{\alpha(n^{2}_{x}+n^{2}_{y})}\right)
     }.
\end{equation}
Hence, $\delta(\alpha)$ is always positive.
For the isotropic limit $\alpha=1$, one finds $\delta(1)\approx 0.1042$.

Alternatively, one can calculate the antiferromagnetic energy directly
in real space using the dipolar form (\ref{dipolarint}).
This can be used to calculate the energies of other antiferromagnetic states
$\mathrm{AF}(m)$, which all scale as
\begin{equation}
\frac{\varepsilon^{\mathrm{AF}(m)}_{\mathcal{L}}}{(\gamma/2)\,n^{\,}_{\mathrm{imp}}}=
-
\frac{c^{\,}_{m}}{J^{\,}_{\perp}}.
\label{eq: evaluation EAFerro if isotropic dipolar}
\end{equation}
From the results (\ref{eq: evaluation EFerro if isotropic dipolar},
\ref{eq: diff energy between F AF(1) for cubic lattice a})
it follows that $c^{\,}_{1}=\delta(1)+2/3$, while one
finds the higher $c^{\,}_{m}$'s to decrease monotonically with
increasing $m$.  From this we conclude that a dilute cubic
superlattice orders antiferromagnetically with layer magnetizations
that alternate in sign ($m=1$).

\subsubsection{Dilute tetragonal, face-centered,
and body-centered tetragonal superlattices}

One readily generalizes the above calculation to tetragonal
superlattices $\mathcal{L}$ with unit vectors
$(A\ell,0,0)^{\mathsf{T}}$, $(0,A\ell,0)^{\mathsf{T}}$,
$(0,0,C\ell)^{\mathsf{T}}$, where $A$ and $C$ are fixed integers while
the integer-valued dilution parameter $\ell$ will be taken to
infinity.  This case is obtained from that of a cubic lattice {by substituting }
\begin{equation}
n^{\,}_{\mathrm{imp}}\to\frac{1}{A^{2}\,C\,\ell^{3}},
\qquad
\alpha\to
\frac{J^{\,}_{\parallel}}{J^{\,}_{\perp}}\,
\frac{C^{2}}{A^{2}},
\label{eq: cubic to tetragonal}
\end{equation}
in Eq.~(\ref{eq: diff energy between F AF(1) for cubic lattice a})
and (\ref{eq: diff energy between F AF(1) for cubic lattice c}).
Independently of the ratio $C/A$ of the tetragonal superlattice,
the Ising antiferromagnetic state
AF(1) is favored over the Ising ferromagnetic state F.

However, similarly as in lattice problems of physical electric or
magnetic dipoles~\cite{LuttingerTisza}
where the interactions have
reversed global sign, a different ground state is found in dilute
body-centered or face-centered tetragonal lattices. The difference
arises because closest neighbors in these lattices have a stronger
tendency to have ferromagnetic interactions than in simple tetragonal
lattices. For the face-centered tetragonal lattice, the basis vectors
are $(A,A,0)$, $(A,0,C)$, and $(0,A,C)$. The corresponding dual basis
vectors in reciprocal space are $\vv{e}^{\,}_{1}=\pi(1/A, 1/A,-1/C)$,
$\vv{e}^{\,}_{2}=\pi(1/A,-1/A,1/C)$, and
$\vv{e}^{\,}_{3}=\pi(-1/A,1/A,1/C)$.  Their linear combinations with
integer coefficients span the reciprocal lattice
$\mathcal{L}^{\star}$.  It is convenient to represent a generic
reciprocal lattice vector $\vv{G}\in\mathcal{L}^{\star}$ as $\vv{G}=
n^{\,}_{1}\,\vv{e}^{\,}_{1}+n^{\,}_{2}\,\vv{e}^{\,}_{2}+n^{\,}_{3}
(\vv{e}^{\,}_{2}+\vv{e}^{\,}_{3})$.  With this choice,
the asymptotic
energy difference between the ferromagnetic and the antiferromagnetic
states in the infinite dilution limit $n^{\,}_{\mathrm{imp}}\to0$ can
be written as 
\begin{subequations}
\begin{equation}
\frac{
	\varepsilon^{\mathrm{F}}_{\mathcal{L}}
	-
	\varepsilon^{\mathrm{AF}(1)}_{\mathcal{L}}
}
{
	(\gamma/2)\,
	n^{\,}_{\mathrm{imp}}
} 
=
-\frac{1}{	J^{\,}_{\perp}       }
-
\sum_{\vv{n}\in \mathbb{Z}^{3}\setminus{\{\vv{0}\}}}
\frac{
	g^{\,}_{\vv{n}}
}
{
	J^{\,}_{\perp}       
}
+
\sum_{\vv{n}\in \mathbb{Z}^{3}}
\frac{
	g^{\,}_{{(n_1,n_2,n_3+\frac{1}{2})}}
}
{
	J^{\,}_{\perp}       
}
, 
\end{equation}
where
\begin{equation}
g^{\,}_{\vv{n}}\:=
\frac{
(n^{\,}_{1}-n^{\,}_{2}-2n^{\,}_{3})^{2}
     }
     {
\alpha
\left[(n^{\,}_{1}+n^{\,}_{2})^{2}+(n^{\,}_{1}-n^{\,}_{2})^{2}\right]
+
(n^{\,}_{1}-n^{\,}_{2}-2n^{\,}_{3})^{2}
     },
\end{equation}
\end{subequations}
where for vanishing wavevector we have to set $g^{\,}_{\vv{0}}=0$.
Carrying out the sum over $n^{\,}_{3}$ one finds 
\begin{equation}
\begin{split}
\frac{
\varepsilon^{\mathrm{F}}_{\mathcal{L}}-\varepsilon^{\mathrm{AF}(1)}_{\mathcal{L}}
     }
     {
(\gamma/2)\, n^{\,}_{\mathrm{imp}}
     }
=&\, 
\sum_{\vv{n}\in\mathbb{Z}^{2}\setminus\{\vv{0}\}}
\frac{
(-1)^{n^{\,}_{1}-n^{\,}_{2}}
     }
     {
J^{\,}_{\perp}
     }
\\
&\,
\times
\frac{
\pi\,\sqrt{2\alpha\,\left(n^{2}_{1}+n^{2}_{2}\right)}
     }
     {
\sinh\left(\pi\sqrt{2\alpha\,\left(n^{2}_{1}+n^{2}_{2}\right)}\right)
     }.
\end{split}
\end{equation}
In the isotropic case $\alpha=1$, the energy difference is negative $J(\varepsilon^{\mathrm{F}}_{\mathcal{L}}-\varepsilon^{\mathrm{AF}(1)}_{\mathcal{L}})/(\gamma/2)\, n^{\,}_{\mathrm{imp}}=-0.3218$, so the ferromagnetic order prevails. 

 \begin{figure}[t!]
\centerline{\includegraphics[width=0.5\textwidth]{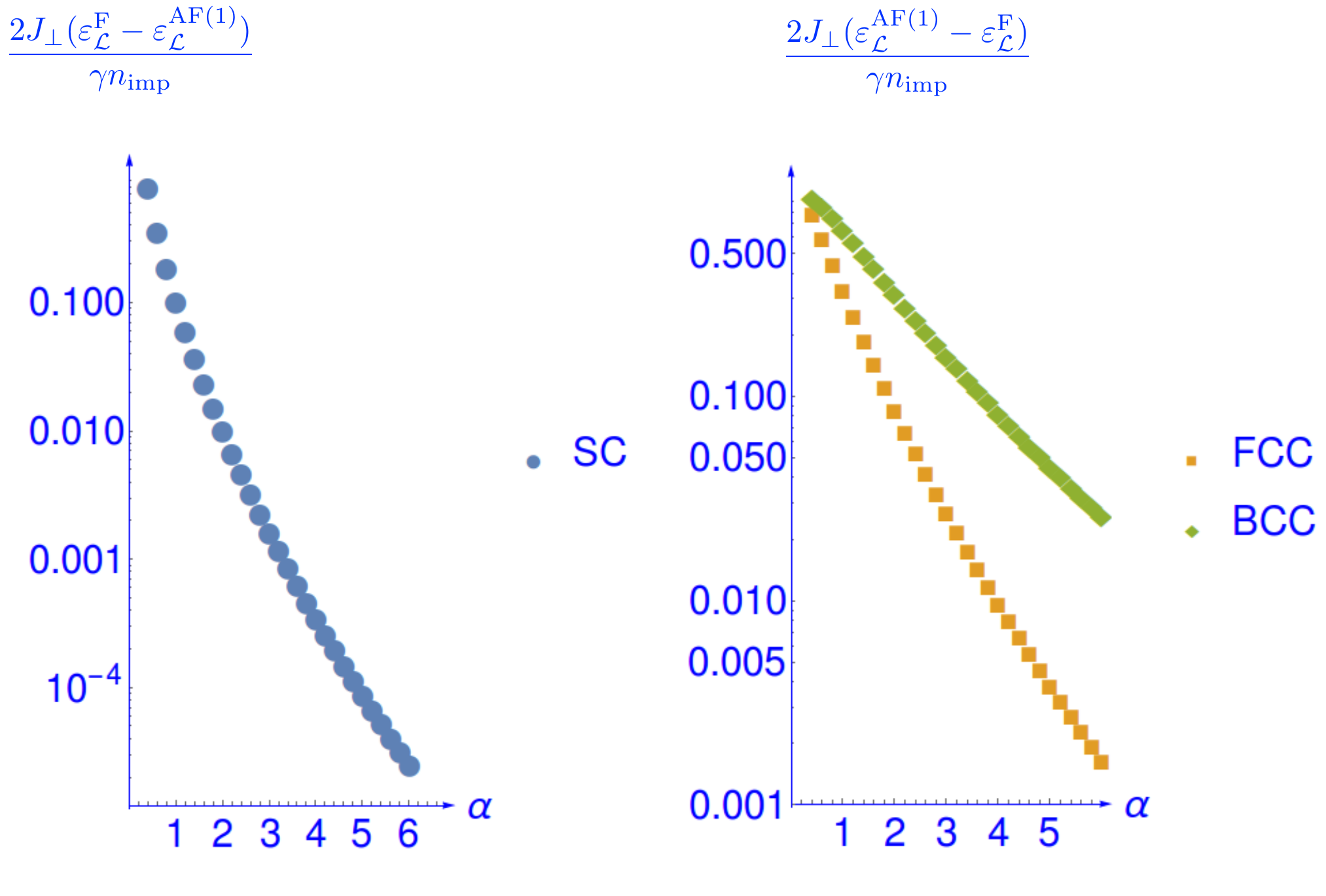}}
\caption{(Color online)
Dependence of the energy difference between the Ising ferromagnetic
state (F) and antiferromagnetic state [AF(1)] as a function of
$\alpha\equiv J^{\,}_{\parallel}/J^{\,}_{\perp}$
[recall Eq.~(\ref{eq: def alpha})]
for a simple cubic superlattice (right panel) in the large
dilution limit. The left panel shows the energy difference between the
Ising antiferromagnetic state [AF(1)] and ferromagnetic state (F) for
face centered superlattices (yellow squares) and a body centered
superlattice (green diamonds). Independent{ly} of the value of $\alpha$,
the antiferromagnetic state has lower energy for simple cubic
superlattices while the ferromagnetic state has lower energy for face
and body centered superlattices.
        }
\label{fig:fcc}
\end{figure}
For body-centered tetragonal lattices one finds the same expression,
but with the replacement $2\alpha\to\alpha$. The energy difference turns
out to be always negative for any value of $\alpha$, as seen in
Fig.\ \ref{fig:fcc}. Thus, in both these types of superlattices
the ferromagnetic state
is  favored over the layered antiferromagnetic state, whatever the tetragonal
aspect ratio.

\section{Random impurities: dilute limit}
\label{sec: Random impurities: dilute limit}

In this section, we study
randomly distributed impurities that occupy
a fraction $n^{\,}_{\mathrm{imp}}$
of the sites of the cubic host lattice $\Lambda$. We assume
again that the relevant contenders for the ground state are given by
 Eqs.~(\ref{eq: def sigma F}) and
(\ref{eq: def sigma AF}). In Eq.~(\ref{eq: def sigma AF}), we set
$\ell=1$, since only the lattice constant of the cubic host lattice
$\Lambda$ is relevant. These configurations are expected to come
reasonably close to the true ground state and the relevant competing
metastable configurations. However, they will differ
in the orientation of a few spins from
the simple configurations (\ref{eq: def sigma F}) and
(\ref{eq: def sigma AF}). The relative fraction of these spins becomes increasingly small as $n^{\,}_{\mathrm{imp}}\to0$, as discussed below.

If the impurities are distributed randomly according to a Poisson
process, the average energy per impurity bond of the trial configurations
$\mathrm{C}=\mathrm{F},\mathrm{AF}(m)$ is given by 
\begin{equation}
\varepsilon^{\mathrm{C}}_{\mathrm{dis}}=
-
\frac{\gamma}{2}\,
n^{\,}_{\mathrm{imp}}
\left(
\sum_{{\vv{r}}\in\Lambda}
\Gamma^{(0)}_{{\vv{r}}}\,
f^{\mathrm{C}}({z})
+
\frac{1}{J^{\,}_{\perp}}\,
\delta^{\,}_{\mathrm{C,F}}
\right),
\label{eq: Eperimp_dis}
\end{equation}
since any site ${\vv{r}}$ of the cubic host lattice $\Lambda$
is the lower end of an impurity bond
with probability $n^{\,}_{\mathrm{imp}}$,
independently of the location of other impurities.
From this observation, one might at first conclude that 
the antiferromagnetic state should dominate again.
However, the above consideration does not treat correctly
impurities located at short distances from each other. On the one hand,
rare pairs of impurities that are located much closer to each other than
the average separation
$n^{-1/3}_{\mathrm{imp}}$ do not follow the pattern (\ref{eq: def sigma F})
and (\ref{eq: def sigma AF}), but simply optimize their mutual interaction
energy, irrespective of the global ordering pattern. Since such pairs
nevertheless contribute a finite fraction to the total energy
estimated above, they must be corrected for, which will turn out to
favor the ferromagnetic ordering. This conclusion will become clear below, as
a corollary to the discussion of another short-distance effect,
which we will consider first. 

Impurity distributions in real materials are usually not simply
governed by a Poisson process, but rather, one should expect them
to exhibit some short-range correlations. For example, in the case of
YBaCuFeO$_5$ impurity bonds arise due to chemical disorder which
occasionally replaces the usual Cu-Fe pairs on bonds along its
crystallographic $c$-axis
by impurity configurations consisting in Fe-Fe or Cu-Cu
pairs. Fe-Fe pairs differ from Fe-Cu pairs by the sign and magnitude of the
resulting magnetic exchange constant. Moreover, both Fe-Fe and Cu-Cu pairs
differ from Fe-Cu pairs in their local charge density. The resulting Coulomb
repulsion between such impurity configurations thus suppresses
the occurence of pairs of impurities at short distances.
In a crude manner, we
can mimic this effect by a hard constraint on the minimal distance
between impurities, excluding distance vectors with $|{\vv{r}}|\leq
R$. With such a constraint the average energy per impurity
(\ref{eq: Eperimp_dis}) is modified to
\begin{equation}
\varepsilon^{\mathrm{C}}_{\mathrm{dis}}(R)=
-
\frac{\gamma}{2}\,
n^{\,}_{\mathrm{imp}}\,
\left(
\sum_{\substack{{\vv{r}}\in\Lambda\\|{\vv{r}}|>R}}
\Gamma^{(0)}_{{\vv{r}}}\,
f^{\mathrm{C}}({z})
+
\frac{1}{J^{\,}_{\perp}}\,
\delta^{\,}_{\mathrm{C,F}}
\right).
\label{eq: Eperimp_dis_R}
\end{equation}

\begin{figure}
\centerline{\includegraphics[width=0.5\textwidth]{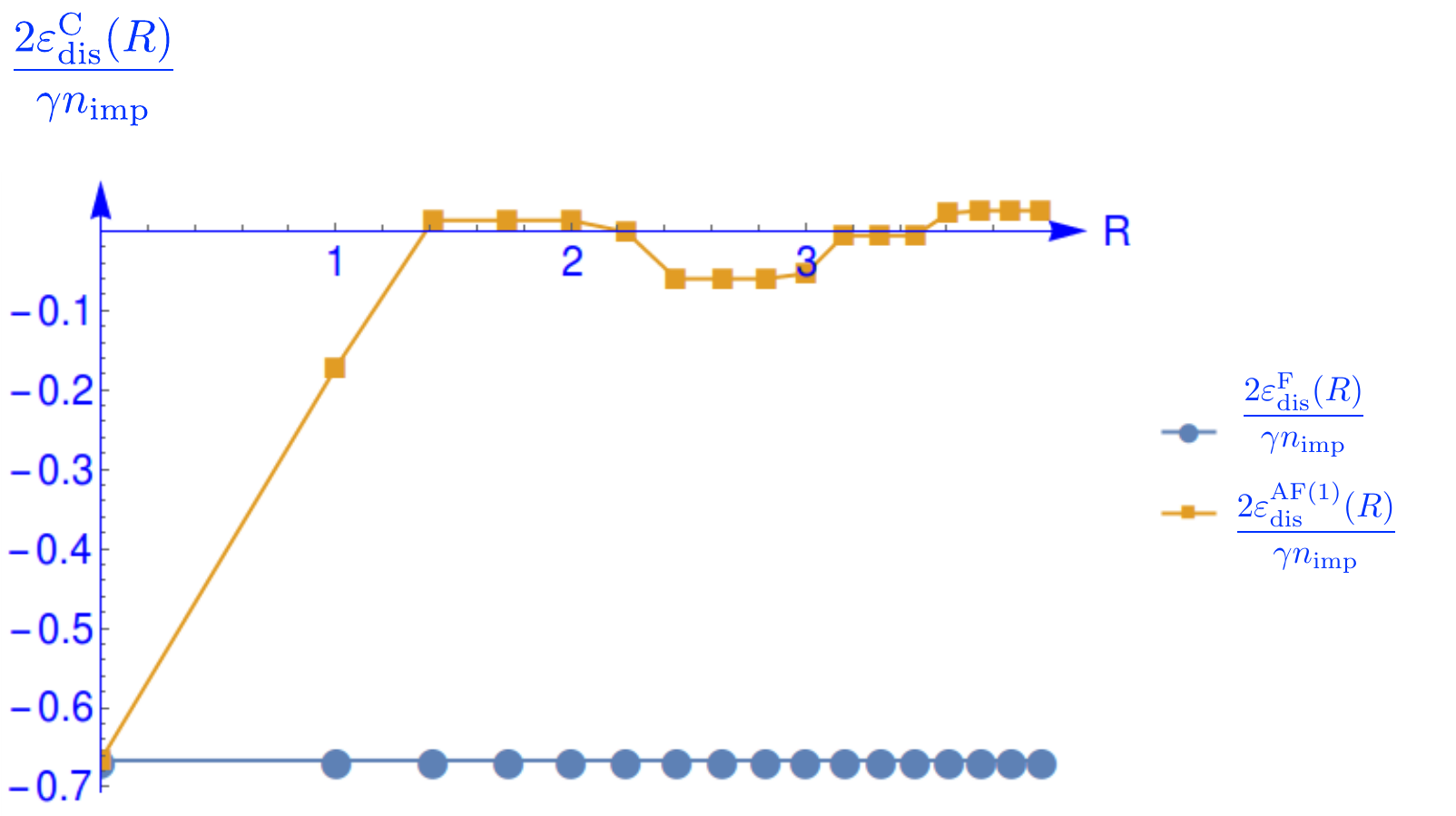}}
\caption{
 (Color online)
Dependence on $R$ of
$\varepsilon^{\mathrm{C}}_{\mathrm{dis}}(R)/(\gamma\,n^{\,}_{\mathrm{imp}})$
defined in Eq.~(\ref{eq: Eperimp_dis_R})
for isotropic couplings $J^{\,}_{\perp}=J^{\,}_{\parallel}\equiv J$
for the ferromagnetic ($\mathrm{C}=\mathrm{F}$, blue dots)
and the layered antiferromagnetic state
($\mathrm{C}=\mathrm{AF}(1)$, yellow squares).
Energies are given in units of $J$.
        }
\label{Fig:constraineddisorderlattice}
\end{figure}

Note that for $R=0$ these energies are simply $n^{\,}_{\mathrm{imp}}$
multiplying the energy per impurity
$\varepsilon_{\mathcal{L}=\Lambda}^{\mathrm{C}}(R)$ of a maximally
dense system of impurities, recall
Eq.~(\ref{eq: shifted energy per impurity bond a}). As we have shown
in the previous section, those energies are all degenerate.  Since the
sum over $\vv{r}$ in Eq.~(\ref{eq: Eperimp_dis_R}) is dominated by
small $|\vv{r}|$, even a small $R$ of the order of one lattice
constant will have a decisive effect and lifts this degeneracy. In
Fig.\ \ref{Fig:constraineddisorderlattice}, we plot as a function of
$R$ the average energies
$2\varepsilon^{\mathrm{F}}_{\mathrm{dis}}(R)/(\gamma\,n^{\,}_{\mathrm{imp}})$ and
$2\varepsilon^{\mathrm{AF}(m=1)}_{\mathrm{dis}}(R)/(\gamma\,n^{\,}_{\mathrm{imp}})$
of the two most relevant competing states.
Already, for the smallest effective exclusion radius of $R\geq
R^{\,}_{\mathrm{c}}=1$ (in units of the host cubic lattice spacing),
we find that the ferromagnetic state (and thus $XY$ spiral order) wins
over the antiferromagnetic state (i.e., $XY$ fan order).  This
numerical result can be understood by recalling that
$\varepsilon^{\mathrm{F}}$ and $\varepsilon^{\mathrm{AF}(1)}$ are
degenerate for $R=0$. Upon barring impurities on nearest-neighbor
sites on the host cubic lattice, the two states receive a relative
energy shift
$4n^{\,}_{\mathrm{imp}}\Gamma^{\,}_{\vv{r}=\vv{z}}=4n^{\,}_{\mathrm{imp}}
\times(\frac{A}{J})$, which stabilizes the ferromagnetic state ($A
\approx 0.123$). Larger exclusion radii tend to reinforce this trend,
as shown in Fig.\ \ref{Fig:constraineddisorderlattice}.  In the limit
of large $R$, the energy per impurity bond of the ferromagnetic state
is more favorable than that of the antiferromagnetic one by $\gamma\,
n^{\,}_{\mathrm{imp}} /(3J)$ in the case of isotropic couplings. This can
be understood as follows. For isotropic couplings, the ferromagnetic
energy per bond, $\varepsilon^{\mathrm{F}} = -\gamma\,n^{\,}_{\mathrm{imp}}
/(3J)$, remains unchanged upon excluding the interactions with a set
of sites that is invariant under the cubic symmetry group, as seen in
Fig.\ \ref{Fig:constraineddisorderlattice}. In contrast, in an
antiferromagnetic state, the interactions with the neighbors in thin
spherical shells of approximately fixed radius $r>R$) come with
alternating signs. Those tend to cancel the more effectively the
larger is $R$, such that
$\varepsilon^{\mathrm{AF}(1)}/(\gamma\,n^{\,}_{\mathrm{imp}})\to 0$
as $R\to \infty$.

Even without any repulsive short-range correlations between impurity
locations, one expects Ising ferromagnetism to prevail at sufficiently low
impurity densities. This is because rare impurities with a neighboring
impurity much closer than $n^{-1/3}_{\mathrm{imp}}$ should effectively be
taken out of the calculation for the average energy. Indeed, if the
close pair is antiferromagnetically coupled, it will anti-align,
have no net moment and thus
essentially decouples from the global ordering pattern. If instead the
pair is ferromagnetically coupled, it forms a bigger spin that can then
be incorporated in the consideration like any other typical spin. The
net effect of treating such close pairs in this way boils down to
considering only original or effective spins with pairwise separations
of the order
of $R^{\,}_{\mathrm{eff}} \gtrsim c\, n^{-1/3}_{\mathrm{imp}}$ with some constant
$c$ of order 1. The competition for the global ordering pattern then becomes
essentially identical to the one of the constrained superlattice
above, with $R^{\,}_{\mathrm{eff}}$ now taking the role of the exclusion
radius in Eq.~(\ref{eq: Eperimp_dis_R}).
From these considerations we predict that for sufficiently
dilute concentrations $n^{\,}_{\mathrm{imp}}\lesssim (c/R)^{3}_{\mathrm{c}}$
the Ising ferromagnetic order prevails.

\section{Finite-temperature transition to the spiral phase}
\label{sec:finiteT}

The effective Ising model (\ref{eq: final effective Ising Hamiltonian})
undergoes ordering at a critical temperature
$T^{\,}_{\mathrm{Ising}}\propto n^{\,}_{\mathrm{imp}}$%
~\footnote{
One might worry that
a critical temperature
$T^{\,}_{\mathrm{Ising}}$
for the effective Ising model
(\ref{eq: final effective Ising Hamiltonian})
is not well defined in the thermodynamic limit in view of the
long-range nature of the kernel
(\ref{eq: final effective Ising Hamiltonian c}). In particular, $T^{\,}_{\mathrm{Ising}}$ might depend on the
aspect ratio of the lattice $\Lambda$ as $|\Lambda|\to\infty$.
We argue that this is not the case as follows.
Since the Hamiltonian
(\ref{eq: def 3d XY model})
has only short-range magnetic interactions,
any ordering temperature that it supports is well-defined (independent of how
the limit $|\Lambda|\to\infty$ is taken) and of order unity, as guaranteed by Griffith's theorem~\cite{Griffith1968}.
We then first rescale the coordinate axes, and then take any reference shape for which a single ferromagnetic domain
is expected (i.e., a prolate rather than a needle-like sample),
so that we can safely assume a single global spiral to emerge. 
We then integrate out the spin waves. In this way, we eventually end up with the global energy scale
$\kappa\,n^{\,}_{\mathrm{imp}}$
multiplying a dimensionless Hamiltonian with unit density of impurity sites,
as is done to obtain Hamiltonians
(\ref{eq: dipolar approximation final effective Ising Hamiltonian}
and \ref{rescHam}),
the critical temperature of which serves as a reference
for all anti-dipolar systems.
          }.
As long as $T^{\,}_{\mathrm{Ising}}$ lies in the range of low temperatures
(\ref{eq: range T's for Ising approximation}), the Ising approximation
is well justified. This is certainly the case for
$n^{\,}_{\mathrm{imp}}\ll 1$. Since the reduction to the Ising model
neglects some fluctuations, we expect $T^{\,}_{\mathrm{Ising}}$ to be
an upper bound to the actual spiral transition temperature
$T^{\,}_{\mathrm{spi}}$. However, the bound should become increasingly
tight as the impurity concentration decreases towards
$n^{\,}_{\mathrm{imp}}\to 0$.

\subsection{Mean-field theory}

We first estimate $T^{\,}_{\mathrm{Ising}}$ using mean-field
theory, which should work well as three-dimensional space is
the upper critical dimension for the Ising model with dipolar
interactions%
~\cite{Larkin1969,CohenPR55,Aharony73I,Aharony73II,Aharony73III}.
However, we will focus on the case of randomly distributed impurities,
where the mean field actually depends on the site that is considered.
This will require a number of additional approximations. In the next
subsection, we will follow an alternative and complementary approach
that instead makes use of the dipolar nature of the interactions and
exploits their covariance under spatial rescalings. This allows us to
predict how $T^{\,}_{\mathrm{Ising}}$ depends on the couplings,
without resorting to a mean field approximation.

To implement a mean field treatment,
we replace the Ising Hamiltonian
(\ref{eq: final effective Ising Hamiltonian})
by the mean-field Hamiltonian
\begin{subequations}
\label{eq: mean-field equations}
\begin{equation}
H^{\mathrm{MF}}_{\mathcal{L}}\:=
-
\sum_{\tilde{\vv{r}}\in\mathcal{L}}
B^{\mathrm{MF}}_{\tilde{\vv{r}}}\,\sigma^{\,}_{\tilde{\vv{r}}},
\label{eq: mean-field equations a}
\end{equation}
where the effects on  $\sigma^{\,}_{\tilde{\vv{r}}}$ from
all the Ising spins $\sigma^{\,}_{\tilde{\vv{r}}'}$
is approximately captured by the mean magnetic field 
\begin{equation}
B^{\mathrm{MF}}_{\tilde{\vv{r}}}\:=
\sum_{
\tilde{\vv{r}}'\in\mathcal{L}\setminus\{\tilde{\vv{r}}\}
     }
J^{(\mathrm{I})}_{\tilde{\vv{r}}-\tilde{\vv{r}}'}\,
\langle\sigma^{\,}_{\tilde{\vv{r}}'}\rangle^{\,}_{\mathrm{MF}}.
\label{eq: mean-field equations b}
\end{equation}
The mean-field magnetic moments
$\langle\sigma^{\,}_{\tilde{\vv{r}}}\rangle^{\,}_{\mathrm{MF}}$ are
subject to the non-linear constraint
{($k^{\,}_{\mathrm{B}}=1$)}
\begin{equation}
\langle\sigma^{\,}_{\tilde{\vv{r}}}\rangle^{\,}_{\mathrm{MF}}=
\tanh
\left(
\frac{1}{T}
\sum_{\tilde{\vv{r}}'\in\mathcal{L}\setminus\{\tilde{\vv{r}}\}}
J^{(\mathrm{I})}_{\tilde{\vv{r}}-\tilde{\vv{r}}'}\,
\langle\sigma^{\,}_{\tilde{\vv{r}}'}\rangle^{\,}_{\mathrm{MF}}
\right).
\label{eq: mean-field equations c}
\end{equation}
\end{subequations}

The mean-field transition temperature is obtained in two steps.
First, we linearize the constraint (\ref{eq: mean-field equations c}),
assuming a small order parameter
\begin{equation}
\langle\sigma^{\,}_{\tilde{\vv{r}}}\rangle^{\,}_{\mathrm{MF}}=
\frac{1}{T}
\sum_{\tilde{\vv{r}}'\in\mathcal{L}\setminus\{\tilde{\vv{r}}\}}
J^{(\mathrm{I})}_{\tilde{\vv{r}}-\tilde{\vv{r}}'}\,
\langle\sigma^{\,}_{\tilde{\vv{r}}'}\rangle^{\,}_{\mathrm{MF}}.
\label{eq: step 1 MF T Ising}
\end{equation}
If translation symmetry held and
$\langle\sigma^{\,}_{\tilde{\vv{r}}}\rangle^{\,}_{\mathrm{MF}}$
were independent of $\tilde{\vv{r}}$, the critical temperature
\begin{equation}
T^{\mathrm{MF}}_{\mathrm{Ising}}=
\sum_{
\tilde{\vv{r}}\in\mathcal{L}\setminus\{\vv{0}\}
     }
J^{(\mathrm{I})}_{\tilde{\vv{r}}}
\end{equation}
would follow. However, translation symmetry breaks down
when the impurity bonds are distributed randomly,
in which case we estimate the critical temperature by the disorder average
\begin{equation}
T^{\mathrm{MF,av}}_{\mathrm{Ising}}\:=
\frac{1}{|\mathcal{L}|}
\sum_{
\quad
\tilde{\vv{r}}\in\mathcal{L}
\vphantom{\in\mathcal{L}\setminus\{\tilde{\vv{r}}\}}
\quad
     }
\sum_{
\tilde{\vv{r}}'\in\mathcal{L}\setminus\{\vv{r}\}
     }
J^{(\mathrm{I})}_{\tilde{\vv{r}}-\tilde{\vv{r}}'}.
\label{eq: def T MF,av Ising} 
\end{equation}

Substituting the definition of $J^{(\mathrm{I})}_{\tilde{\vv{r}}-\tilde{\vv{r}}'}$
in Eq.~(\ref{eq: def T MF,av Ising}) yields
\begin{subequations}
\begin{align}
T^{\mathrm{MF,av}}_{\mathrm{Ising}}=&\,
\frac{\gamma}{|\mathcal{L}|}
\sum_{
\quad
\tilde{\vv{r}}\in\mathcal{L}
\vphantom{\in\mathcal{L}\setminus\{\vv{r}\}}
\quad
     }
\sum_{
\tilde{\vv{r}}'\in\mathcal{L}\setminus\{\tilde{\vv{r}}\}
     }
\left(
\Gamma^{(0)}_{\tilde{\vv{r}}-\tilde{\vv{r}}^{\prime}}\,
+
\frac{1}{J^{\,}_{\perp}\,|\Lambda|}
\right),\nonumber
\\
=&\,
\gamma\,
n^{\,}_{\mathrm{imp}}
\left[
\sum_{
{\vv{r}}\in\Lambda\setminus\{\vv{0}\}
     }
\Gamma^{(0)}_{{\vv{r}}}\,
+
\frac{1}{J^{\,}_{\perp}}
+
\mathcal{O}\left(\frac{1}{|\Lambda|}\right)
\right],
\nonumber\\
=&\,
\gamma\,
n^{\,}_{\mathrm{imp}}
\left[
-\Gamma^{(0)}_{{\vv{r}}={\vv{0}}}\,
+
\frac{1}{J^{\,}_{\perp}}
+
\mathcal{O}\left(\frac{1}{|\Lambda|}\right)
\right] \nonumber\\
=&\, 
n^{\,}_{\mathrm{imp}}\, (\Delta\theta)^{2}\,
\frac{J^{\,}_{c}}{J^{\,}_{\perp}}(J^{\,}_{c}+J^{\,}_{\perp})
+ 
\mathcal{O}\left(\frac{1}{|\Lambda|}\right).
\label{eq: def T MF,av Ising bis}
\end{align}
\end{subequations}
To reach the second equality,
we have used the fact that in a random, uncorrelated
set of points $\mathcal{L}$ the distance vectors
$\tilde{\vv{r}}-\tilde{\vv{r}}^{\prime}$ appear with the same relative
frequency as in the translationally invariant host lattice
$\Lambda$. More precisely, we used 
\begin{equation}
\frac{1}{|\mathcal{L}|}
\sum_{
\quad
\tilde{\vv{r}}\in\mathcal{L}
\vphantom{\in\mathcal{L}\setminus\{\vv{r}\}}
\quad
     }
\sum_{
\tilde{\vv{r}}'\in\mathcal{L}\setminus\{\tilde{\vv{r}}\}
     } (\cdots) = 
\frac{ n^{\,}_{\mathrm{imp}} }{|\Lambda|}
\sum_{
\quad
{\vv{r}}\in\Lambda
\vphantom{\in\Lambda\setminus\{\vv{r}\}}
\quad
     }
\sum_{
{\vv{r}}'\in\Lambda\setminus\{{\vv{r}}\}
     } (\cdots).
\end{equation}
To reach the third equality,
we have used that
\begin{equation}
\sum_{{\vv{r}}\in\Lambda}
\Gamma^{(0)}_{{\vv{r}}}=
\hat{\Gamma}^{(0)}_{{\vv{k}}=\vv{0}}=0.
\end{equation}
The fourth equality follows from the relations
\begin{equation}
\gamma\:=
(\Delta\theta)^{2}\,
\left(J^{\,}_{\mathrm{c}}+J^{\,}_{\perp}\right)^{2}, 
\qquad
\frac{1}{J^{\,}_{\perp}}-\Gamma^{(0)}_{{\vv{r}}={\vv{0}}}=
\frac{J^{\,}_{\mathrm{c}}/J^{\,}_{\perp}}{J^{\,}_{\mathrm{c}}+J^{\,}_{\perp}},
\qquad
\label{gamma2}
\end{equation}
[see Eq.~(\ref{eq: final saddle point for one impurity b})
for $\Gamma^{(0)}_{{\vv{r}=\vv{0}}}$].

Next we compare the transition temperature $T^{\mathrm{MF,av}}_{\mathrm{Ising}}$ with  
the absolute value of the spiral twist rate at zero temperature
\begin{equation}
|Q|=
\Delta\theta\,\frac{
J^{\,}_{\mathrm{c}}+J^{\,}_{\perp}}{J^{\,}_{\perp}}\,
\frac{1}{|\Lambda|}
\sum_{
\tilde{\vv{r}}\in\mathcal{L}
     }
\sigma^{\,}_{\tilde{\vv{r}}}
\leq
\Delta\theta\, \frac{
J^{\,}_{\mathrm{c}}+J^{\,}_{\perp}}{J^{\,}_{\perp}}\,n^{\,}_{\mathrm{imp}},
\end{equation}
as follows from Eq.~(\ref{Qsaddle}). Equality holds when the
canting degrees of freedom, $\sigma^{\,}_{\tilde{\vv{r}}}$ order
ferromagnetically. In that case we find that both the transition
temperature and the twist rate of the spiral are proportional to the
impurity concentration $n^{\,}_{\mathrm{imp}}$, with a ratio
\begin{equation}
\frac{T^{\mathrm{MF,av}}_{\mathrm{Ising}}}{|Q|}=
\Delta\theta\, J^{\,}_{\mathrm{c}}.
\label{ratioTQ}
\end{equation}
Note that this ratio is \textit{independent} of
$n^{\,}_{\mathrm{imp}}$. It only depends on the coupling strengths
$J^{\,}_{\perp}$, $J^{\,}_{\parallel}$ and $J^{\,}_{\mathrm{imp}}$ via
$ J^{\,}_{\mathrm{c}}$
[recall Eq.~(\ref{eq: Heff_1imp relative angle only b})]
and $\Delta\theta$
[recall Eq.~(\ref{eq: saddle point b})].
In experiments, this ratio can be measured without knowing
the density of impurity bonds~\cite{ShangMedarde2018}.

\subsection{Dipolar approximation}

The mean-field theory of the previous section has at least two
drawbacks. As usual, the neglect of fluctuations will lead to an
overestimate of the critical temperature by a certain factor $\mathcal{O}(1)$,
which might itself be a function of the ratios between the
couplings. This makes it difficult to predict the precise dependence
of $T^{\,}_{\mathrm{Ising}}$ on the couplings. A second and more serious
drawback of these approximations is the fact that the site-averaged mean
field of Eq.~(\ref{eq: def T MF,av Ising}) receives rare, but large
contributions from pairs of sites that are nearest neighbors on the
underlying lattice $\Lambda$. This contribution represents a nonvanishing
fraction of the resulting mean field. However, physically it is clear
that the Ising spins on very close pairs of sites will
lock strongly together and act either as an effective spin
with a doubled moment for ferromagnetic pairs,
or they essentially decouple from the rest for antiferromagnetic
coupling. In either case, these strong short-range couplings have
essentially no influence on the long-range ordering, and thus it seems
unphysical that such strong couplings should enter
in our mean-field estimate of $T^{\,}_{\mathrm{Ising}}$ at all.

Here, we follow a different approach to establish the dependence of
$T^{\,}_{\mathrm{Ising}}$ on the couplings.
Let $\xi$ be the length scale beyond which
we can approximate the interactions
$J^{(\mathrm{I})}_{\tilde{\vv{r}}-\tilde{\vv{r}}'}$
as being anti-dipolar, i.e., given by
Eq.~(\ref{dipolarint}).
We assume that we can safely neglect
pairs of Ising spins that are within a distance
of order $\xi$ of each other
[the probability to find another Ising spin a distance $\xi$
from a given one is of order $\mathcal{O}(\xi^{d}\,n^{\,}_{\mathrm{imp}})$,
a negligible probability as $n^{\,}_{\mathrm{imp}}\to0$].
If so, we may replace the Ising Hamiltonian
(\ref{eq: final effective Ising Hamiltonian})
with the effective Ising Hamiltonian given by
\begin{subequations}
\label{eq: dipolar approximation final effective Ising Hamiltonian}
\begin{equation}
\begin{split}
H^{(\mathrm{eff})}_{\mathcal{L}}[\sigma^{\,}_{\tilde{\vv{r}}}]\:=&\,
-
\frac{1}{2}
\sum_{\tilde{\vv{r}},\tilde{\vv{r}}'\in\mathcal{L}}
\sigma^{\,}_{\tilde{\vv{r}}}\,
J^{(\mathrm{adip})}_{\tilde{\vv{r}}-\tilde{\vv{r}}'}\,
\sigma^{\,}_{\tilde{\vv{r}}'}
\\
&\,
-
\kappa\,n^{\,}_{\mathrm{imp}}\,\frac{1}{|\mathcal{L}|}
\sum_{\tilde{\vv{r}},\tilde{\vv{r}}'\in\mathcal{L}}
\sigma^{\,}_{\tilde{\vv{r}}}\,
\sigma^{\,}_{\tilde{\vv{r}}'}.
\label{eq: dipolar approximation final effective Ising Hamiltonian a}
\end{split}
\end{equation}
The  parameter 
\begin{equation}
\kappa\:=
\frac{1}{2}\,
\left(
\Delta\theta\,
\frac{J^{\,}_{\mathrm{c}}+J^{\,}_{\perp}}{J^{\,}_{\perp}}
\right)^{2}\,
J^{\,}_{\perp}
\label{eq: dipolar approximation final effective Ising Hamiltonian e}
\end{equation}
determines the characteristic energy due to the coupling to the spiral twist
(the infinite-range contribution to the Hamiltonian).
The anti-dipolar interaction is
\begin{equation}
J^{(\mathrm{adip})}_{\tilde{\vv{r}}}\:=
\frac{J^{\,}_{0}}{2\pi}\,
\frac{
\tilde{r}^{2}_{x}+\tilde{r}^{2}_{y}-2\alpha\,\tilde{r}^{2}_{z}
     }
     {
\left(\tilde{r}^{2}_{x}+\tilde{r}^{2}_{y}+\alpha\,\tilde{r}^{2}_{z}\right)^{5/2}
     },
\label{eq: dipolar approximation final effective Ising Hamiltonian b}
\end{equation}
with the anisotropy of exchange couplings,
$\alpha\equiv J^{\,}_{\parallel}/J^{\,}_{\perp}$
[recall Eq.~(\ref{eq: def alpha})],
and the prefactor
\begin{equation}
J^{\,}_{0}\:=
\frac{1}{2}\,
\sqrt{\frac{J^{\,}_{\parallel}}{J^{\,}_{\perp}}}\,
\left(
\Delta\theta\,\frac{J^{\,}_{\mathrm{c}}+J^{\,}_{\perp}}{J^{\,}_{\perp}}
\right)^{2}\,J^{\,}_{\perp}=
\sqrt{\alpha}\,\kappa.
\label{eq: dipolar approximation final effective Ising Hamiltonian c}
\end{equation}
\end{subequations}

We expect that, for the purpose of
determining the critical temperature,
replacing Hamiltonian
(\ref{eq: final effective Ising Hamiltonian})
with Hamiltonian
(\ref{eq: dipolar approximation final effective Ising Hamiltonian})
is an excellent approximation.

We now claim that the critical temperature
$T^{\,}_{\mathrm{Ising}}$
is well approximated by
\begin{equation}
\label{Tcfromrescaling}
T^{\,}_{\mathrm{Ising}}\approx
c\,\kappa\,n^{\,}_{\mathrm{imp}}
+
\mathcal{O}(n^{\,}_{\mathrm{imp}})
\end{equation}
with $c$ a number of order $\mathcal{O}(1)$, independent of
$J^{\,}_{\perp}$, $J^{\,}_{\parallel}$, and $\Delta\theta$.
Indeed, by assumption
$T^{\,}_{\mathrm{Ising}}$
is well approximated by the critical temperature of
the Hamiltonian
(\ref{eq: dipolar approximation final effective Ising Hamiltonian}).
Now, we may trade the scaling transformation
(\ref{xyz_with_bars})
for the scaling transformation
\begin{equation}
\begin{pmatrix}
\tilde{r}^{\,}_{x}
\\
\tilde{r}^{\,}_{y}
\\
\sqrt{\alpha}\,\tilde{r}^{\,}_{z}
\end{pmatrix}\=:
\left(\frac{\sqrt{\alpha}}{n^{\,}_{\mathrm{imp}}}\right)^{1/3}\,
\begin{pmatrix}
\mathsf{r}^{\,}_{x}
\\
\mathsf{r}^{\,}_{y}
\\
\mathsf{r}^{\,}_{z}
\end{pmatrix},
\label{rescaling}
\end{equation}
which preserves the Poissonian nature of the impurity distribution
and is equivalent to replacing
$J^{\,}_{0}$ by
$J^{\,}_{0}\,n^{\,}_{\mathrm{imp}}/\sqrt{\alpha}= \kappa\,n^{\,}_{\mathrm{imp}}$ in
Eq.~(\ref{eq: dipolar approximation final effective Ising Hamiltonian}).
After this rescaling, we can factorize out the common energy scale
$\kappa\,n^{\,}_{\mathrm{imp}}$
from both the dipolar and the spiral twist contributions
to the Hamiltonian. The putative ordering temperature is then encoded
in the dimensionless Hamiltonian
\begin{equation}
\label{rescHam}
\frac{1}{2}
\sum_{\mathsf{r},\mathsf{r}'\in\mathcal{L}}
\sigma^{\,}_{\mathsf{r}}\,
\left[
-
\frac{1}{2\pi}
\frac{
\mathsf{r}^{2}_{x}+\mathsf{r}^{2}_{y}-2\mathsf{r}^{2}_{z}
     }
     {
\left(\mathsf{r}^{2}_{x}+\mathsf{r}^{2}_{y}+\mathsf{r}^{2}_{z}\right)^{5/2}
     }
-
\frac{2}{|\mathcal{L}|} 
\right]
\sigma^{\,}_{\mathsf{r}'}
\end{equation}
which has a dimensionless ordering temperature $c=\mathcal{O}(1)$,
and in turn confirms our claim in Eq.~(\ref{Tcfromrescaling}).
Numerical simulations of the anti-dipolar Ising model yield an estimate
of the dimensionless prefactor to be $c\approx1.5$%
~\footnote{
We caution that rather large system sizes are necessary
to reach the thermodynamic limit, as was already observed in
Ref.~\onlinecite{Scaramucci_2016},
where the simulated system sizes for the full $XY$ model were insufficient
to reach the thermodynamic limit. Indeed, the apparent finite-size transition
temperature exhibited a very significant size dependence. Here, we have
directly simulated the effective Ising model. While we reproduced the results
of the full model for small samples,
we were now able to reach much bigger sizes,
where the transition temperature was found to saturate eventually,
as expected.
That saturation value was taken to estimate the value of $c$.
          }. 

The above prediction for $T^{\,}_{\mathrm{Ising}}$ differs from the
mean-field theory result (\ref{eq: def T MF,av Ising bis}) by a factor
of $(J^{\,}_\perp+J^{\,}_{\mathrm{c}})/J^{\,}_{\mathrm{c}}$ and
additional numerical factors that in the case of mean-field theory,
might depend on the ratio of couplings. The deviation between the two
approaches traces back to the various approximations made in
the mean-field theory.

From the result (\ref{Tcfromrescaling}) we deduce that the ratio of
the transition temperature to the spiral twist rate has the following
dependence 
\begin{equation}
\frac{T^{\,}_{\mathrm{Ising}}}{|Q|}=
\frac{c}{2}\Delta\theta\,\left(J^{\,}_{\mathrm{c}}+ J^{\,}_\perp\right)
\label{ratioTQ_resc}
\end{equation}
on the exchange couplings,
where we recall that $\Delta\theta$ depends on all couplings
$J^{\,}_\perp$, $J^{\,}_\parallel$, and $J^{\,}_{\rm imp}$
through the solution of Eq.~(\ref{eq: saddle point b}).

\subsection{Comparison to simulations in $XY$ model and to experiments}

We can now compare our theoretical predictions with experimental findings.
Reference \onlinecite{MorinNatComm2016} reports a ratio%
~\footnote{
References~\onlinecite{MorinNatComm2016,ShangMedarde2018}
use a different convention for the spiral wavevector.
Their wavevector $q^{\,}_{G}$ is related to our
$Q$ via the conversion $Q=\pi\,q^{\,}_{G}$.
          }
$T^{\,}_{\mathrm{spi}}/|Q|\approx 60\,\mathrm{meV}$ in YBaCuFeO$_5$
while our theory predicts
$T^{\,}_{\mathrm{Ising}}/|Q|\approx\,68\,\mathrm{meV}$
in the limit of low impurity density, upon using the couplings given
in Eq.~(\ref{values of J in YBaCuFeO5}), see Fig.\ \ref{Fig:expth}.
It is encouraging that our theory overestimates
$T^{\,}_{\mathrm{spi}}/|Q|$ only by $\approx 13\%$,
considering the simplifications that go into the modelling of the spin system
and the uncertainty in the value of the exchange couplings.
As noted above, the ordering temperature is proportional to
the concentration of impurity bonds.
For the concrete case of YBaCuFeO$_5$, our theory predicts
that a small fraction of $1\%$ of the oxygen bipyramids realizing
the strongly frustrating Fe-Fe magnetic interactions induces a transition
to the spiral phase at an ordering temperature
of approximately $85\,\mathrm{K}$,
see Fig.\ \ref{Fig:expth}.
Note that at those temperatures the constraint of Eq.%
~(\ref{eq: range T's for Ising approximation})
is satisfied by a large margin and the mapping to the effective Ising model
is thus well controlled.

A subsequent experimental study investigated a family of chemically
modified compounds related to YBaCuFeO$_5$~\cite{ShangMedarde2018}.
In those materials the lattice parameters could be altered, which
affects the exchange constants, and the concomitant changes to
observables such as $T^{\,}_{\mathrm{spi}}/|Q|$ were recorded.
Examples of such modifications are the application of uniaxial
pressure, or chemical substitution that replaces the atoms between the
layers containing the impurity bonds. The latter modifies the
interlayer spacing and thus the perpendicular coupling
$J^{\,}_{\perp}$. The experiments of
Ref.~\onlinecite{ShangMedarde2018} shows that the ratio
$T^{\,}_{\mathrm{spi}}/|Q|$ is only very weakly sensitive to the
modification of the interlayer spacing and thus $J^{\,}_{\perp}$.

These empirical findings can be rationalized by analyzing
Eqs.~(\ref{ratioTQ}) and (\ref{ratioTQ_resc}).  In layered
materials such as YBaCuFeO$_5$, the exchange anisotropy between
intra- and inter-layer couplings is large.
We model this empirical fact by requiring that
$\alpha\equiv J^{\,}_{\parallel}/J^{\,}_{\perp}\gg 1$
[recall Eq.~(\ref{eq: def alpha})].
Furthermore, the
impurity coupling strength turns out to be large as well,
$|J^{\,}_{\mathrm{imp}}|/J^{\,}_{\parallel}\gg1$
[recall Eq.~(\ref{values of J in YBaCuFeO5})].
In this limit,
$J^{\,}_{\mathrm{c}}\approx2\pi J^{\,}_{\parallel}
/(\ln\alpha+2.47)\gg J^{\,}_{\perp}$
[recall the approximation mentioned
in the caption of
Fig.~\ref{fig: critical Jc for single impurity}~].
The canting angle between the $XY$ spins on either
end of an impurity bond comes close to $\Delta\theta\approx\pi$
[recall Eq.~(\ref{eq: saddle point b})]. More precisely,
the deviation from $\pi$ is
\begin{equation}
\pi-\Delta\theta\approx
2\pi^2
\frac{
J^{\,}_{\parallel}
     }
     {
J^{\,}_{\mathrm{imp}}
\left[\ln\left(J^{\,}_\parallel/J^{\,}_\perp\right) + 2.47\right]
     }.
\end{equation}
After dropping this correction, to a first approximation, the ratio between
the critical temperature and the spiral twist rate at zero temperature
can be approximated by
\begin{align}
\frac{T^{\,}_{\mathrm{Ising}}}{|Q|}\approx\,
\frac{
c\,\pi^{2} \,J^{\,}_{\parallel}
     }
     {
\ln\left(J^{\,}_{\parallel}/J^{\,}_{\perp}\right)
+
2.47
     },
\end{align}
with the constant $c\approx1.5$.

The degree to which the ratio $T^{\,}_{\mathrm{Ising}}/|Q|$
depends on the coupling $J^{\,}_{\perp}$
can be quantified by the logarithmic derivative
\begin{equation}
\label{logderiv}
\frac{
\partial\ln(T^{\,}_{\mathrm{Ising}}/|Q|)
     }
     {
\partial\ln J^{\,}_{\perp}
     }
\approx
\frac{1}{\ln\left(J^{\,}_\parallel/J^{\,}_\perp\right)+2.47}.
\end{equation}
For large anisotropy $\alpha$, this becomes small. For the
experimental values of Eq.~(\ref{values of J in YBaCuFeO5}), the
logarithmic derivative of Eq.~(\ref{logderiv}) evaluates to
approximately $0.2$, implying that a $50\%$-change in
$J^{\,}_{\perp}$ only results in a $10\%$-change of the ratio
$T^{\,}_{\mathrm{Ising}}/|Q|$, in qualitative agreement with the
experimental observations in Ref.~\onlinecite{ShangMedarde2018}.

\section{Conclusion and outlook}
\label{Sec:Conclusions}

Any three-dimensional lattice hosting $XY$ spins that 
interact through ferromagnetic nearest-neighbor exchange interactions
display a ferromagnetic long-range order below some critical temperature.
We have given sufficient conditions under which the replacement of a dilute
fraction of the ferromagnetic bonds by antiferromagnetic bonds destabilizes
the ferromagnetic order in favor of non-collinear long-range
order in the form of a spiral phase. 
A necessary but not sufficient condition for spiral order
is that the  antiferromagnetic  exchanges along the impurity bonds  
be sufficiently larger than the ferromagnetic couplings.
{This induces local canting,
which lowers the energy close to the frustrating bond}.
If this condition is met, 
a sufficient condition for spiral order is a strong correlation between the
impurity bonds such that 
(i) they all point along a preferred direction
and
(ii) they are distributed in space such that ferromagnetic
interactions dominate between the Ising degrees of freedom associated
with the local canting patterns around the impurities.  We showed
rigorously that (ii) is satisfied for impurities located on Bravais
superlattices whose shortest lattice vectors tend to point in directions in
which the effective Ising interactions are ferromagnetic, while
neighboring impurities along the $z$-axis, for which the interactions
are antiferromagnetic, appear only at larger distance. Small
distortions away from a perfectly regular Bravais lattice will not destroy
the spiral order. We also argued that completely randomly distributed
impurities are prone to stabilize spiral order at low enough impurity
density. At higher impurity densities, a short-ranged repulsion among
impurity bonds, e.g., due to Coulomb constraints in real materials, has
the main effect of reducing the stability of fan states (layered
antiferromagnetic orderings of the canting degrees of freedom), and
thus also stabilizes spiral order.  Hence, once the orientational
correlation (i) is ensured, the tendency towards spiral order is
rather strong.

On the other hand, if the impurity bonds 
\textit{and} their orientations are white-noise correlated in space, 
the microscopic $XY$ Hamiltonian belongs to the family
of three-dimensional $XY$ gauge glasses introduced by Villain.
Those host amorphous, glassy order. 
From this it follows that the zero-temperature phase diagram of
two-dimensional $XY$ magnets (as characterized by the strength of the
frustrating antiferromagnetic interactions and their spatial
correlations) contains at least four stable phases: The ferromagnetic
phase, the spiral phase, the fan phase (i.e., ferromagnetic in plane
order with  orientation oscillating from plane to plane), and the gauge
glass phase.

From the perspective of the original microscopic $XY$ spins in
Hamiltonian (\ref{eq: def 3d XY model}), the phenomenology for small
concentrations $n^{\,}_{\mathrm{imp}}\ll1$ is the following. Upon
lowering the temperature in the $XY$ paramagnetic phase, a continuous
phase transition takes place in the three-dimensional $XY$
universality class to a ferromagnetic phase at the temperature
$T^{\,}_{XY}$. This ferromagnetic phase becomes further unstable at
the temperature $T^{\,}_{\mathrm{spi}}\ll T^{\,}_{XY}$ (as estimated
by $T^{\,}_{\mathrm{Ising}}$ in Eq.~(\ref{Tcfromrescaling})), where
a $XY$ spiral phase emerges via a continuous phase transition. It is
driven by the dilute concentration $n^{\,}_{\mathrm{imp}}\ll1$ of impurity
bonds that are orientationally correlated. The spiral wavevector
$Q$
may serve as an order parameter for this Ising
transition. The associated critical exponents are expected to take
mean-field values, given the dimensionality and the long-range nature
of the dipolar interactions.

What happens as $n^{\,}_{\mathrm{imp}}$ is increased, so that
$T^{\mathrm{MF}}_{\mathrm{spi}}\sim T^{\,}_{XY}$?  In this limit, the
effective Ising model (\ref{eq: final effective Ising Hamiltonian}) is
not a valid approximation of Hamiltonian (\ref{eq: def 3d XY model})
anymore, so that at this stage we cannot make controlled predictions.
However, it seems very likely that at large enough
$n^{\,}_{\mathrm{imp}}\lesssim1$, the impurity bonds will dominate the coupling
between adjacent $a,b$-planes, inducing a layered antiferromagnetic
state. Upon increasing $n^{\,}_{\mathrm{imp}}$ this state might be reached
either continuously, with the spiral wavevector saturating at
 $Q=\pi$,
or it appears discontinuously,
via a first order transition at some critical value of $n^{\,}_{\mathrm{imp}}$.
In the regime of smaller $n^{\,}_{\mathrm{imp}}\lesssim1$, the critical
temperature of the ferromagnetic instability will decrease with
increasing $n^{\,}_{\mathrm{imp}}$, while the spiral instability
temperature is expected to continue to increase. They might merge into
a single direct transition, if this is not
pre-empted by the emergence of a layered antiferromagnetic phase. An
approach that is non-perturbative in $n^{\,}_{\mathrm{imp}}$ is needed to
address these questions.  It remains an open challenge to determine
optimal combinations of exchange couplings that would allow to
maximize $T^{\,}_{\mathrm{spi}}$ by increasing $n^{\,}_{\mathrm{imp}}$, and
thereby extend the regime of the incommensurate spiral magnetic phase
to high temperatures.

In our earlier companion paper Ref.\onlinecite{Scaramucci_2016},
it was argued that YBaCuFeO$_5$ unites all the essential ingredients of
the Hamiltonian discussed in this work, and thus could realize
the spiral $XY$ phase described above. The supporting evidence
is as follows. On the one hand,
Monte Carlo simulations for realistic values
of the magnetic exchange couplings in YBaCuFeO$_5$ 
yield transition temperatures to the magnetic spiral phase
as high as $250$~K. On the other hand,
it was reported in Ref.\ \onlinecite{MorinNatComm2016}
that tuning the degree of occupational disorder
by changing the annealing procedure of YBaCuFeO$_5$ affects
the transition temperature and the wavevector of the spiral
in a way that is qualitatively and quantitatively consistent
with Eq.~(\ref{ratioTQ_resc}).
Finally, we point out that a mechanism very similar to the one
described here might be at work in hole-doped cuprates, where pairs of
holes might take the role of the frustrating impurity
bonds~\cite{Capati}.

\subsection{Applications to other systems}

The main physical mechanism we discussed in this work applies to
other systems as well. First, we point out that the restriction to $XY$ spins
is not essential. Indeed, we expect that Heisenberg spins with an
$O(3)$ symmetry (or any other set of continuous degrees of freedom
undergoing spontaneous symmetry breaking) would exhibit essentially
the same phenomenology: At low temperatures the unfrustrated system
will order ferromagnetically. Frustrating antiferromagnetic impurity
bonds induce local canting patterns that are subject to effective
pairwise interactions upon integrating out spin waves. A ferromagnetic
order of the canting degrees of freedom again imply spiral order for
the original Heisenberg spins.  If the canting induced by a local
impurity bond preserves the coplanarity of the background
ferromagnetic order, the problem simply reduces to an effective XY
model. This is what we found to happen in the presence of nearest
neighbor Heisenberg interactions.  However, with more complex
interactions, it might occur that the local canting pattern is
non-planar.  This would imply that the canting does not only have a
discrete Ising degree of freedom, but rather a continuous $XY-$like
degree of freedom. Indeed, for an isolated impurity, any rotation of
all spins around the direction of the background ferromagnetic
magnetization yields an energetically equivalent canting pattern.
Upon integrating out spin waves, these effective $XY$ canting degrees
of freedom will be coupled through dipole-like interactions, and their
ferromagnetic order will again induce a spiral of the
original Heisenberg spins.

The phenomenology of magnetic $XY$ spins immediately carries over to
superconducting systems, too. There, the role of $XY$ spins is taken by
the phase of superconducting islands with a well established amplitude of the
superconducting order parameter, and Josephson couplings replace the
magnetic exchange couplings. Frustration could be induced by Josephson
couplings with a negative sign (based on ferromagnetic materials for
example). However, a much simpler way to achieve frustration consists
in threading a homogenous magnetic flux through a Josephson junction
array. The recent advances in fabrication techniques and
nanolithography for such devices should allow to artificially design
and control $XY$ systems with a desired spatial pattern of frustrated
plaquettes that emulate the presence of the antiferromagnetic
impurity bonds in the magnetic analogue.  A magnetic spiral phase with
ferromagnetic order of the Ising degrees of freedom of the canting
patterns then translates into a system of vortices of the same vorticity
(sense of circulation), entailing a global supercurrent in the
system. This will be explored in future work.

\begin{acknowledgments}
This research was partially supported by NCCR MARVEL, funded by the
Swiss National Science Foundation. We would like to thank
N. Spaldin, M. Troyer, M. Medarde, M. Kenzelman, and M. Morin for
useful discussions.  H.S. acknowledges support from the DFG via FOR
1346, the SNF Grant 200021E-149122, ERC Advanced Grant SIMCOFE, ERC
Consolidator Grant CORRELMAT (project number 617196).  This work was
supported by JSPS KAKENHI Grant Numbers 16H01064 (J-Physics),
16K17735.
\end{acknowledgments}

\appendix

\section{Degeneracy of all configurations with ferromagnetic order
in the planes}
\label{appsec: Lattice sum}

Let us consider a maximally dense lattice of impurity bonds, i.e.,
$\mathcal{L}=\Lambda$, with $\Lambda$ the host cubic lattice.
By comparing the interaction energies of various candidates for ground
states we will establish that, in the dense limit, an infinite family
of spin configurations are degenerate. These degenerate
configurations are such that the Ising degrees of freedom take values
that depend solely on the $z$ component of their position
$\vv{r}=(x,y,z)$,
\begin{equation}
\sigma^{\,}_{\vv{r}}\equiv s^{\,}_{z}=\pm 1.
\label{appeq: Ansatz xy independent}
\end{equation}
Within any $xy-$plane of the cubic lattice the Ising degrees of
freedom are ferromagnetically ordered, but they are uncorrelated among
different planes.  According to
Eq.~(\ref{eq: final effective Ising Hamiltonian}),
up to a global constant, the total energy per lattice
site of such a configuration is
\begin{align}
\varepsilon[s^{\,}_{z}]=\,
-
\frac{1}{2}
\times
\frac{1}{|\Lambda|}
\sum_{{\vv{r}',\vv{r}''}\in\Lambda}
J^{(\mathrm{I})}_{{\vv{r}'}-{\vv{r}}''}\,
s^{\,}_{z'}\,
s^{\,}_{z''}.
\label{appeq: layered energy a}
\end{align}

\begin{widetext}
We now focus on the interaction $E^{\,}_{z''|z'}$ between two layers with
$z$ coordinates $z'$ and $z''$, respectively.  It is proportional to
$s^{\,}_{z'}\,s^{\,}_{z''}$, with
\begin{align}
-s^{\,}_{z'}\,
s^{\,}_{z''}\,
E^{\,}_{z''|z'}\:=
\frac{1}{2}
\sum_{x',y',x'',y''}
J^{(\mathrm{I})}_{{\vv{r}'}-{\vv{r}}''}=
\sum_{x',y',x'',y''}
\,
\frac{\gamma}{2|\Lambda|}
\sum_{\vv{k}\in\mathrm{BZ}(\Lambda)\setminus\{\vv{0}\}}
\hat\Gamma^{(0)}_{\vv{k}}
e^{\mathrm{i}\vv{k}\cdot({\vv{r}'}-{\vv{r}}'')}
+
\frac{\gamma\,L^{\,}_{x}\,L^{\,}_{y}}{2J^{\,}_{\perp}\,L^{\,}_{z}},
\end{align}
where $L^{\,}_{x}$, $L^{\,}_{y}$ and $L^{\,}_{z}$ are the number of
lattice sites along the $x$, $y$ and $z$-direction, respectively, and
$|\Lambda|=L^{\,}_{x}\times L^{\,}_{y}\times L^{\,}_{z}$.
With the help of [recall Eq.~(\ref{eq: def Gammas and hat Gammas b})]
\begin{align}
\hat\Gamma^{(0)}_{(k^{\,}_{x}=0,k^{\,}_{y}=0,k^{\,}_{z}\neq 0)^{\mathsf{T}}}=
\frac{1}{J^{\,}_{\perp}},
\end{align}
we can perform the sums over $x$- and $y$-coordinates. This sum gives
\begin{align}
-s^{\,}_{z'}\,s^{\,}_{z''}\,E^{\,}_{z''|z'}=
\frac{\gamma\,L^{\,}_{x}\,L^{\,}_{y}}{2\,L^{\,}_{z}}
\sum_{
\substack{
k^{\,}_{z}\\
k^{\,}_{z}\vv{e}^{\,}_{z}\in\mathrm{BZ}(\Lambda)\setminus\{\vv{0}\}
         }
     }
\frac{e^{\mathrm{i}k^{\,}_{z}(z'-z'')}}{J^{\,}_{\perp}}
+
\frac{\gamma\,L^{\,}_{x}\,L^{\,}_{y}}{2J^{\,}_{\perp}\,L^{\,}_{z}}=
\frac{\gamma\,L^{\,}_{x}\,L^{\,}_{y}}{2J^{\,}_{\perp}}\,\delta^{\,}_{z',z''}.
\end{align}
The energy per spin in all configurations of arbitrarily layered,
ferromagnetically ordered planes is thus
$-\gamma/(2J^{\,}_{\perp})$,
independently of the magnetization structure $s^{\,}_{z}$.
\end{widetext}

\bibliography{references}

\end{document}